**Exploring the Potential of Ternary Blending for Two and Three-Junction RAINBOW Solar Cells**

*Francesc Xavier Capella-Guardià, Jolanda Simone Müller, Muhammad Ahsan Saeed, Xabier Rodríguez-Martínez, Miquel Casademont-Viñas, Albert Harillo-Baños, Jaime Martín, Jenny Nelson, Alejandro R. Goñi, Mariano Campoy-Quiles**

F. X. Capella-Guardià, Dr. M. A. Saeed, Dr. M. Casademont-Viñas, Dr. A. Harillo-Baños, Prof. A. R. Goñi, Prof. M. Campoy-Quiles
Institut de Ciència de Materials de Barcelona, ICMAB-CSIC, Campus de la UAB, 08193 Bellaterra, Spain
E-mail: mcampoy@icmab.es

Dr. J. S. Müller, Prof. J. Nelson
Department of Physics and Centre for processable Electronics, Imperial College London, Blackett Laboratory, Prince Consort Road, London SW7 2AZ, United Kingdom

Dr. X. Rodríguez-Martínez, Dr. J. Martín
Universidade da Coruña, Campus Industrial de Ferrol, CITENI, Campus de Esteiro S/N, 15471 Ferrol, A Coruña, Spain

Dr. J. Martín
Oportunius Program, Axencia Galega de Innovacion, Xunta de Galicia, Rua Airas Nunes, S/N, 15702 Santiago de Compostela, Spain

Prof. A. R. Goñi
ICREA, Passeig Lluís Companys 23, 08010 Barcelona, Spain

Funding: PRE2022-104632, PID2021-128924OB-I00, PID2024-163010OB-I00, TED2021-131911B-I00, MSCA-101104491, ERC CoG-101086805, EP/Z533361/1 POTENtIAl.

## Abstract

The efficiency of organic photovoltaics (OPV) has been steadily increasing over the past decade until reaching the 20% milestone. Multijunction architectures provide a promising approach to further enhance performance. Here we explore the potential of a spectral splitting geometry, referred to as RAINBOW, in which subcells are placed side-by-side and externally connected, thus minimizing the fabrication and current matching challenges found in vertically stacked configurations. First, we tested 7 different binaries with bandgaps spanning from 1.98 to 1.16 eV. The systems with the widest and narrowest gaps suffered greater losses and so we evaluate if ternary mixing could help to overcome these limitations by comparing 5 different ternaries. We found that ternary mixing tunes the $V_{oc}$, and when morphology and energy levels are well aligned, the overall PCE can be boosted in the spectral region where the subcell is absorbing, as is the case for PTB7-Th:COTIC-4F:BTP-eC9 when operating as red subcell. Device simulations help to identify the 2-junction and 3-junction configurations with highest PCEs, all of which include ternaries. We fabricate proof-of-concept RAINBOW devices using scalable methods in which the subcells are deposited by meniscus-guided blade coating. The efficiency improves from 12.9% in single-junction devices to 15.9% in 2-junction devices (16.4% in simulations) and 17.3% in 3-junction devices (17.7% in simulations), confirming the viability of the RAINBOW architecture for scalable, high-efficiency OPVs. Finally, detailed balance analysis indicates that the potential of this geometry can be very high provided that high efficiency wide bandgap (2–2.5 eV) materials become available.

## 1. Introduction

Over the last few years, the development of various non-fullerene acceptors (NFAs) has made organic photovoltaics (OPV) a promising option for producing clean energy. Indeed, OPVs have recently exceeded 20% efficiency for single-junction devices.[1–9] Moreover, device fabrication is compatible with industrial scaling,[10–12] its energy payback time is short, and its economic and environmental impact is low.[13,14] Additionally, the different properties of NFAs, due to their bandgaps, make them suitable for a variety of applications, including agrivoltaics,[15,16] underwater PV,[17,18] wearable electronics,[19] and indoor PV.[20,21]

One of the fundamental challenges in photovoltaics is the mismatch between the incident spectrum and the bandgap of the photoactive material, which limits the maximum power conversion efficiency (PCE). Photons with energies below the bandgap are not absorbed, while absorbed photons with energies exceeding the bandgap have part of their energy dissipated, as generated charge carriers thermalize to lower energy levels. Therefore, efficiency in single-junction PV exhibits a theoretical limit under one-sun illumination, as determined by the thermodynamical detailed balance.[22,23] Several strategies have been proposed to overcome these losses, including concentrated PV and up- or down-conversion.[24–27] Yet, the most practical and explored option is multijunction architectures, particularly due to its compatibility with well-established silicon-based PV technologies.[28]

Multijunction devices consist of multiple subcells with different bandgaps each designed to absorb a specific region of the solar spectrum. This way, both thermalization and absorption losses are reduced, leading to an increased theoretical upper limit of efficiency.[27,29,30] In a tandem configuration, the subcells are vertically stacked, with the top cell absorbing high-energy photons, remaining transparent for the low-energy ones, which will be absorbed by the following subcell. This concept has been successfully implemented with various PV technologies, including silicon, perovskites and organic systems.[31–35] Nevertheless, tandem architecture faces significant fabrication challenges, such as current matching between subcells, deposition of interconnective layers, increased device complexity, and limitations for scalability.[36,37] In the specific case of OPV, the reported efficiencies of all-OPV tandems do not exceed those of record single-junction devices .[35,38]

An alternative multijunction strategy called RAINBOW has been proposed, in which the subcells are deposited laterally and connected independently, while an optical element splits the incident light laterally onto the subcells.[39,40] In this configuration, each subcell is illuminated only with a spectral section in between two dividing wavelengths closely matching the main absorption region of every photoactive material. Such materials and their assigned spectral windows are optimized to ensure that the sum of all partial efficiencies is the highest. The RAINBOW concept has previously been proposed and experimentally tested with inorganic-based PV technologies, employing optical elements such as dichroic mirrors, holographic lenses, and more complex, theoretical, optical structures.[41–49]

In the field of OPV, the RAINBOW architecture has been explored through numerical simulations as well as through experimentally measured devices using a custom setup specifically developed to characterize solar cells under the conditions of a split spectrum.[50,51] These studies established the feasibility of the RAINBOW approach with a 2-junction champion device (simulation PCE = 13.7%, experimental PCE = 13.4%) composed of a narrow-bandgap (NBG) subcell based on PTB7-Th:COTIC-4F (single PCE = 6.71%), and a mid-bandgap (MBG) subcell based on PM6:Y6 (single PCE = 11.9%). In addition, the combination of the NBG with a wide-bandgap (WBG) subcell based on PM6:IO-4Cl (single

PCE = 5.75%) produced a smaller, yet considerable, efficiency gain (simulation PCE = 10.1%, experimental PCE = 9.36%).[51]

Peters et al. discussed the dependence on the number of junctions of overall efficiency.[30] They theoretically established that when taking into account optical losses, the RAINBOW efficiency will increase up to a maximum of 6 junctions, covering a bandgap range from 0.9 to 2.5 eV. Part of this wide range can be achieved due to the remarkable variety of OPV materials. However, the availability of highly efficient systems for NBG (< 1.3 eV) and WBG (> 1.7 eV) regimes is still limited compared to much more investigated and highly efficient MBG systems (1.3–1.7 eV).[52–54] By identifying new combinations and improving the performance of existing limited systems, the efficiency in 2-junction RAINBOW devices can be increased and potentially be extended to 3 or more junctions.

OPV devices are typically processed through thin-film deposition of a bulk heterojunction (BHJ), which consists of a binary blend of donor (D) and acceptor (A) materials. A popular strategy to increase efficiency in OPVs is the use of ternary blends, in which a third component, either an additional donor or acceptor, is incorporated into the active layer. The use of ternary blends has emerged as an effective method to broaden absorption coverage, enhance device stability, improve transport properties (thus FF), tune the $V_{oc}$ and ultimately improve power conversion efficiency (PCE). Several factors can be considered regarding the selection of a suitable third component, including backbone similarity, absorption complementarity, and energy level matching. [55–59] Notably, some of the highest efficiency reports on OPV, including tandem structures, rely on this technique to increase the total current due to a synergistic effect on the blend or due to an extended absorption towards other parts of the spectrum.[60,61]

One of the narrowest bandgap NFAs is COTIC-4F (1.1 eV), which in combination with donor PTB7-Th, produces the OPV devices that absorb up to 1100 nm.[62] As a result, they are great candidates for the lowest bandgap subcell in a RAINBOW configuration. Nevertheless, this system suffers from high non-radiative recombination losses and a low open-circuit voltage ($V_{oc}$).[63] The addition of a third material can help improve the $V_{oc}$, along with the overall efficiency.

The objective of this work is to explore the use of ternary blends for NBG and WBG subcells in RAINBOW devices. Hence, blends with different bandgaps are selected according to their performance within specific absorption windows. Device fabrication is carried out through high-throughput blade-coating active layer deposition to determine the optimal thickness for each blend. First, 7 binary systems covering a broad span of bandgaps are identified. For systems exhibiting limited efficiency, ternary blends are explored as a strategy to surpass the performance of established NBG and WBG systems. This approach should simultaneously enhance desired current-voltage (JV) parameters while preserving favorable donor-acceptor interactions and their localized absorption characteristics. With the obtained experimental results, a simulation study is conducted to estimate the achievable efficiencies of 2-junction and 3-junction RAINBOW devices. Then, the simulations are validated with the fabrication of monolithic multijunction devices. Finally, detailed balance is used to evaluate the potential of organic semiconductors in the RAINBOW geometry and suggest avenues to increase performance.

## 2. Results and Discussion

The active layer of OPV devices consists of a blend of donor and acceptor materials and is responsible for absorbing photons and converting them into charge carriers. The effective bandgap of the mix is typically determined by the component that can absorb the lowest-energy photons, generally the acceptor. Proper interaction between these species will lead to effective charge generation and extraction. Both types of materials are defined by two characteristic energy levels, the lowest unoccupied molecular orbital (LUMO) and the highest occupied molecular orbital (HOMO). Excitons generated between them are difficult to dissociate into free charge carriers on their own, thus, the small offset between the energy levels of the donor and the acceptor is used to induce dissociation and drive charge transfer.[64] **Figure 1** displays the molecular structures of the three donors and the nine acceptors used in this study to fabricate devices as well as their HOMO and LUMO energy levels, extracted from cyclic voltammetry (FCC-Cl) or photoelectron spectroscopy (the rest).[65–70]

The general architecture of a RAINBOW device is shown in Figure 1c. High overall efficiency requires each subcell to operate as optimally as possible. Each subcell is denoted by color depending on the relative spectral region it is intended to absorb. In a 2-junction configuration, higher-energy photons are directed to the blue subcell, while lower-energy photons are absorbed by the red subcell. The introduction of a third junction results in a green subcell between the two existing ones. The bandgap limits for each subcell are not strictly defined, therefore, a material operating as a blue cell in a 2-junction device may also serve as a green cell in a 3-junction device.

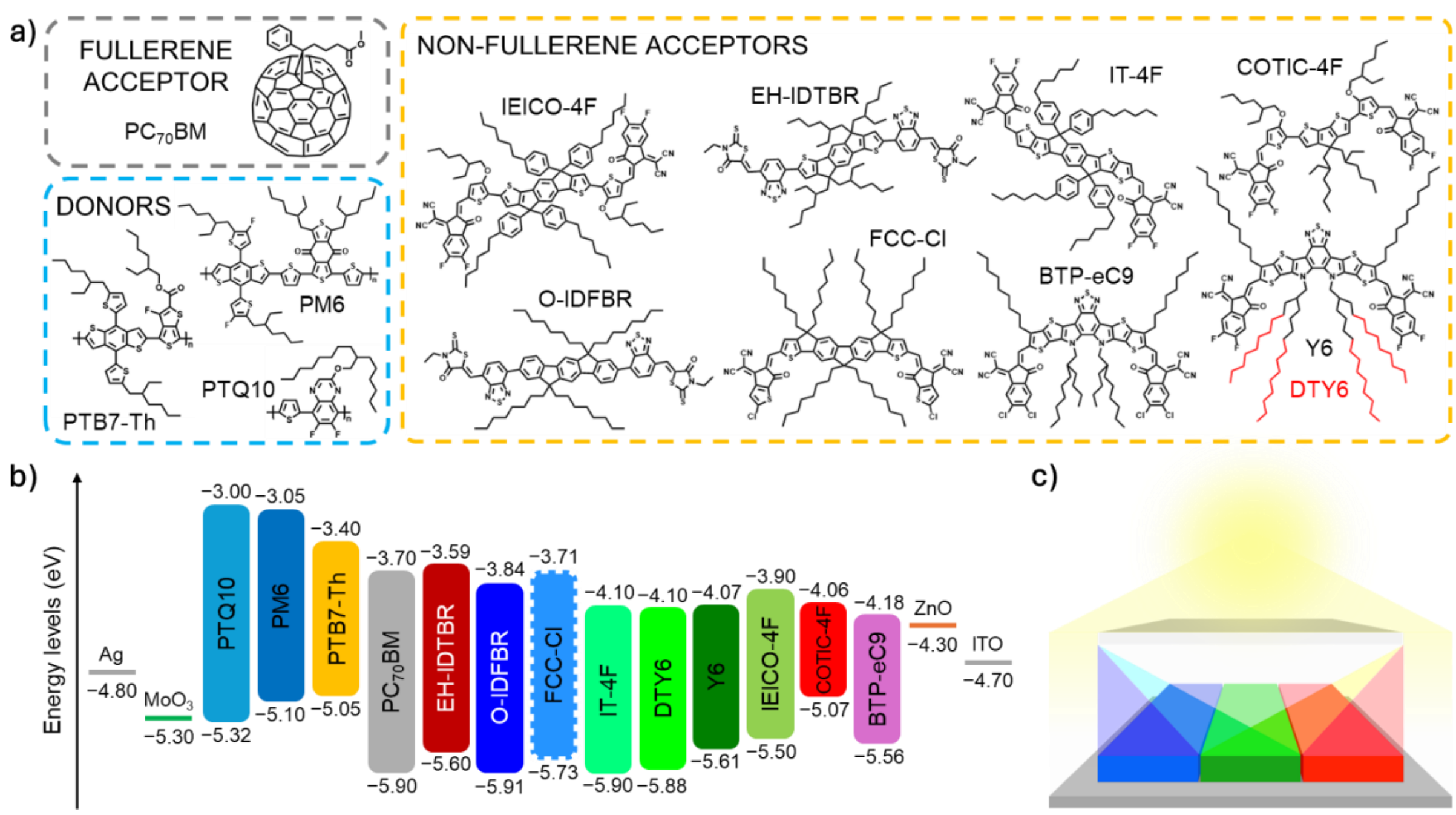

**Figure 1.** a) Chemical structure of the materials used in this study and b) their energy levels;[65–70] c) RAINBOW architecture showing an optical element to divide the incident spectrum laterally over three subcells.

### 2.1. Binary blends

Firstly, the performance and the absorption properties of single-junction binary blend systems are evaluated to identify suitable candidates for each subcell. Devices are fabricated using a thickness gradient high-throughput approach, with each substrate containing 24 devices of 12 different active-layer thicknesses. JV curves are measured under AM1.5G illumination to extract the performance parameters. The results of the champion cell for each blend are summarized in Table 1. External quantum efficiency (EQE) profiles of the champion cells, shown in **Figure 2a**, are also measured to obtain information about the efficiency of charge collection depending on the absorption characteristics of each system.

Among the WBG systems, we use the acceptor O-IDFBR combined with the donor PTQ10, yielding a very large $V_{oc}$ of 1.36 V, one of the highest reported values for a single-junction binary blend after optimization.[71] Although its short-circuit current density ($J_{sc}$) and overall efficiency are relatively low under the solar spectrum, it exhibits highly localized absorption in the high-energy spectral region; more specifically, with a bandgap of 626 nm, it stretches over most of the visible region. Another tested WBG NFA, FCC-Cl, previously reported with PM6 and D18,[68] is here evaluated with PTQ10 and achieves a $V_{oc}$ exceeding 1 V. This blend, with a bandgap of 713 nm, covers the entire visible spectrum.

For mid-bandgap absorbers, we selected PTQ10:IT-4F, PM6:DTY6, PM6:Y6, and PTB7-Th:IEICO-4F. All these combinations absorb in the visible region and span into the near infrared (NIR), with bandgaps of 775 nm, 861 nm, 879 nm, and 932 nm, respectively. PTQ10:IT-4F exhibits $V_{oc}$ below 1 V, yet its high $J_{sc}$ and fill factor (FF) makes it a competitive system. PM6:Y6 is the workhorse material of OPV with high PCE, and here yields high $J_{sc}$.[72] Being similar to Y6, DTY6 is employed. PTB7-Th:IEICO-4F extends its absorption tail up to 1000 nm. Finally, for narrow-bandgap applications, we choose PTB7-Th:COTIC-4F, with a bandgap of 1069 nm and an absorption tail extending up to 1100 nm.

Across all investigated systems, there is a clear trade-off between the $V_{oc}$ and the $J_{sc}$ depending on the bandgap of the materials, as observed in Figure 2b. Higher bandgaps display larger $V_{oc}$ values despite reduced $J_{sc}$ due to narrower spectral absorption, whereas lower bandgaps absorb a greater number of photons, thus improving their total $J_{sc}$ at the expense of $V_{oc}$. Following this trend, the $V_{oc}$ of the NBG PTB7-Th:COTIC-4F goes down to 0.54 V; nevertheless, the $J_{sc}$ does not scale up accordingly.

As evidenced by their EQE profile, most materials achieve a charge carrier generation and extraction efficiencies around 70% within their main absorption spectral region. In contrast, the WBG PTQ10:O-IDFBR gets a reduced value, nearly 60%, and in the case of the NBG PTB7-Th:COTIC-4F, further down to around 50%. These results suggest less efficient charge transport and extraction in both systems, especially for the NBG case.

Despite its limitations, the extended absorption into NIR of PTB7-Th:COTIC-4F makes it an extremely promising composition to be used as a red cell in a RAINBOW device, being able to harvest photons that would otherwise be lost with other blends. However, its reduced $V_{oc}$ and inefficient charge extraction currently limit its performance. Therefore, we propose improving these parameters through the addition of a second acceptor in the system, making it a ternary blend.

**Table 1.** JV curve parameters of different bandgap binary systems.

| Binary blend D:A | Bandgap (eV) | $V_{oc}$ (V) | $J_{sc}$ (mA cm$^{-2}$) | FF (%) | Efficiency (%) |
|---|---|---|---|---|---|
| PTQ10:O-IDFBR | 1.98 | 1.36 | 9.27 | 57.5 | 6.46 |
| PTQ10:FCC-Cl | 1.74 | 1.06 | 16.35 | 69.9 | 12.08 |
| PTQ10:IT-4F | 1.60 | 0.90 | 18.77 | 76.46 | 12.93 |
| PM6:DTY6 | 1.44 | 0.83 | 20.17 | 72.4 | 12.07 |
| PM6:Y6 | 1.41 | 0.82 | 21.82 | 64.0 | 11.49 |
| PTB7-Th:IEICO-4F | 1.33 | 0.70 | 23.02 | 61.4 | 9.92 |
| PTB7-Th:COTIC-4F | 1.16 | 0.54 | 20.51 | 59.7 | 6.62 |

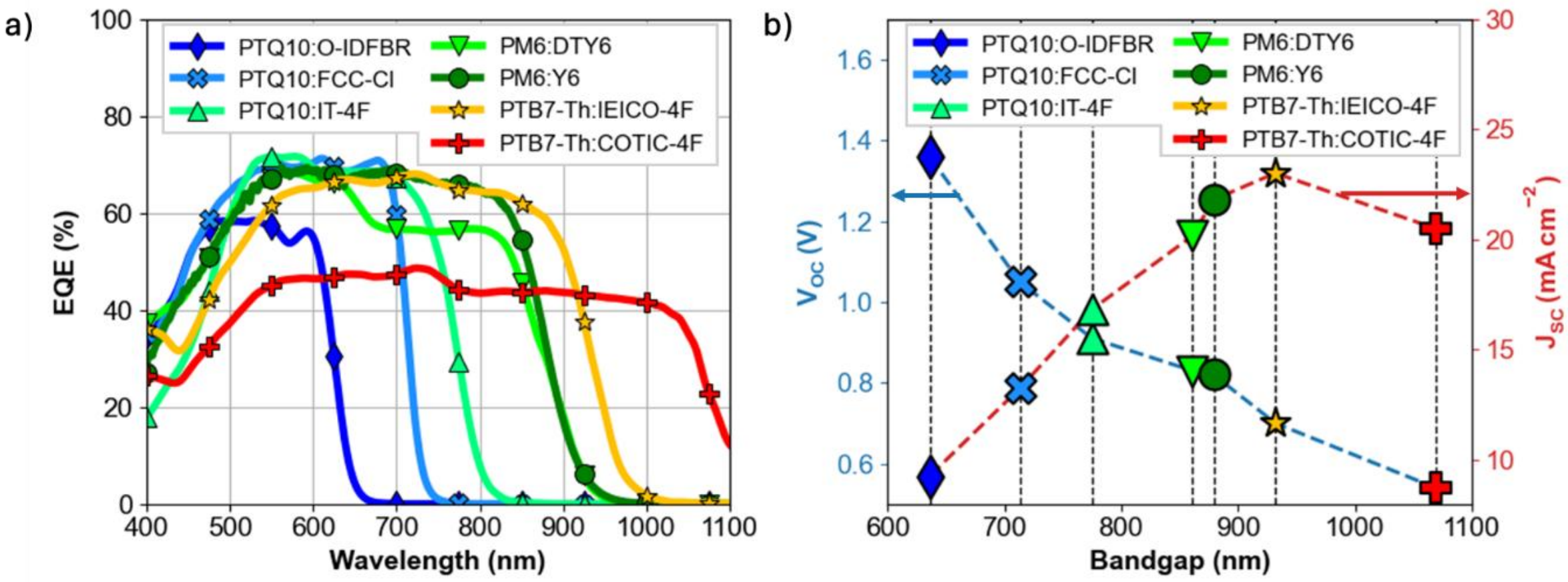

**Figure 2.** a) EQE of different BHJ across the solar spectrum and b) their corresponding $V_{oc}$ and $J_{sc}$ values over bandgap distribution.

## 2.2. Study of Ternaries

We first focus on PTB7-Th:COTIC-4F as it is the lowest bandgap blend available whose FF, $V_{oc}$ and $J_{sc}$ are clearly lower than expected according to the gap, possibly due to transport limitations. Adding another acceptor with higher gap as a third component may help to increase $V_{oc}$ and transport properties. The materials tested for a ternary blend with PTB7-Th:COTIC-4F include fullerene-based acceptor $PC_{70}BM$ and NFAs IEICO-4F and BTP-eC9. $PC_{70}BM$ does not absorb in the NIR, as opposed to COTIC-4F, but instead exhibits a low absorption profile in the visible spectrum that gradually increases into the UV region. It has been previously used with other NFAs due to its higher charge carrier mobility and stability in ternary combinations.[55,57] IEICO-4F is chosen due to its absorption up to 1000 nm and its efficient performance with donor PTB7-Th. BTP-eC9 is a Y-series NFA, which has reached an efficiency around 19% in the literature in combination with PM6.[73] In a blend with PTB7-Th, it does not perform exceptionally well, achieving a PCE around 10%. In this case, the advantage of using this NFA is the favorable energy level alignment of all the materials in the blend (Figure 1b).

We decided to keep the usual donor-to-acceptor weight ratio of 1:1.5 for the ternary blends thus the combined amount of the two acceptors $A_1 + A_2$ remains fixed at 1.5. **Figure 3a** shows the $V_{oc}$ trend of the different binary and ternary combinations depending on the proportion of the second acceptor within the total acceptor quantity. The values to the left correspond to the binary PTB7-Th:COTIC-4F, which has a $V_{oc}$ of 0.54 V, while the ones to the right are for the binaries of PTB7-Th:$A_2$, with values of 0.69 V (BTP-eC9), 0.70 V (IEICO-4F), and 0.80 V ($PC_{70}BM$). In between them, the $V_{oc}$ values of the diverse ternary combinations increase monotonically with the addition of the second acceptor and removal of COTIC-4F, following an alloying-like behavior. All ternary values of $V_{oc}$ are confined within their corresponding binaries, yet they rise at different rates, as observed in the inset of the picture, which displays the $V_{oc}$ normalized to the minimum and maximum values of the binaries. The devices with IEICO-4F have the most similar trend to a proportional line, followed by BTP-eC9 with a similar shape. In the case with $PC_{70}BM$, the $V_{oc}$ does not practically change until COTIC-4F becomes scarce in the blend.

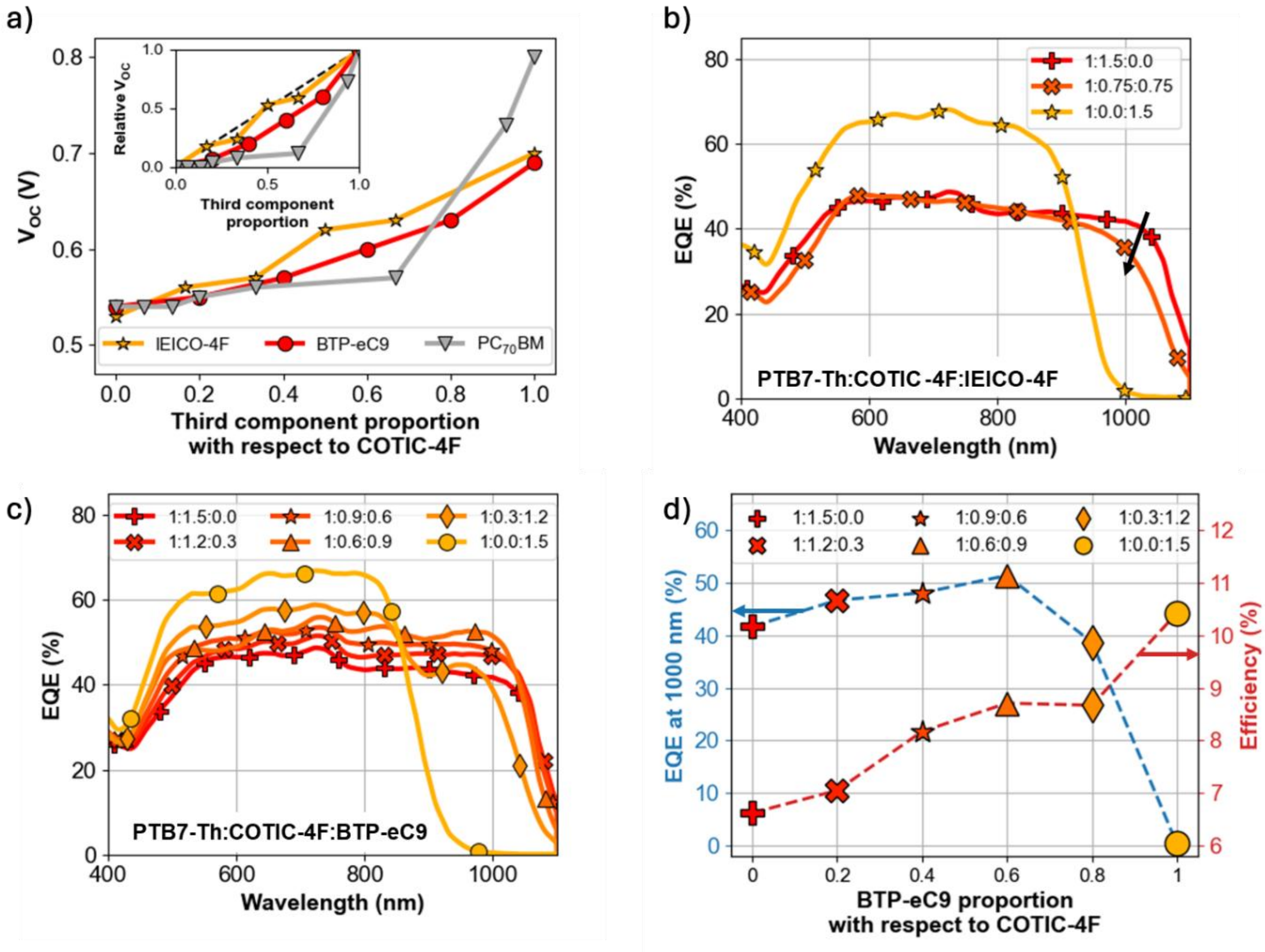

**Figure 3.** a) $V_{oc}$ depending on the addition of a third component ($A_2$) to PTB7-Th:COTIC-4F. The inset shows the relative increase of $V_{oc}$ between the ternaries (PTB7-Th:COTIC-4F:$A_2$) and the two binaries (left, PTB7-Th:COTIC-4F; right, PTB7-Th:$A_2$); b) EQE of PTB7-Th:COTIC-4F:IEICO-4F; c) EQE of the best devices of every ratio of PTB7-Th:COTIC-4F:BTP-eC9; d) trend of EQE at 1000 nm and efficiencies of the best devices for each ratio of PTB7-Th:COTIC-4F:BTP-eC9.

**Table 2.** Parameters of the best devices for each ternary blend ratio of PTB7-Th:COTIC-4F:BTP-eC9.

| Blend ratio PTB7-Th:COTIC-4F:BTP-eC9 | $V_{oc}$ (V) | $J_{sc}$ (mA cm$^{-2}$) | FF (%) | Efficiency (%) |
|---|---|---|---|---|
| 1:1.5:0.0 | 0.54 | 20.51 | 59.7 | 6.62 |
| 1:1.2:0.3 | 0.55 | 21.15 | 60.7 | 7.05 |
| 1:0.9:0.6 | 0.57 | 23.38 | 61.0 | 8.17 |
| 1:0.6:0.9 | 0.6 | 22.69 | 64.2 | 8.71 |
| 1:0.3:1.2 | 0.63 | 22.39 | 61.5 | 8.67 |
| 1:0.0:1.5 | 0.69 | 22.36 | 67.2 | 10.42 |

The use of ternary blends for a NBG clearly results in an improvement in $V_{oc}$, especially when adding NFAs. For the overall performance, the other parameters need to be considered as well. For $PC_{70}BM$ and IEICO-4F, the incorporation of a small quantity of a second acceptor brings only a small improvement over the NBG binary. Nevertheless, the subsequent replacement of COTIC-4F leads to a drop in $J_{sc}$ and FF (Figure S1 and S2 in Supplementary Information), which we tentatively assign to a bad microstructure ($PC_{70}BM$-based blend) or to energy level misalignment (IEICO-4F).

The EQE for the ternary blend PTB7-Th:COTIC-4F:IEICO-4F (1:0.75:0.75) is compared to those of their respective binaries in Figure 3b. In this case, the blend shows no increase of values with respect to PTB7-Th:COTIC-4F in the shared absorption region of the binaries, whereas the region comprised between 900 and 1100 nm exhibits a reduced EQE value due to less efficient charge extraction. In ternary blends with IEICO-4F, removing COTIC-4F reduces exciton generation near 1100 nm yet preserves its unfavorable transfer characteristics.

In comparison, a notable improvement was achieved with the ternary system PTB7-Th:COTIC-4F:BTP-eC9, as summarized in Table 2 with the champion device parameters for each blend. The best performing ternary ratios correspond to those where BTP-eC9 is in a higher amount than COTIC-4F, although they don't exceed PTB7-Th:BTP-eC9 binary values. This overall result would not represent an improvement for single-junction devices; however, it is very promising for multijunction solar cells, as we will show below. EQEs of the best devices for each blend are shown in Figure 3c, revealing that ternaries combine both of their acceptors' spectral responses. The binary-based devices show different absorption onsets linked to their bandgap value, which also reflects on their $V_{oc}$. On one hand, PTB7-Th:COTIC-4F absorption extends far into NIR up to 1100 nm, albeit with low EQE, barely reaching a maximum of 50%. On the other hand, PTB7-Th:BTP-eC9 has an absorption onset centered around 886 nm, but the system is more efficient in extracting photogenerated charges, yielding an EQE maximum of nearly 70%. In the case of ternaries, removal of COTIC-4F and incorporation of BTP-eC9 derives into a sequential increase of EQE between those of binaries. Interestingly, absorption up to 1100 nm is not compromised in any ternary combination, in addition to EQE increase in the shared absorption region, except for blend 1:0.3:1.2. In Figure 3d, the value of EQE at 1000 nm for each blend is compared to each's efficiency champion. Seeing this, the best-performing ternary blend is 1:0.6:0.9, which has an increased $V_{OC}$ although not the maximum, but the best performance in absorption in the region of interest and an improved FF (see Table 2).

The different behavior observed for the different ternaries may be the result of their specific energy level alignment, dissimilar microstructures, or both. To understand this in more detail we performed synchrotron-based Grazing Incidence Wide Angle X-ray Scattering (GIWAXS) measurements. The measured samples were films of the single materials (PTB7-Th, COTIC-4F, and IEICO-4F); of binary blends (PTB7-Th:COTIC-4F, PTB7-Th:BTP-eC9, and PTB7-Th:IEICO-4F); and of exemplary ternary blends, namely PTB7-Th:COTIC-4F:BTP-eC9 (ratios of 1:1.2:0.3, 1:0.9:0.6, 1:0.6:0.9, and 1:0.3:1.2) and PTB7-Th:COTIC-4F:IEICO-4F (ratio of 1:0.75:0.75). The diffractograms and the extracted values of peak positions, interplanar spacings, peak widths, X-ray coherence lengths ($L_c$), and lattice distortion parameters, *g*-parameters, are summarized in Section S3 of Supplementary Information. The first observation is that the degree of order of PTB7-Th seems to increase when mixed with the acceptors in binary blends. That is suggested, exemplarily, by the larger X-ray coherence length ($L_c$) and lower g-parameter observed for the characteristic first-order lamellar reflection of PTB7-Th in the neat ($L_c$ = 3.4 nm, g = 31%) and the blend films (typically, $L_c$ > 8 nm, g < 20%). Moreover, binaries containing BTP-eC9 and IEICO-4F (but not COTIC-4F) show multiple peaks in the in-plane (IP) and out-of-plane (OOP) linecuts that were not originally observed in the single-component film measurements. This suggests that in these binary blends (BTP-eC9 and IEICO-4F) the NFA could be crystallizing strongly upon blending with PTB7-Th, while in COTIC-4F blends their microstructure is mostly preserved with respect to what is found in neat PTB7-Th and NFA films. This could help to understand why the former blends (BTP-eC9, IEICO-4F) offer higher photocurrents (and FF) than COTIC-4F binaries, even when the latter exhibit an extended absorption range. Importantly, the microstructure of all ternary blends appears to be dominated by COTIC-4F, as the ternaries show a very similar GIWAXS pattern with respect to that of the COTIC-4F binary regardless the fraction added of second acceptor.

The strong microstructural similarity between high performing and low performing ternaries suggests that a proper energy level alignment of the materials in the blend dictates the transfer and collection of charge carriers. Binary systems extract holes from the HOMO of PTB7-Th and electrons from the LUMO of the NFAs. High EQE values indicate efficient charge transfer and extraction, as well as reduced recombination. Accordingly, binaries with BTP-eC9 and IEICO-4F show lower recombination losses and more effective charge transfer than those with COTIC-4F. For the ternary blends, EQE data suggests that charge transfer and subsequent extraction are more efficient in systems containing BTP-eC9 than in those with IEICO-4F or $PC_{70}BM$. As shown in Figure 1b, holes are extracted from the donor, while electrons cascade to the LUMO of COTIC-4F when it is paired with IEICO-4F, and to the LUMO of BTP-eC9 when BTP-eC9 is present. Photons absorbed exclusively by COTIC-4F generate electrons that are either extracted inefficiently through COTIC-4F or transferred to BTP-eC9, the latter improving overall performance. In systems incorporating BTP-eC9, excitons formed in COTIC-4F can transfer to BTP-eC9, maintaining absorption around 1100 nm while enabling more efficient extraction.

**Figure 4** summarizes the photovoltaic performance of working devices of all binary and ternary ratios of PTB7-Th:COTIC-4F:BTP-eC9. For comparison, binary blends with COTIC-4F and the second acceptor are shown at the left and right ends of each graph, respectively. The symbol sizes correspond to different thicknesses (d), bigger for thicker and smaller for thinner films. As shown in Figure 4a, the open-circuit voltage ($V_{oc}$) increases nearly linearly between the two binary extremes, practically independent on the thickness value. The short-circuit current density ($J_{sc}$, Figure 4b) clearly improves in the ternary blends compared to either binary system, indicating more effective absorption and charge generation. Within the range of thickness explored, there is a significant dependance of $J_{sc}$ on d, with the optimum thickness being around 100 nm (Table S1). FF increases with increasing BTP-eC9, which supports the idea that this component improves the transport properties of the blend. Moreover, for each composition FF strongly depends on thickness, which again indicates the relatively poor transport properties of COTIC-4F based devices. The combination of all these factors turns into an improvement in efficiency (Figure 4d) with respect to PTB7-Th:COTIC-4F, which was the system we aimed to improve.

Taken together, these results highlight the potential of dual-acceptor, low-bandgap ternary blends as a viable strategy to surpass binary device performance in NBG OPVs. These combinations suggest the importance of a proper energy level alignment when tailoring a ternary system. A good ternary with COTIC-4F should display a cascading set of LUMO and HOMO levels.

Given the successful ternary blending for NBG, we explored the potential of this strategy also for wide bandgap PV, more specifically to improve PTQ10:O-IDFBR in overall efficiency. Attempts to identify effective wide-bandgap ternary systems by incorporating a second acceptor did not yield well-working devices. We tried two ternaries, namely PTQ10:FCC-Cl:O-IDFBR, with the purpose of increasing absorption of high-energy photons and generating higher current while not compromising the $V_{OC}$, and PTQ10:EH-IDTBR:O-IDFBR, due to the similar molecular structure of EH-IDTBR with O-IDFBR and their possible synergy. $V_{oc}$ scaled linearly between the values for the binary. However, the $J_{sc}$ and FF were similar or worse than the binaries, leading to an overall worse performance than any of the two binaries (see Figure S3 and S4 in Supplementary Information). This suggests a suboptimal microstructure and/or energy misalignment.

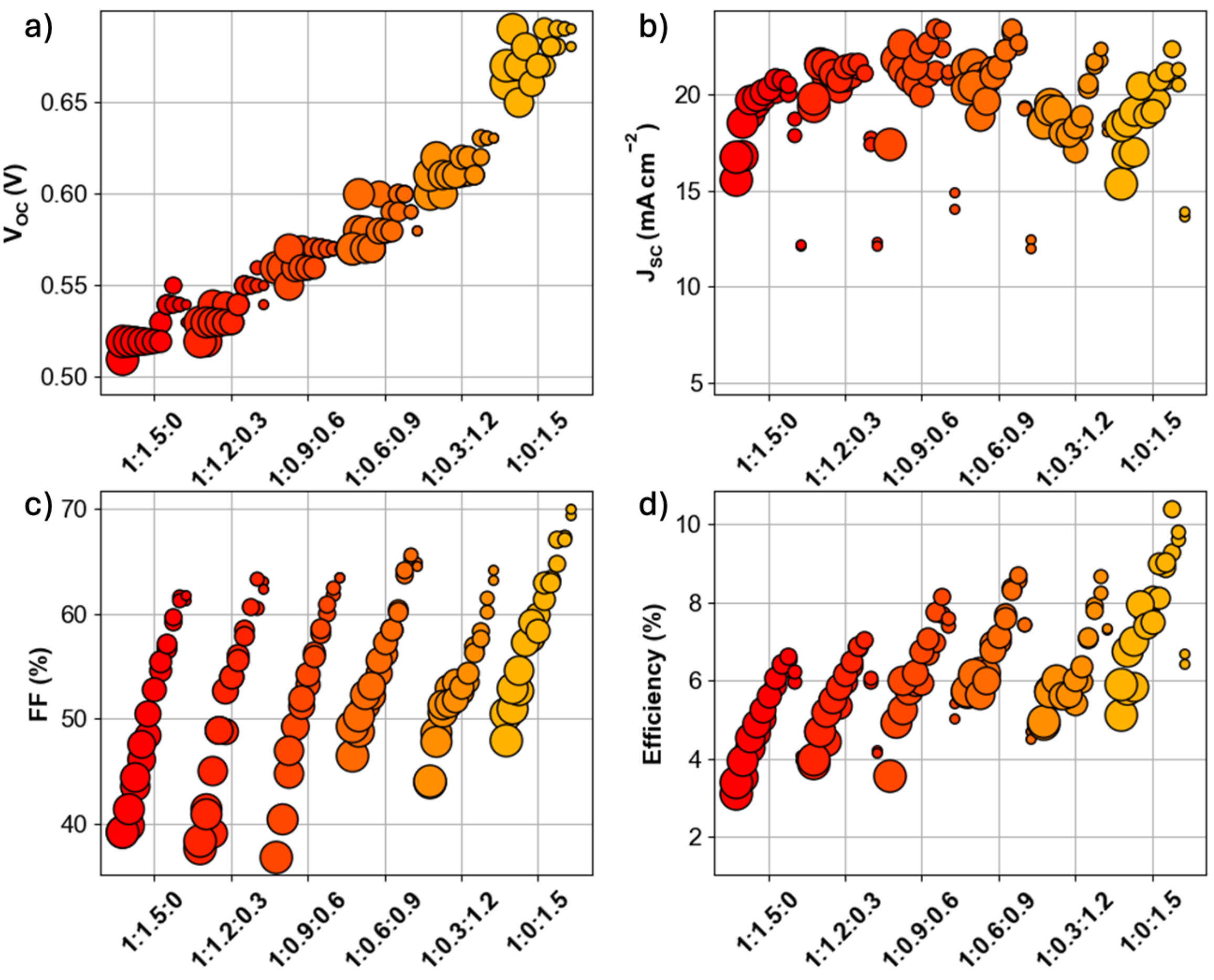

**Figure 4.** JV curve performance parameters of PTB7-Th:COTIC-4F:BTP-eC9 under the solar simulator for different active layer thicknesses (marker size) and ratios (x-axis label): a) $V_{oc}$, b) $J_{sc}$, c) FF, and d) Efficiency.

## 2.3. RAINBOW Performance Projections

In this section, we evaluate the potential of the binary and ternary systems discussed above as subcells in a multijunction RAINBOW geometry. The performance projections, hereafter referred as simulations, use experimental JV parameters and the EQE of single-junction devices. $J_{sc}$ can be obtained from the integration of EQE over the solar spectrum irradiance with the relation

$$J_{SC}(\lambda) = q \int_{\lambda_1}^{\lambda_2} \Phi_{sun}(\lambda)\, EQE(\lambda)\, d\lambda$$

In this equation, $q$ is the elemental charge, $\lambda$ is wavelength, and $\Phi_{sun}(\lambda)$ is the incident photon flux, typically AM1.5G. The integration of the EQE of a certain material under the totality of AM1.5G spectrum ($\lambda_1 = 0, \lambda_2 = \infty$), will result in its $J_{sc}$ under standard illumination conditions.

This definition is used in RAINBOW simulations to find the partial $J_{sc}$ of each subcell depending on the spectral region that they absorb. For a 2-junction device, only one dividing wavelength is employed to assign a blue region to the blue cell within the integration limits of $\lambda_1 = 0$ and $\lambda_2 = \lambda_{div}$ , and a red region to the red cell, between $\lambda_1 = \lambda_{div}$ and $\lambda_2 = \infty$. For a 3-junction, two dividing wavelengths would divide the spectrum into a blue region $\lambda_1 = 0, \lambda_2 = \lambda_{div1}$, a green region $\lambda_1 = \lambda_{div1}, \lambda_2 = \lambda_{div2}$ and a red region $\lambda_1 = \lambda_{div2}, \lambda_2 = \infty$. The partial $J_{sc}$ values, along with their corresponding experimental $V_{oc}$ and FF, are used to calculate the partial efficiencies of each subcell. Then, the total efficiency is calculated by adding the different partial efficiencies.[51]

### *2.3.1. 2-Junction Simulations*

For the 2-junction RAINBOW simulations, the red subcells are based on the binary and ternary combinations of PTB7-Th:COTIC-4F:BTP-eC9, while the blue subcells considered in this study are the binary systems PM6:Y6, PM6:DTY6, PTQ10:IT-4F, PTQ10:FCC-Cl, and PTQ10:O-IDFBR. **Figure 5a** presents the simulated performance of a 2-junction RAINBOW device consisting of a PTB7-Th:COTIC-4F:BTP-eC9 ternary blend (1:0.6:0.9) as the red subcell and PTQ10:FCC-Cl as the blue subcell, which yields the best overall performance of all tested combinations of subcells. The upper panel shows the partial PCE of each subcell as a function of $\lambda_{div}$. For a given $\lambda_{div}$, the blue subcell (blue curve) is illuminated by photons with wavelengths shorter than $\lambda_{div}$; when this value is higher than the bandgap, the partial PCE becomes constant due to the absence of further absorption. In contrast, the red subcell (red curve) is illuminated by photons with wavelengths longer than $\lambda_{div}$. Its partial PCE is zero below its bandgap and increases as $\lambda_{div}$ moves toward shorter wavelengths, eventually reaching its maximum value once most of the spectrum is absorbed. The total RAINBOW efficiency, shown by the black curve, is obtained by adding the partial PCEs of both subcells at each $\lambda_{div}$, with the maximum defining the optimal dividing wavelength. The lower panel of Figure 5a shows the corresponding partial $J_{sc}$ to each subcell. The red and blue curves represent the cumulative $J_{sc}$ of the red and blue subcells, respectively, as a function of the spectral region defined by $\lambda_{div}$ they get to absorb. The black line represents the RAINBOW $J_{sc}$ given by the sum of the partial values.

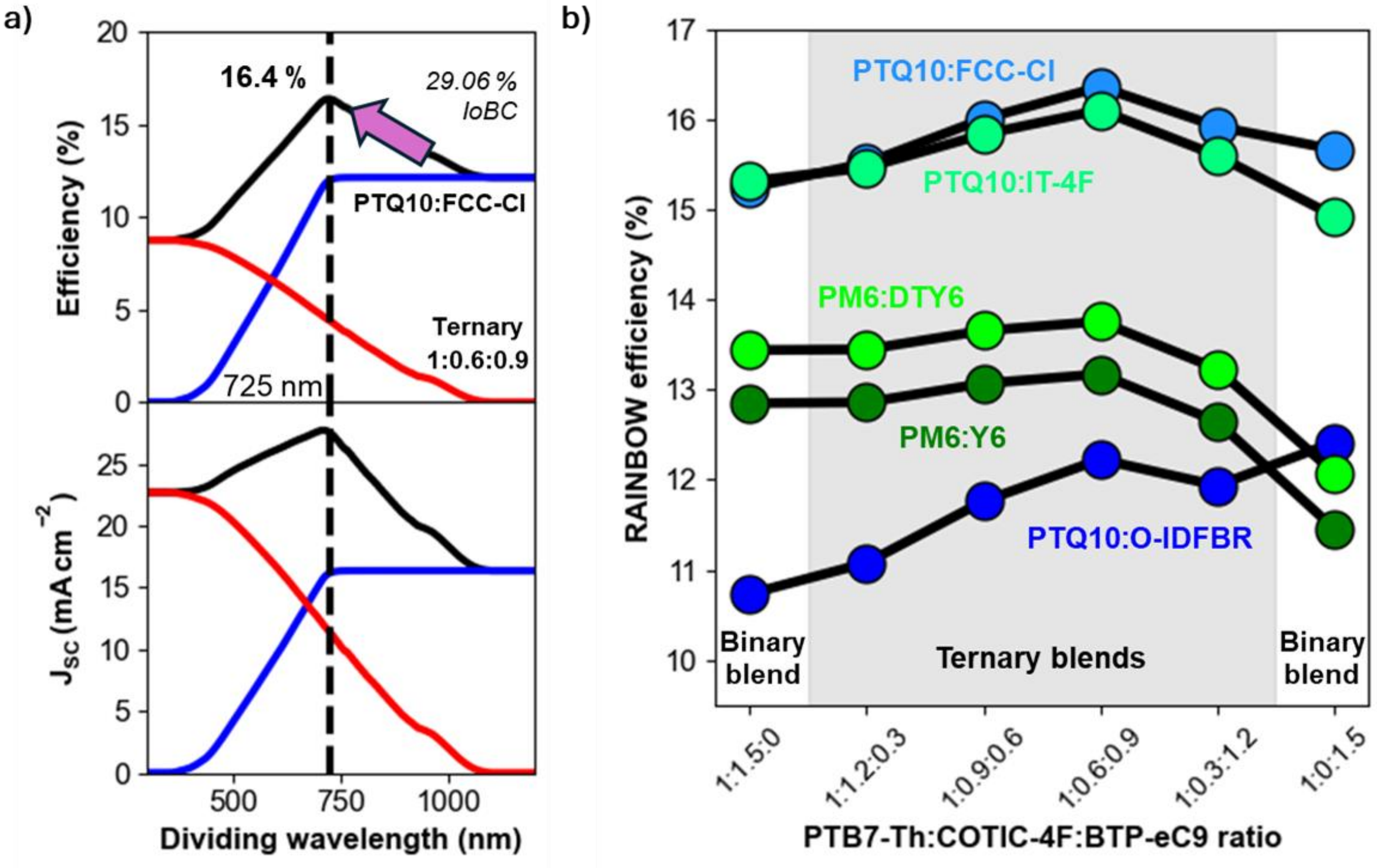

**Figure 5.** Simulation results of a 2-junction RAINBOW. a) Determination of the best cutting wavelength based on the maximum achievable RAINBOW efficiency with IoBC (top) and its corresponding RAINBOW $J_{sc}$ (bottom); b) RAINBOW efficiencies using PTB7-Th:COTIC-4F:BTP-eC9 binary and ternary blends as red cell in combination with different binary systems as blue cell.

Figure 5b summarizes the results of all RAINBOW combinations at their optimal $\lambda_{div}$. Replacing the PTB7-Th:COTIC-4F binary red subcell with a PTB7-Th:COTIC-4F:BTP-eC9 ternary leads to improved RAINBOW performance, mainly due to enhanced partial efficiency. A similar improvement is observed compared to the PTB7-Th:BTP-eC9 binary, except when PTQ10:O-IDFBR is used as the blue subcell. In this specific case, a red subcell with higher overall efficiency but limited NIR absorption performs better than a combination with minimal spectral overlap. For all other blue subcells, the use of ternary red subcells results in higher RAINBOW PCEs.

Among the evaluated combinations, the highest efficiencies are obtained for ternary red subcells paired with PTQ10:FCC-Cl, followed by PTQ10:IT-4F, PM6:DTY6, and PM6:Y6. The optimal configuration reaches a maximum RAINBOW PCE of 16.4%, compared to 12.98% for the best-performing single-junction subcell (PTQ10:IT-4F). This corresponds to an increase over the best cell (IoBC) of 35.5% at a dividing wavelength of 725 nm. In comparison, RAINBOW devices using binary red subcells achieve lower efficiencies, with 15.2% for PTB7-Th:COTIC-4F and 15.7% for PTB7-Th:BTP-eC9.

The results for all other 2-junction configurations not involving PTB7-Th:COTIC-4F:BTP-eC9-based blends as red subcells are presented in Figure S5. Apart from systems employing PTQ10:O-IDFBR as the blue subcell, 2-junction RAINBOW devices incorporating NBG ternary red subcells perform better than all alternative red subcell configurations examined in this study.

*2.3.2. 3-Junction Simulations*

The simulations for a 3-junction RAINBOW consider the incorporation of another subcell, here called green cell. The following materials are assigned to each subcell:

- Red cell: PTB7-Th:COTIC-4F:BTP-eC9 binaries and ternaries, PTB7-Th:IEICO-4F, PM6:Y6, PM6:DTY6.
- Green cell: PTB7-Th:IEICO-4F, PTB7-Th:BTP-eC9, PM6:Y6, PM6:DTY6, PTQ10:IT-4F.
- Blue cell: PTQ10:IT-4F, PTQ10:FCC-Cl, PTQ10:O-IDFBR.

The results of the obtained highest efficiency are shown in **Figure 6a**, corresponding to the combination of PTQ10:FCC-Cl + PM6:DTY6 + Ternary 1:0.6:0.9 (blue-green-red subcells, respectively). The upper panel represents the EQEs of the employed subcells, and the vertical lines indicate the two optimal dividing wavelengths for maximum PCE, more specifically, 715 nm and 865 nm. The AM1.5G spectral irradiance curve appears divided into three spectral regions, each displayed in the color of their assigned subcells. This is a direct consequence of having active materials with similarly intense hut-shaped EQE curves, as dictated by the design rules derived in a previous study.[51] Redirecting higher-energy photons to these subcell not only reduces thermalization losses but also improves their capture due to improved EQE. The role of the red cell is to bring extra current to the device, reducing the absorption losses of photons that no other system can get.

The lower panel corresponds to the simulation results. The two axes are the dividing wavelengths, either between blue and green (1) or between green and red (2). The color scale bar corresponds to the combined RAINBOW PCE values obtained from the simulations with the lowest numbers in blue, while the maximum ones appear in dark red. The vertices of the plot represent the single-junction values for the red (top, 8.7%), for the green (bottom left, 11.8%), and for the blue subcell (bottom right, 12.1%). The sides of the triangle depict the simulation results for the different 2-junction RAINBOW combinations, i.e., red–green (left side, 13.8%), green–blue (bottom side, 16.0%), and red–blue (diagonal side, 16.4%). Finally, the colored part shows the results of all the possible ways in which the two dividing wavelengths can split the spectrum in a 3-junction configuration. Given their optimization, the maximum RAINBOW efficiency obtained is at a staggering 17.7%

Table S2 in Supplementary Information shows the results of all 3-junction combinations studied. Again, most devices using ternary blends of PTB7-Th:COTIC-4F:BTP-eC9 achieve higher efficiencies than when using binary PTB7-Th:COTIC-4F. In general, the blends with ratio 1:0.9:0.6 and 1:0.6:0.9 yield the best results for most combinations.

As seen in Figure 6b, the boost in efficiency of a RAINBOW device is mainly determined by a 2-junction. Nevertheless, efficiency can still be increased with the addition of a third subcell, as highlighted in the picture. The general trend shows the best single-junction, with a value of 12.9%, scales up in a 2-junction to 16.4%, and for a 3-junction to 17.7%. This partial convergence is expected according to detailed balance limit. By analyzing the overall number of results, the spread of efficiencies reduces, which means that all combinations achieve higher PCE in 3-junctions compared to single-junctions. Combinations that started with very limited PCEs increased from 6.2% (1-junction) to 10.7% (2-junction), and up to 13.2% (3-junction). This suggests that despite having starting systems with low efficiencies, it is possible to achieve high PCEs using this geometry, given the proper optical conditions.

The smaller jump in efficiency increase from a 2-junction device to a 3-junction device may seem small for practical applications. Indeed, this would be the case for other multijunction technologies that would have an extra cost for each additional subcell incorporated in the geometry. Nevertheless, for organic based RAINBOW cells, the incorporation of subcells does not significantly increase the overall cost due to the easiness of side-by-side deposition, as we will next show.

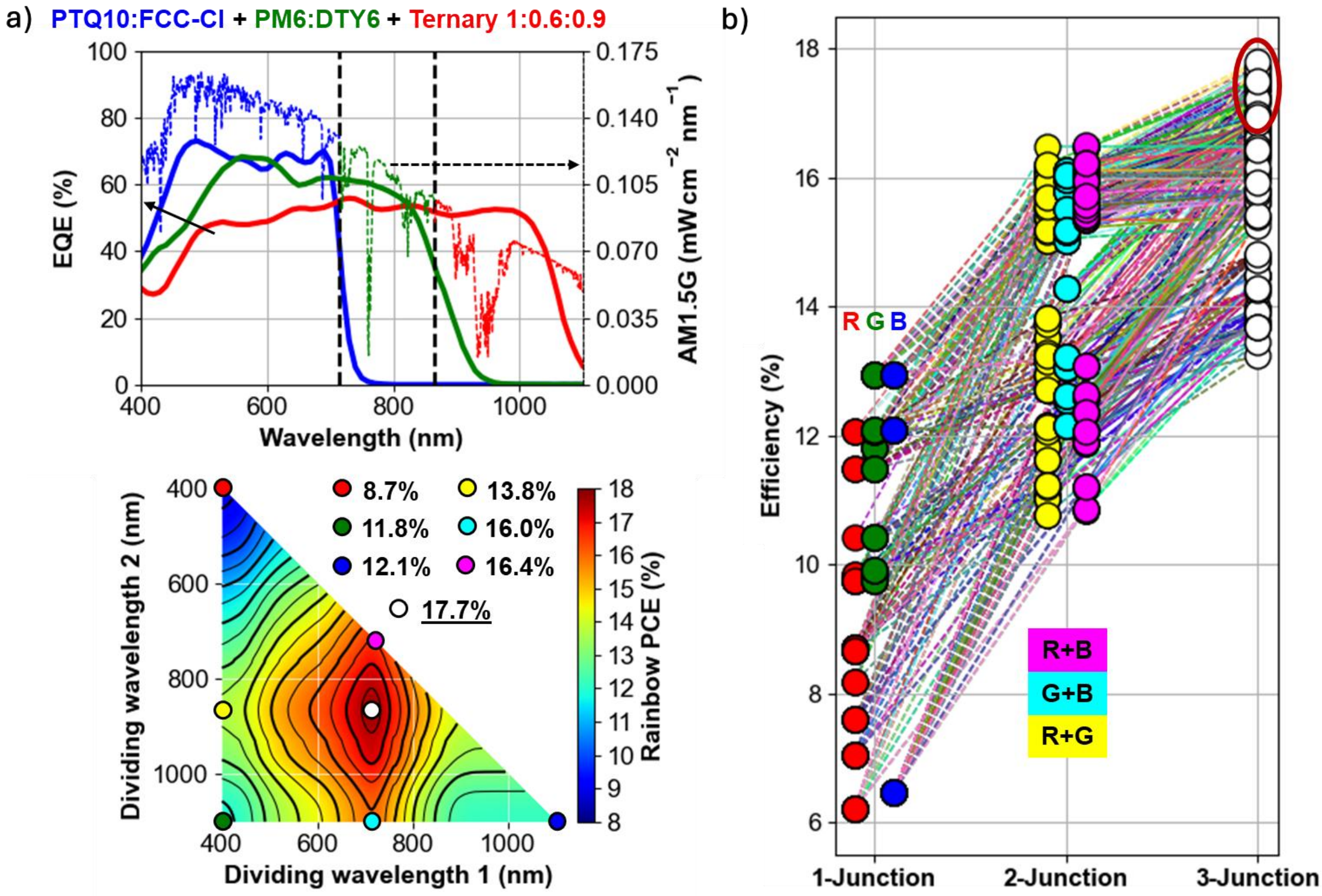

**Figure 6.** a) 3-junction RAINBOW simulation of PTQ10:FCC-Cl + PM6:DTY6 + Ternary 1:0.6:0.9. The upper panel shows the EQE of the subcells over the spectrum at the optimal dividing wavelengths. The lower panel shows the RAINBOW PCE depending on 2 dividing wavelengths; b) increase of efficiency with increasing junctions for all possible subcell combinations. The red circle marks the 3-junction combinations that exhibit PCEs above any 2-junction device.

### 2.4. Proof-of-concept multijunction RAINBOW device

Experimental devices were fabricated to prove lateral deposition by blade coating. Up to 6 different BHJ are deposited side-by-side on one single substrate, as seen in **Figure 7a**.

To prove the validity of the simulations, pixels of each material were characterized under desired partial spectra tailored with the customizable LEDs of the simulator. The tested 2-junction combinations use a red cell of either PTB7-Th:COTIC-4F or Ternary 1:0.6:0.9, while blue cell is either PTQ10:O-IDFBR, PTQ10:FCC-Cl, or PTQ10:IT-4F. For 3-junction combinations, red and blue cells may be the previous ones with an added green subcell of either PTQ10:IT-4F or PM6:DTY6.

According to the simulations, the best 2-junction is PTQ10:FCC-Cl + Ternary 1:0.6:0.9 and for 3-junction is PTQ10:FCC-Cl + PM6:DTY6 + Ternary 1:0.6:0.9, with dividing wavelength values of 725 nm for the 2-junction and 715 nm and 865 nm for the 3-junction. The resulting custom spectra with these conditions for each subcell are shown in Figure 7b as produced with an individually addressable LED solar simulator. In the case of the 2-junction, the simulations achieved 16.4% with an IoBC of 35.5%. The sum of the partial experimental efficiencies resulted in 15.9% with an IoBC of 32.7%. For the 3-junction, the simulations reached 17.7% and an IoBC with an IoBC 46.6%. Finally, the sum of the partial experimental results is 17.3 with an IoBC of 38.3%. Table S3 in Supplementary Information compares all the partial JV parameters obtained under partial spectra measurements with the previously obtained simulation values.

Again, it is worth noting the improvement in using a ternary blend as the red cell instead of the binary blend, which provides lower values overall. For a 2-junction, 15.2% was achieved in simulations and 15.1% in experimental, and for a 3-junction, 17.3% in simulations and 17.0% in experimental.

Table 3 shows the resulting experimental RAINBOW efficiencies of all tested combinations and their value from the simulations. The partial spectra profiles used to characterize each subcell can be found in Figure S6 of Supplementary Information.

**Table 3.** Experimental RAINBOW efficiencies compared to simulations.

| Blue Cell | Green Cell | Red Cell | Exp / Sim | RAINBOW PCE (%) | IoBC (%) |
|---|---|---|---|---|---|
| PTQ10:FCC-Cl | | Ternary | Exp | 15.91 | 32.69 |
| | | | Sim | 16.37 | 35.47 |
| | | PTB7-Th: COTIC-4F | Exp | 15.12 | 26.11 |
| | | | Sim | 15.24 | 26.18 |
| PTQ10:IT-4F | | Ternary | Exp | 15.86 | 21.16 |
| | | | Sim | 16.11 | 24.55 |
| | | PTB7-Th: COTIC-4F | Exp | 15.22 | 16.27 |
| | | | Sim | 15.32 | 18.44 |
| PTQ10:O-IDFBR | | Ternary | Exp | 12.11 | 40.81 |
| | | | Sim | 12.23 | 40.37 |
| | | PTB7-Th: COTIC-4F | Exp | 10.70 | 40.05 |
| | | | Sim | 10.73 | 62.13 |
| PTQ10:FCC-Cl | PM6:DTY6 | Ternary | Exp | 17.34 | 38.28 |
| | | | Sim | 17.71 | 46.58 |
| | | PTB7-Th: COTIC-4F | Exp | 17.00 | 35.57 |
| | | | Sim | 17.37 | 43.83 |
| PTQ10:IT-4F | PM6:DTY6 | Ternary | Exp | 16.72 | 27.73 |
| | | | Sim | 16.89 | 30.64 |
| | | PTB7-Th: COTIC-4F | Exp | 16.38 | 25.13 |
| | | | Sim | 16.57 | 28.15 |
| PTQ10:O-IDFBR | PTQ10:IT-4F | Ternary | Exp | 15.52 | 20.50 |
| | | | Sim | 16.44 | 27.1 |
| | | PTB7-Th: COTIC-4F | Exp | 14.88 | 15.53 |
| | | | Sim | 15.63 | 20.91 |

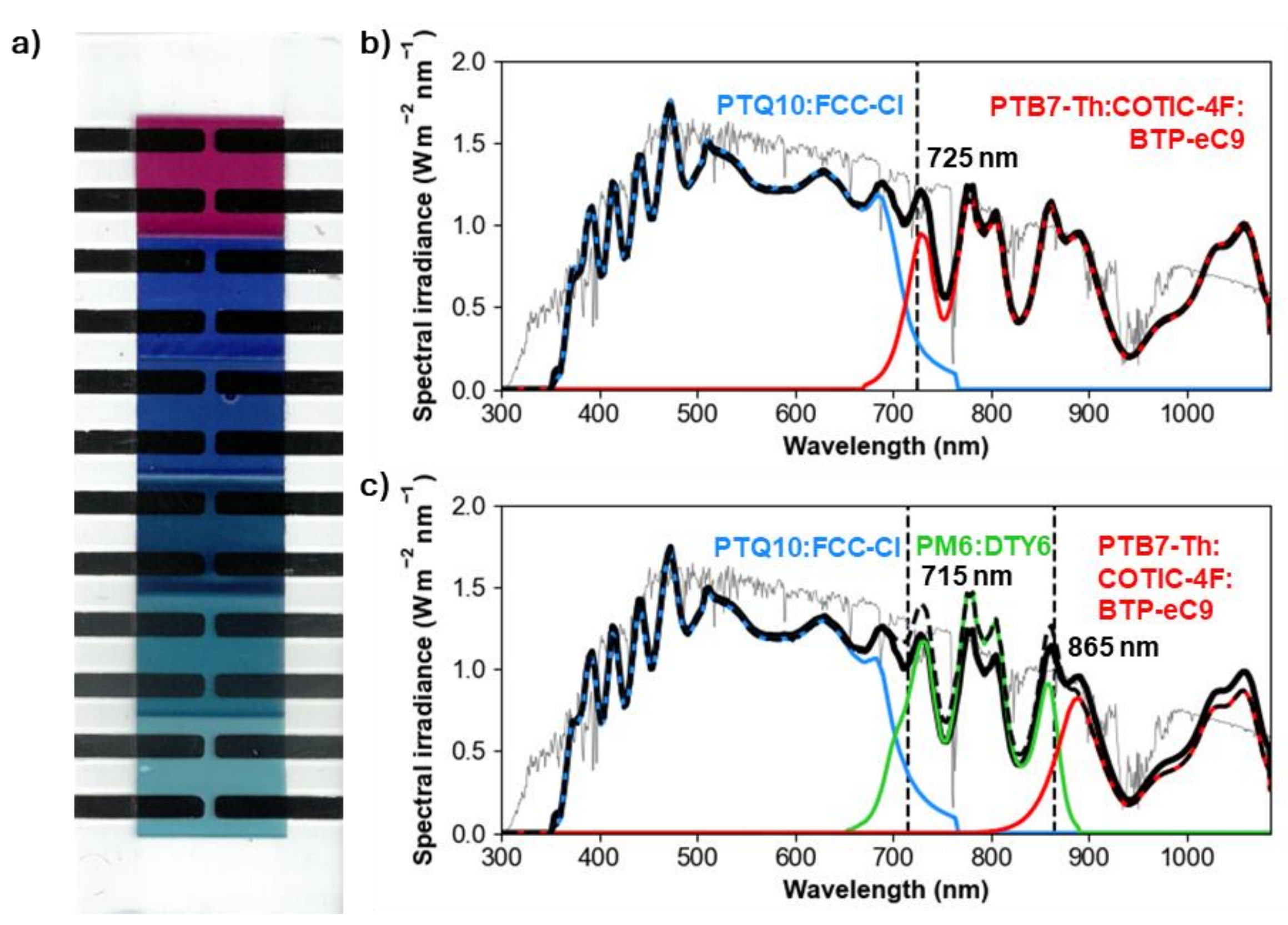

**Figure 7.** a) Monolithic RAINBOW device with 6 different materials deposited by blade coating side-by-side. From top to bottom: PTB7-Th:COTIC-4F, Ternary 1:0.6:0.9, PM6:DTY6, PTQ10:IT-4F, PTQ10:FCC-Cl, and PTQ10:O-IDFBR; partial spectra used for each subcell in a b) 2-junction of PTQ10:FCC-Cl + Ternary 1:0.6:0.9 and c) 3-junction of PTQ10:FCC-Cl + PM6:DTY6 + Ternary 1:0.6:0.9.

**2.5. Theoretical Limitations of Multijunction Organic Solar Cells**

We next evaluate the theoretical potential of the proposed rainbow architecture, while considering the unique limitations of OPVs compared to inorganic photovoltaic technologies. In OPVs, the characteristic molecular structure and charge transfer processes lead to additional intrinsic loss mechanisms, with non-radiative recombination generally dominating over radiative recombination. Moreover, achieving complete charge separation (CS) is more challenging compared to inorganics due to inherently strong exciton binding energies limiting to the local exciton (LE) dissociation and the subsequent charge transfer (CT) rates.[74,75] Therefore, a typical detailed balance model is not sufficient to estimate the efficiency limits that represent the conditions in OPVs. A recalculation is needed to find not only realistic PCE limits in OPV multijunction cells, but also to determine the optimal bandgaps and spectral slicing for a given number of subcells. To address this, we use a previously established approach that combines a semi-classical rate model (Marcus-Levich-Jortner), with a zero-dimensional kinetic model (three-state system with LE, CT, and CS state), and with a one-dimensional semiconductor device model (drift-diffusion).[76]

With this modeling approach, we can investigate a large range of fundamental properties that govern the loss mechanisms in OPVs, including reorganization energies (inner and outer), the energetic offset between the local exciton (LE) and the charge transfer (CT) state, exciton and charge transfer dissociation rates, transport mobilities, energetic disorder, and more.

This type of model allows us to explore how specific material properties impact the ideal bandgap of an OPV cell. For example, the ideal bandgap of materials with a larger inner reorganization energy is shifted to higher energies as shown in Figure S7 of Supplementary Information. For this set of example parameters, we find that the maximum achievable efficiency is around 27% at 1.5 eV (in contrast to the well-known 33% at 1.34 eV in inorganic cells. Consequently, the efficiency limit of multijunction cells will also differ from the limits established in inorganic multijunction cells. In addition, the limits in OPVs will not only depend on bandgaps of the subcells but also on their specific material properties.

To optimize the multi-junction solar cell configurations, we incorporated the physical model into a dynamic programming (DP) framework that reduces the computational complexity of spectral slice allocation from combinatorial to polynomial scaling.[77] This approach, combined with the superposition principle and pre-calculated lookup libraries, enabled high-resolution optimization of 12-cell architectures. **Figure 8a** shows the optimal bandgaps and spectral slicing for all subcells depending on the desired total number of junctions for a given material parameter set (i.e. this assumes that apart from the bandgap, all other parameters remain the same in each subcell). The resulting maximum combined efficiency as a function of the number of junctions is shown in Figure 8b. Even adding just one additional junction significantly increases the maximum PCE from 27% to 34%, when optimal bandgaps of 1.4 eV and 2 eV are chosen for the subcells. For an increasing number of junctions, the lowest bandgap decreases deeper into the gap to around 1.2 eV, while the upper bandgap continues to grow above 3 eV. Table S5 in Supplementary Information contains the performance parameters of each subcell depending on the number of junctions.

For single-junction devices, the optimal bandgap suggested by the simulation lies well within the range of experimentally available bandgaps of best performing materials. However, we find that to fully leverage the potential of multijunction solar cells, it will be necessary to develop novel organic semiconductor materials with wider bandgaps above 2 eV. A representative material combination close to this range is PTQ10:O-IDFBR with a bandgap of nearly 2 eV and exceptional $V_{oc}$. Nevertheless, its efficiency is still heavily restrained by poor current and FF. Given the material constraints and expected optical losses, we propose a multijunction with 3 to 5 junctions as a realistic architecture for the RAINBOW design.[30]

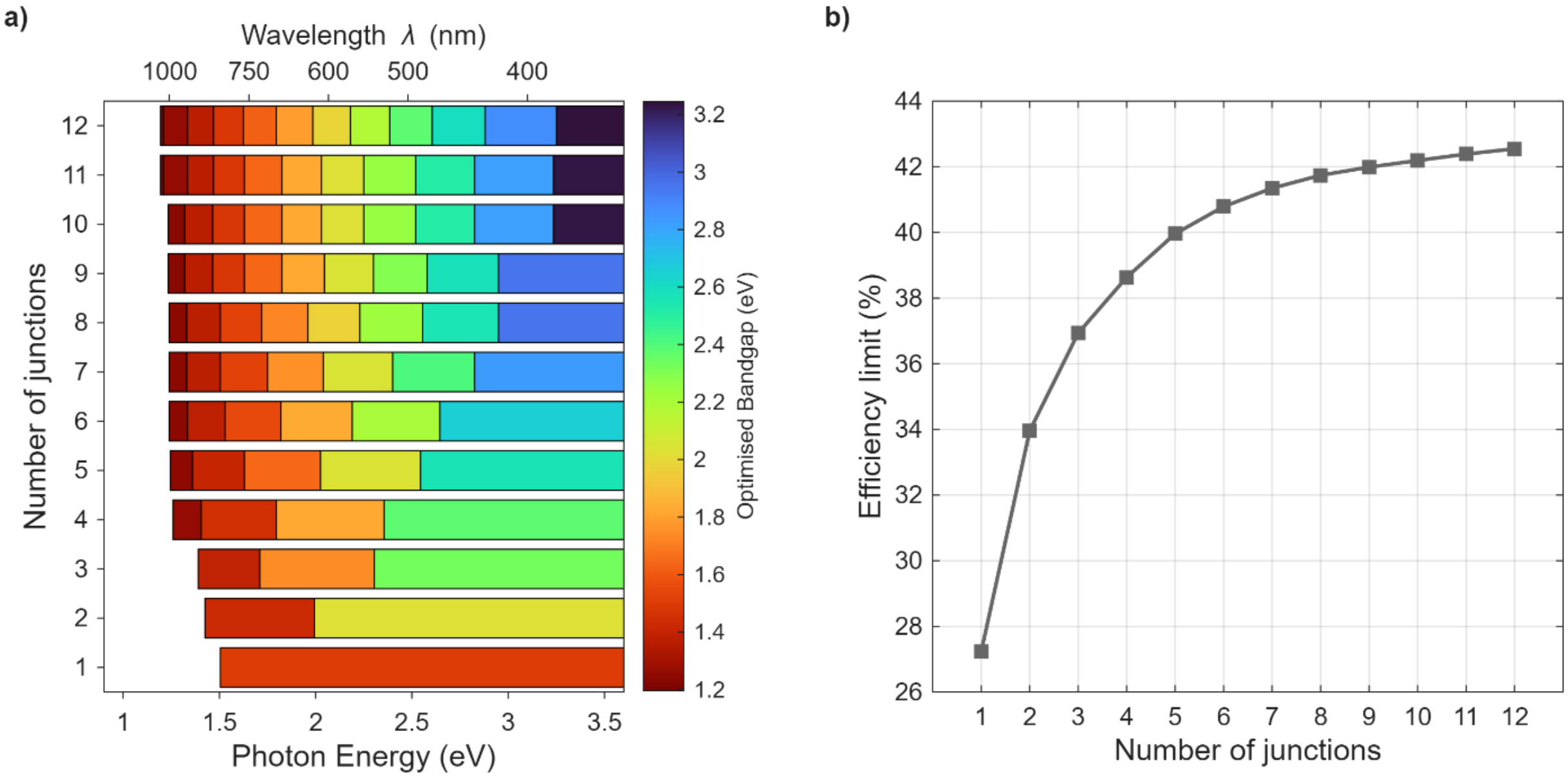

**Figure 8.** a) Optimal subcell bandgaps for maximum multijunction efficiency; b) maximum efficiency for each number of junctions.

## 3. Conclusion

In this study, we explored the RAINBOW multijunction architecture with 2 and 3 junctions as a scalable way to improve OPV performance. To enhance the efficiency of the narrow bandgap binary system PTB7-Th:COTIC-4F, used as a red subcell due to its extended infrared absorption, a second acceptor was incorporated to form a ternary blend. After a material screening, the resulting PTB7-Th:COTIC-4F:BTP-eC9 system increased efficiency from 6.6% in the binary to 8.7% in the ternary with a ratio of 1:0:6:0:9, while maintaining the absorption range and improving photon-to-electron conversion.

Additional subcells based on mid- and wide-bandgap materials were evaluated, with the best single-junction performance reaching 12.9% for the PTQ10:IT-4F system. Simulations of RAINBOW configurations identified the ternary blend along with a high-efficiency PTQ10:FCC-Cl system for the blue subcell as the optimal 2-junction combination, achieving 16.4%, compared to 15.2% with the binary. Extending to a 3-junction by incorporating a green subcell of PM6:DTY6 yielded simulated efficiencies of 17.7% for the ternary and 17.4% for the binary.

Experimental validation through blade-coated, side-by-side printed devices confirmed the feasibility of the approach, reaching 15.9% efficiency for the 2-junction (15.1% for the binary) and 17.3% for the 3-junction (17.0% for the binary). Finally, a revised detailed balance limit was proposed, which incorporates fundamental mechanisms inherent to OPV to account for additional loss pathways such as non-radiative losses. Further progress will require the development of wide-bandgap materials exceeding 2 eV with improved efficiency to optimize overall multijunction performance.

## 4. Experimental Section

*Materials:* organic semiconductors were commercially obtained from 1-Material and Ossila.

*Device fabrication*: OPV devices were fabricated on top of patterned glass/ITO substrates from Ossila (100 nm, 20 Ω square$^{-1}$) following an inverted structure of glass/ITO/ZnO/AL/$MoO_3$/Ag. The substrates were consecutively cleaned in a sonic bath for 5 minutes in acetone, 2% Hellmanex soap water solution and isopropanol, and 10 minutes in 2M NaOH water solution. After drying the substrates with compressed air, the electron transport layer (ETL) was deposited by blade coating (ZAA 2300 Automatic Film Applicator and ZUA 2000 Universal

Film Applicator, Zehntner) using a ZnO nanoparticle dispersion (N-10, Avantama) with the hot bed at 40 ºC and a speed of 5 mm $s^{-1}$, a height of 1150 nm, followed by a 10 min thermal annealing at 120 ºC in air. The substrates were then transferred to a controlled $N_2$ atmosphere glovebox where active layers were deposited along the long side, also by blade coating, with a height of 1200 nm and temperature 80 ºC, a gradient speed of deposition from 90 to 10 mm $s^{-1}$. After such variable speed deposition, each substrate comprises 24 devices of 12 different thicknesses. The samples were then annealed at different temperatures. Finally, 15 nm of $MoO_3$ and 150 nm of Ag were deposited by thermal evaporation at high vacuum, corresponding to the hole transport layer (HTL) and back contact, respectively.

*Laterally deposited device fabrication:* laterally deposited devices were fabricated in the same way as single-material devices except in the active layer deposition step. To avoid full meniscus coverage over the substrate, 4 layers of solvent-resistant tape were put at the edge of the applicator with a total width of 1 cm and a thickness of 220 nm. Solutions were deposited over the short side of the substrate thus the meniscus formed under the tape covered two pixels per side, four in total. First, one single material substrates were sequentially deposited side-by-side at a constant velocity that provides an optimized thickness under partial spectra measurements. Then, proof-of-concept RAINBOW devices were prepared by depositing different materials side-by-side after optimization.

*Device characterization*: *J-V* curves were measured with a Keithley 2401 source meter. The 24 pixels of a substrate were sequentially measured thanks to a demultiplexer that changes the connection to each device automatically. The employed solar simulator is SINUS-70 ADVANCED from WAVELABS, consisting of 21 customizable LEDs capable of reproducing a class-AAA AM1.5G spectrum as well as the partial spectra determined through simulations for monolithic RAINBOW devices. A calibrated reference solar cell (91150V, Newport) and a thermopile (S401C, Thorlabs) were used to calibrate the spectrum.

EQE is also measured with a custom-built setup using a steady-state, monochromatic illumination system based on a Xe arc lamp (SLS401, Thorlabs), which provides a broadband spectral output covering from UV to NIR. The lamp radiation is first collimated and directed into a monochromator (MicroHR, Horiba), which selects the desired wavelength between 350 nm and 1100 nm. A mechanical optical chopper is placed in the beam path after the monochromator to modulate the light intensity at a known reference frequency. The chopped beam allows for phase-sensitive detection using a lock-in amplifier, significantly improving the signal-to-noise ratio by suppressing background and dark current contributions. The monochromatic, modulated beam is alternately directed onto a calibrated silicon (Si) photodiode positioned in an equivalent optical plane as the sample (using a broadband beam splitter on the main monochromatic beam). The Si photodiode has a known wavelength-dependent responsivity traceable to a standard (NIST), enabling accurate determination of the incident photon power density at each wavelength. The quantum efficiency of the sample is then calculated by comparing its photocurrent to the photon flux determined from the calibrated Si photodiode.

GIWAXS measurements were performed at the BL11 NCD-SWEET beamline at the ALBA Synchrotron Radiation Facility in Spain. An incident X-ray beam energy of 12.4 keV was set using a channel-cut Si (111) monochromator and collimated with an array of Be Compound Refractive Lenses. The angle of incidence was set to 0.12°, the exposure time set to 1 s, and the diffraction patterns detected with a Rayonix LX255-HS. The scattering wavevector data were calibrated using $Cr_2O_3$ as the standard (NIST) resulting in a sample-to-detector distance of 202.37 mm.

*RAINBOW simulations*: semi-experimental simulations were carried out with *J-V* parameter values and EQE curves. The product between the AM1.5G photon flux and the device EQE over the whole spectrum is equivalent to $J_{SC}$. Partial $J_{SC}$ is calculated by adjusting the integration limits, and this result is used along with previously measured $V_{OC}$ and FF to compute partial PCE.

**Acknowledgements**

The authors acknowledge the financial support from the Spanish Ministerio de Ciencia, Innovación y Universidades and Agencia Estatal de Investigación (MICIU/AEI/10.13039/501100011033) under grants PRE2022-104632, PID2021-128924OB-I00, PID2024-163010OB-I00, and TED2021-131911B-I00, and the Severo Ochoa program for Centres of Excellence CEX2023-001263-S. The authors thank CSIC for funding through the JAE-Chair program via project DOMMINO. F.X.C.G. acknowledges the PhD program in Materials Science from Universitat Autònoma de Barcelona. M.A.S. thanks the European Union for its fellowship with grant number MSCA-101104491. J.N. and J.S.M thank the UKRI for project EP/Z533361/1 POTENtIAl under the ERC underwrite scheme. J.S.M. thanks the EPSRC for award of a Doctoral Prize Fellowship and J.N. thanks the Royal Society for the award of a Research Professorship. J.R.M. and J.M. acknowledge financial support from the European Research Council under grant ERC CoG-101086805. GIWAXS experiments were performed at NCD-SWEET beamline at ALBA Synchrotron with the collaboration of ALBA staff.

# Supporting Information

**S1. Active layer materials names**

**PTQ10**: poly[(thiophene)-alt-(6,7-difluoro-2-(2- hexyldecyloxy)quinoxaline)]

**PM6**: poly[(2,6-(4,8-bis(5-(2-ethylhexyl-3-fluoro)thiophen-2-yl)-benzo[1,2-b:4,5-b’]dithiophene))-alt-(5,5-(1’,3’-di-2-thienyl-5’,7’-bis(2-ethylhexyl)benzo[1’,2’-c:4’,5’-c’]dithiophene-4,8-dione)]

**PTB7-Th**: poly[4,8-bis(5-(2-ethylhexyl)thiophen-2-yl)benzo[2-b;4, 5-b']dithiophene-2,6-diyl-alt-(4-(2-ethylhexyl)-3-fluorothieno[4-b]thiophene-)-2-carboxylate-2-6-diyl]

**$PC_{70}BM$**: [6,6]-phenyl-C71-butyric acid methyl ester

**O-IDFBR**: (5Z,5'Z)-5,5'-((7,7'-(6,6,12,12-tetraoctyl-6,12-dihydroindeno[1,2-b]fluorene-2,8-diyl)bis(benzo[c][1,2,5]thiadiazole-7,4-diyl))bis(methanylylidene))bis(3-ethyl-2-thioxothiazolidin-4-one)

**EH-IDTBR**: (5Z)-3-ethyl-2-sulfanylidene-5- [[4-[9,9,18,18-tetrakis(2-ethylhexyl)-15-[7-[(Z)-(3-ethyl-4-oxo-2-sulfanylidene-1,3-thiazolidin-5-ylidene)methyl]-2,1,3-benzothiadiazol-4-yl]-5,14-dithiapentacyclo[10.6.0.03,10.04,8.013,17]octadeca-1(12),2,4(8),6,10,13(17),15-heptaen-6-yl]-2,1,3-benzothiadiazol-7-yl]methylidene]-1,3-thiazolidin-4-one

**FCC-Cl**: 2-(2-chloro-6-oxo-5,6-dihydro-4Hcyclopenta[b]thiophen-4-ylidene)-malononitrile

**IT-4F**: 3,9-bis(2-methylene-((3-(1,1-dicyanomethylene)-6,7-difluoro)-indanone))-5,5,11,11-tetrakis(4-hexylphenyl)-dithieno[3-d:2', 3'-d']-s-indaceno[2-b:5, 6-b']dithiophene

**DTY6**: 2,2'-((2Z,2'Z)-((12,13-bis(2-decyltetradecyl)-3,9-diundecyl-12,13-dihydro-[1,2,5]thiadiazolo[3,4-e]thieno[2",3":4',5']thieno[2',3':4,5]pyrrolo[3,2-g]thieno[2',3':4,5]thieno[3,2-b]indole-2,10-diyl)bis(methanylylidene))bis(5,6-difluoro-3-oxo-2,3-dihydro-1H-indole-2,1-diylidene))dimalononitrile

**Y6**: 2,2'-((2Z,2'Z)-((12,13-bis(2-ethylhexyl)-3,9-diundecyl-12,13-dihydro-[1,2,5]thiadiazolo[3,4-e]thieno[2",3":4',5']thieno[2',3':4,5]pyrrolo[3,2-g]thieno[2',3':4,5]thieno[3,2-b]indole-2,10-diyl)bis(methanylylidene))bis(5,6-difluoro-3-oxo-2,3-dihydro-1H-indene-2,1-diylidene))dimalononitrile

**BTP-eC9**: 2,2'-[[12,13-bis(2-butyloctyl)-12,13-dihydro-3,9-dinonylbisthieno[2", 3":4', 5']thieno[2', 3':4]pyrrolo[2-e:2', 3'-g]benzothiadiazole-2,10-diyl]bis[methylidyne(5, 6-chloro-3-oxo-1H-indene-2, 1(3H)-diylidene)]]bis[propanedinitrile]

**IEICO-4F**: 2,2′-((2Z,2′Z)-(((4,4,9,9-tetrakis(4-hexylphenyl)-4,9-dihydro-s-indaceno[2-b:5, 6-b′]dithiophene-2,7-diyl)bis(4-((2-ethylhexyl)oxy)thiophene-5,2-diyl))bis(methanylylidene))bis(5,6-difluoro-3-oxo-2,3-dihydro-1H-indene-2,1-diylidene))dimalononitrile

**COTIC-4F**: 2,2'-((2Z,2'Z)-(((4,4-bis(2-ethylhexyl)-4H-cyclopenta[1-b:3, 4-b']dithiophene-2,6-diyl)bis(4-(2-ethylhexyloxy)thiophene-5,2-diyl))bis(methanylylidene))bis(5,6-difluoro-3-oxo-2,3-dihydro-1H-indene-2,1-diylidene))dimalononitrile

## S2. Ternary blend combinations

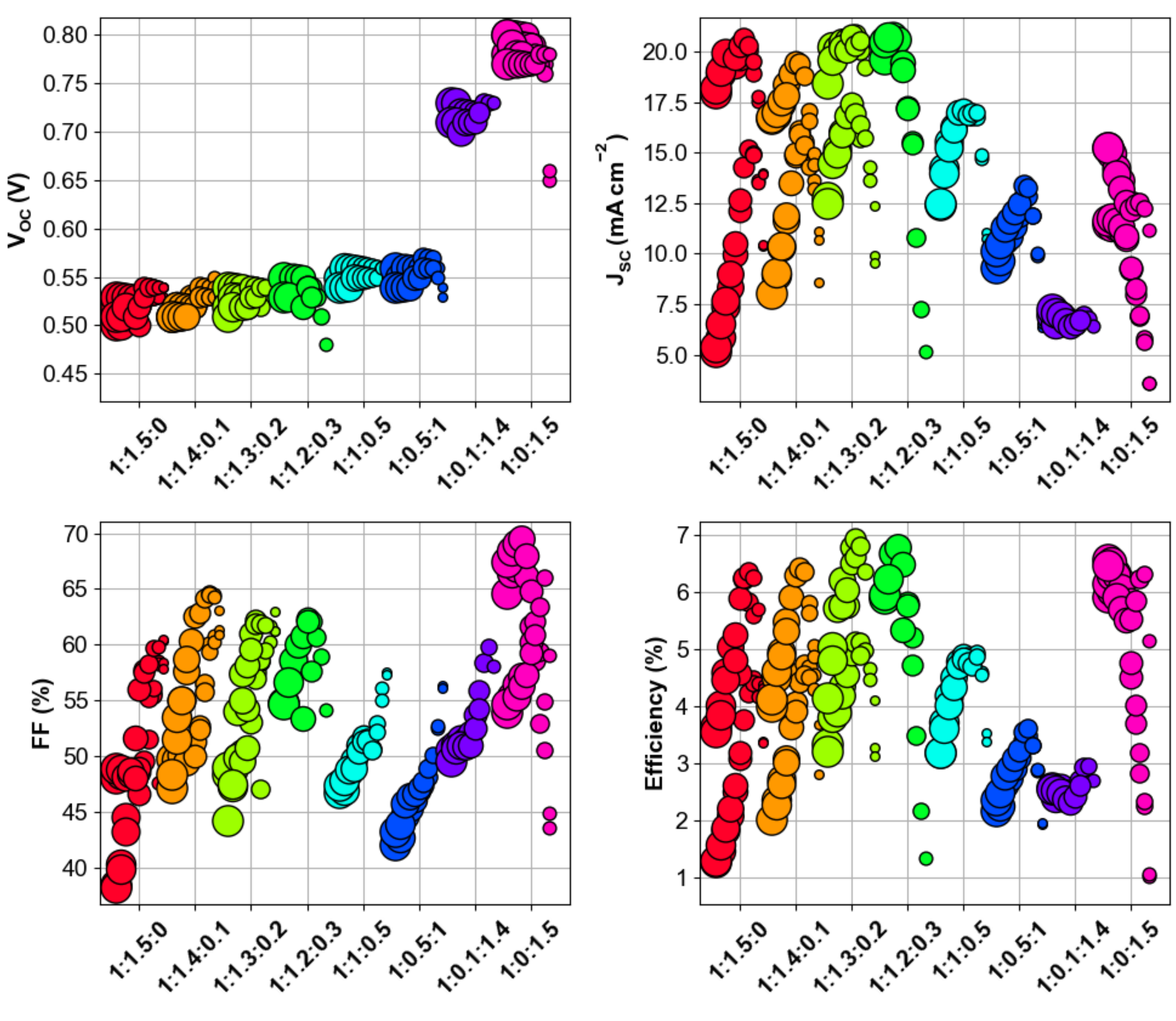

**Figure S1.** JV curve parameters of ternary blend PTB7-Th:COTIC-4F:PC70BM.

The size of the symbols scales up with the active layer thickness. A modest improvement in overall PCE is found for the ternary blend 1:1.3:0.2, mainly given by FF increase.

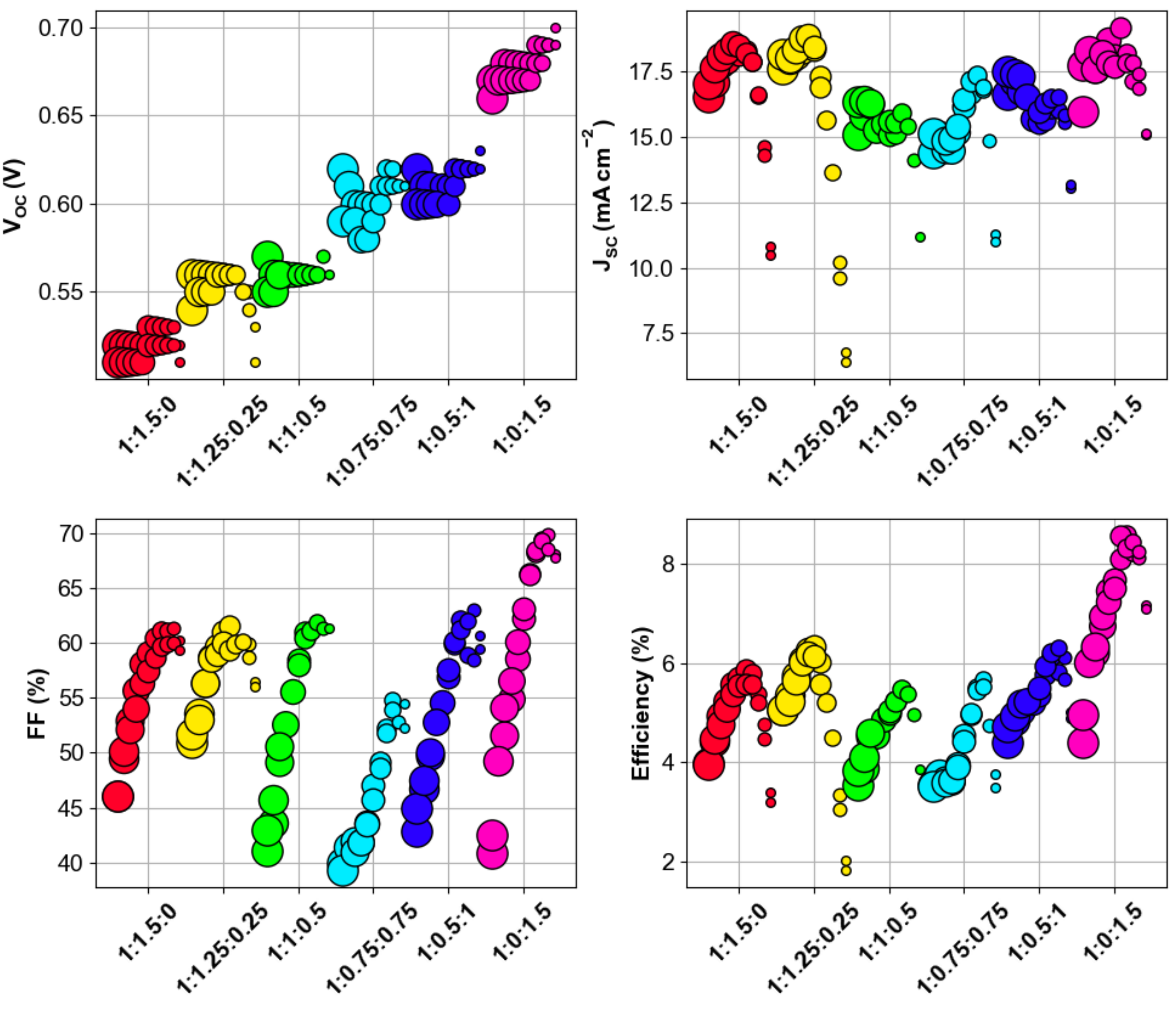

**Figure S2.** JV curve parameters of ternary blend PTB7-Th:COTIC-4F:IEICO-4F.

The size of the symbols scales up with the active layer thickness. A small PCE improvement is found for the 1:1.25:0.25 ternary blend, which comes from an improvement in $V_{OC}$. Further addition of IEICO-4F results in an assumable loss of $J_{SC}$.

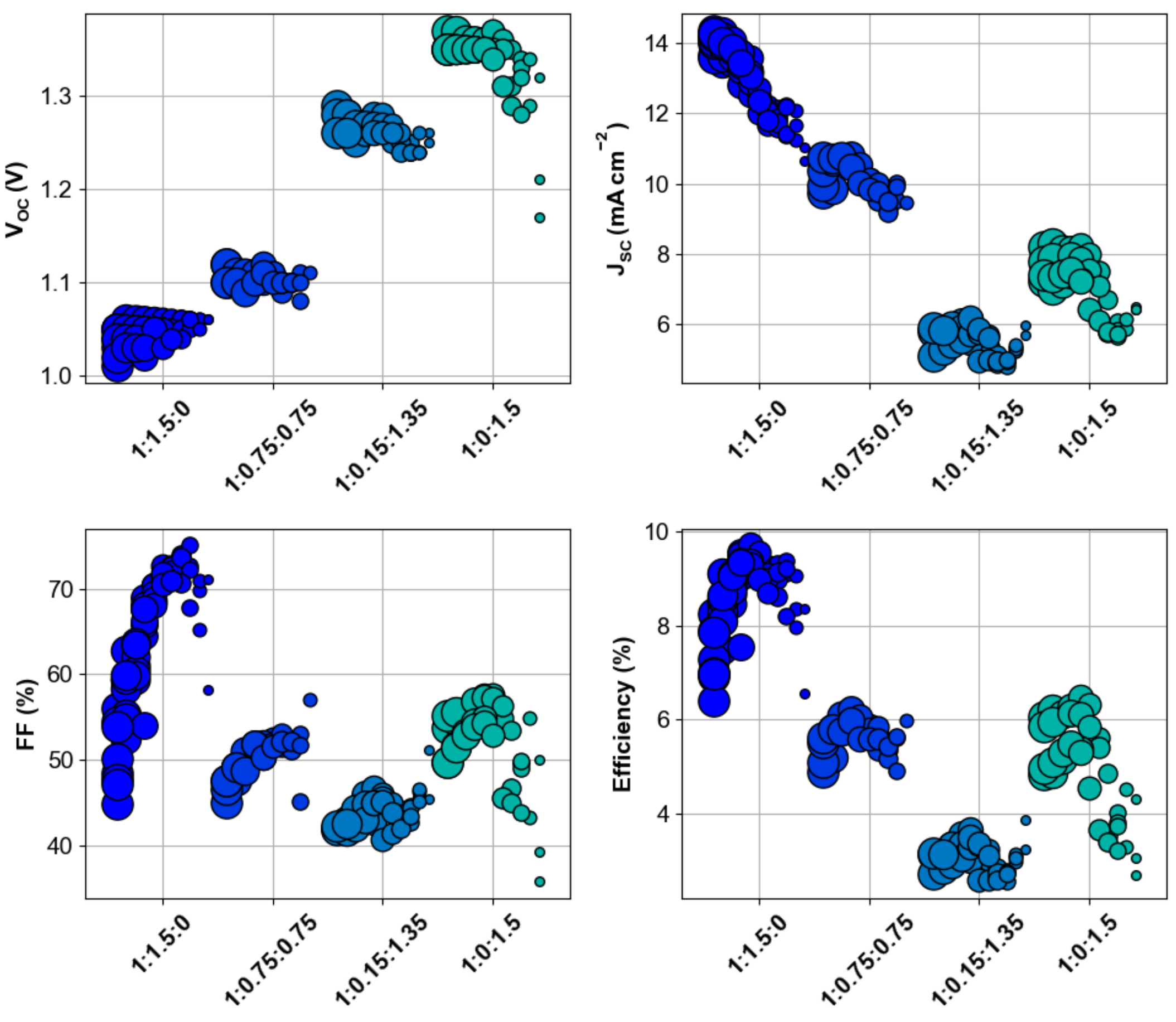

**Figure S3.** JV curve parameters of ternary blend PTQ10:FCC-Cl:O-IDFBR.

The size of the symbols scales up with the active layer thickness. $V_{OC}$ ternary-blend values fall between those of the two binaries but $J_{SC}$ and FF in ternary combinations yield worse results.

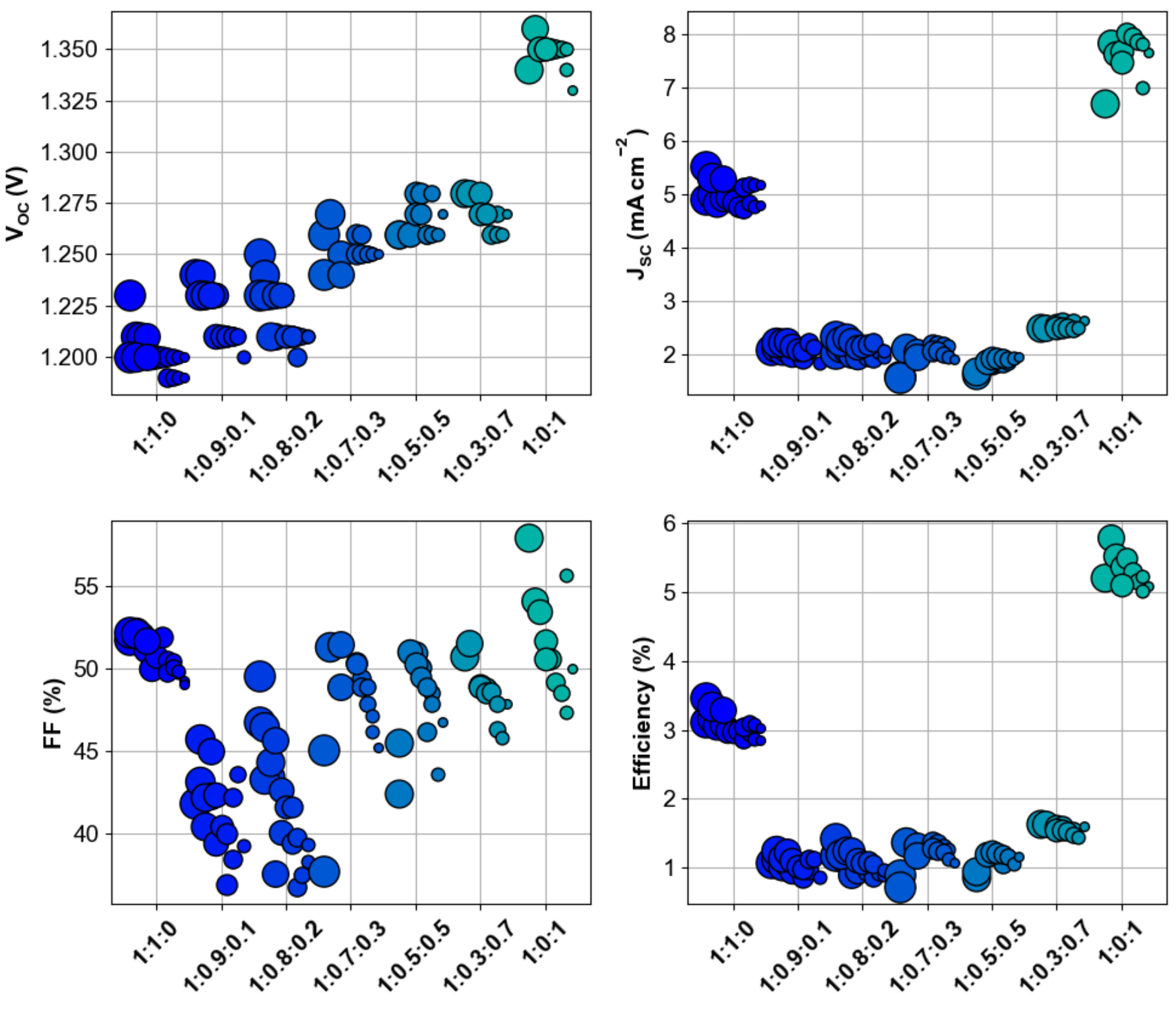

**Figure S4.** JV curve parameters of ternary blend PTQ10:EH-IDTBR:O-IDFBR.

The size of the symbols scales up with the active layer thickness. $V_{OC}$ ternary-blend values fall between those of the two binaries but $J_{SC}$ and FF in ternary combinations yield worse results.

**Table S1.** Thicknesses for every pixel and every ratio of PTB7-Th:COTIC-4F:BTP-eC9. The champion pixels under 1 sun illumination are highlighted.

| Blend ratio PTB7-Th:COTIC-4F:BTP-eC9 \ Pixel position | 1:1.5:0 | 1:1.2:0.3 | 1:0.9:0.6 | 1:0.6:0.9 | 1:0.3:1.2 | 1:0:1.5 |
|---|---|---|---|---|---|---|
| 1 | 340 | 300 | 333 | 273 | 340 | 331 |
| 2 | 303 | 298 | 267 | 265 | 301 | 284 |
| 3 | 267 | 270 | 254 | 228 | 295 | 257 |
| 4 | 248 | 213 | 239 | 210 | 243 | 226 |
| 5 | 225 | 213 | 210 | 179 | 236 | 194 |
| 6 | 191 | 169 | 187 | 159 | 180 | 176 |
| 7 | 186 | 145 | 160 | 139 | 152 | 149 |
| 8 | 138 | 119 | 143 | 115 | **119** | **117** |
| 9 | **113** | **108** | **100** | **93** | 109 | 96 |
| 10 | 95 | 102 | 81 | 68 | 72 | 73 |
| 11 | 49 | 47 | 50 | 42 | 58 | 49 |
| 12 | 28 | | 27 | 28 | | 29 |

## S3. GIWAXS measurements

GIWAXS measurements include the 2D diffractograms obtained at an incidence angle of 0.12° and linecuts of the in-plane (IP) and out-of-plane (OOP) directions. Analysis is performed by fitting the peaks of the linecuts.

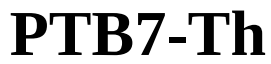
PTB7-Th

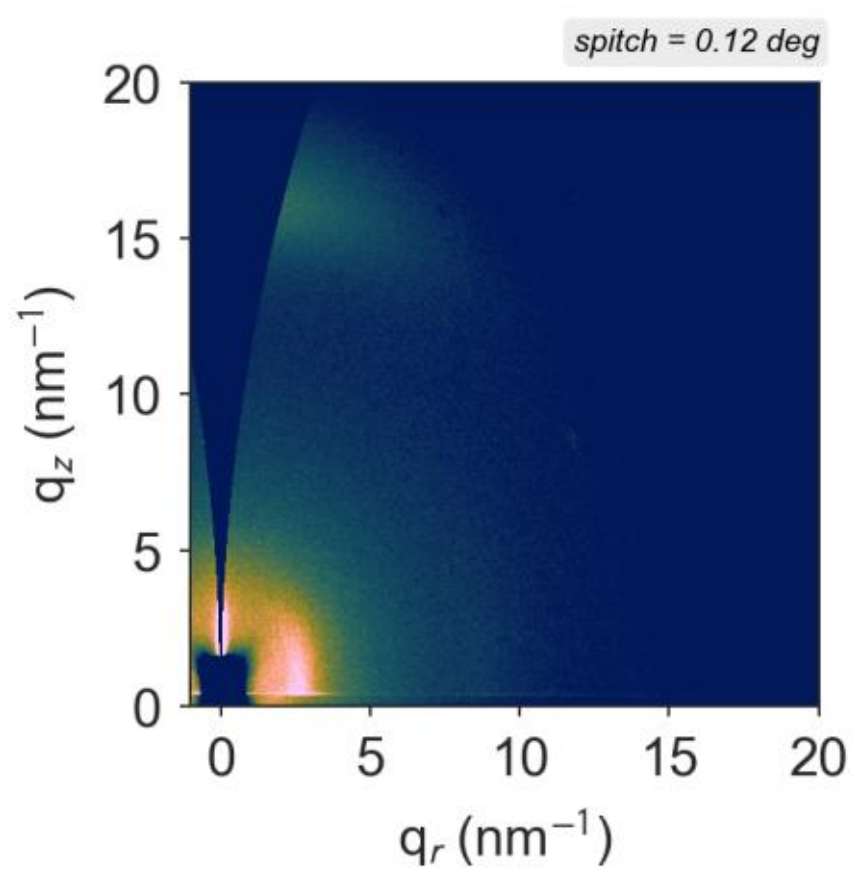

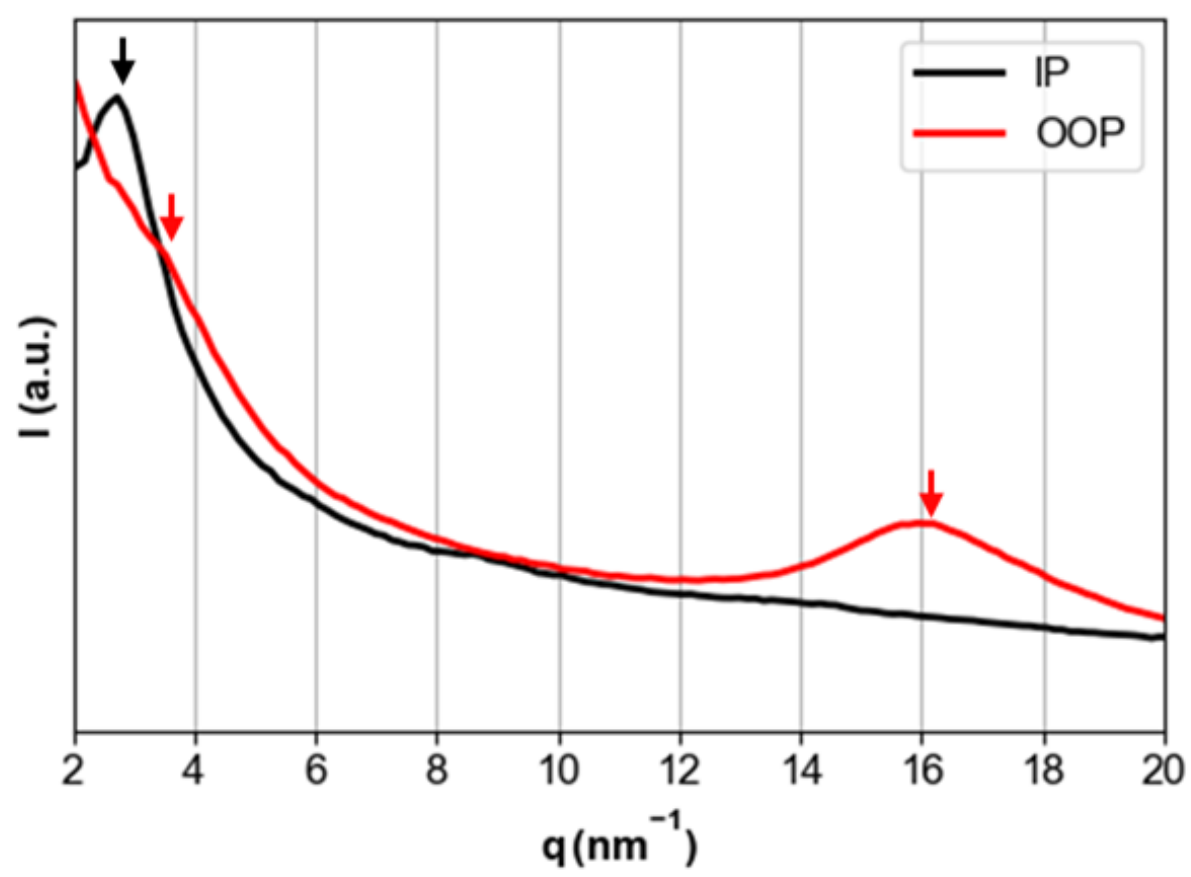

**COTIC-4F**

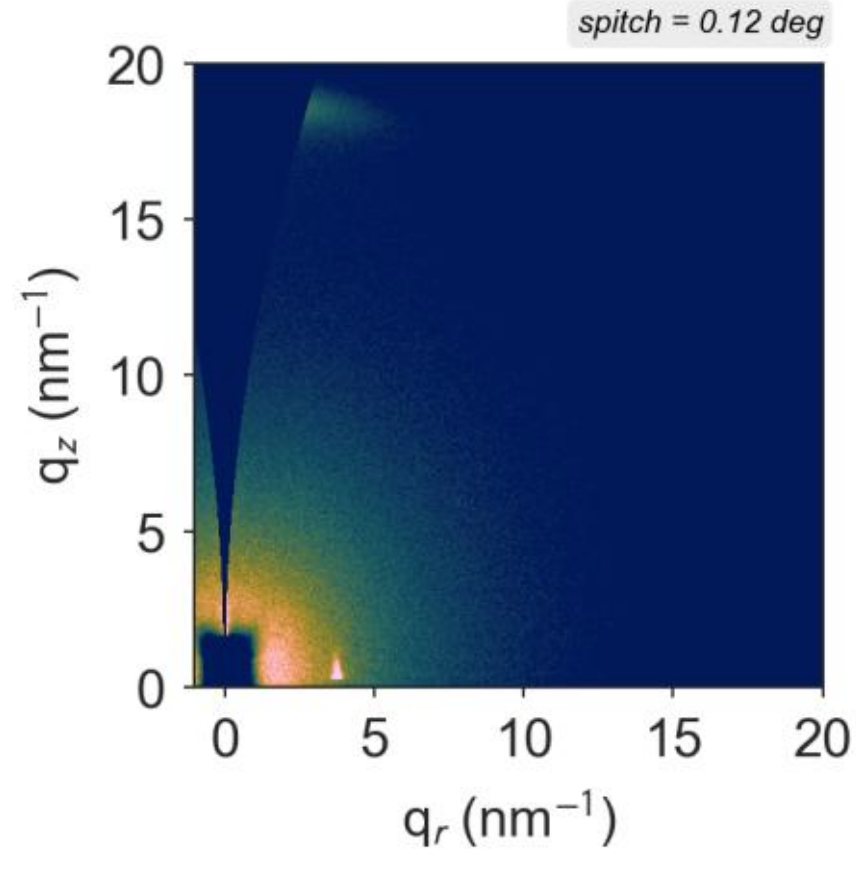

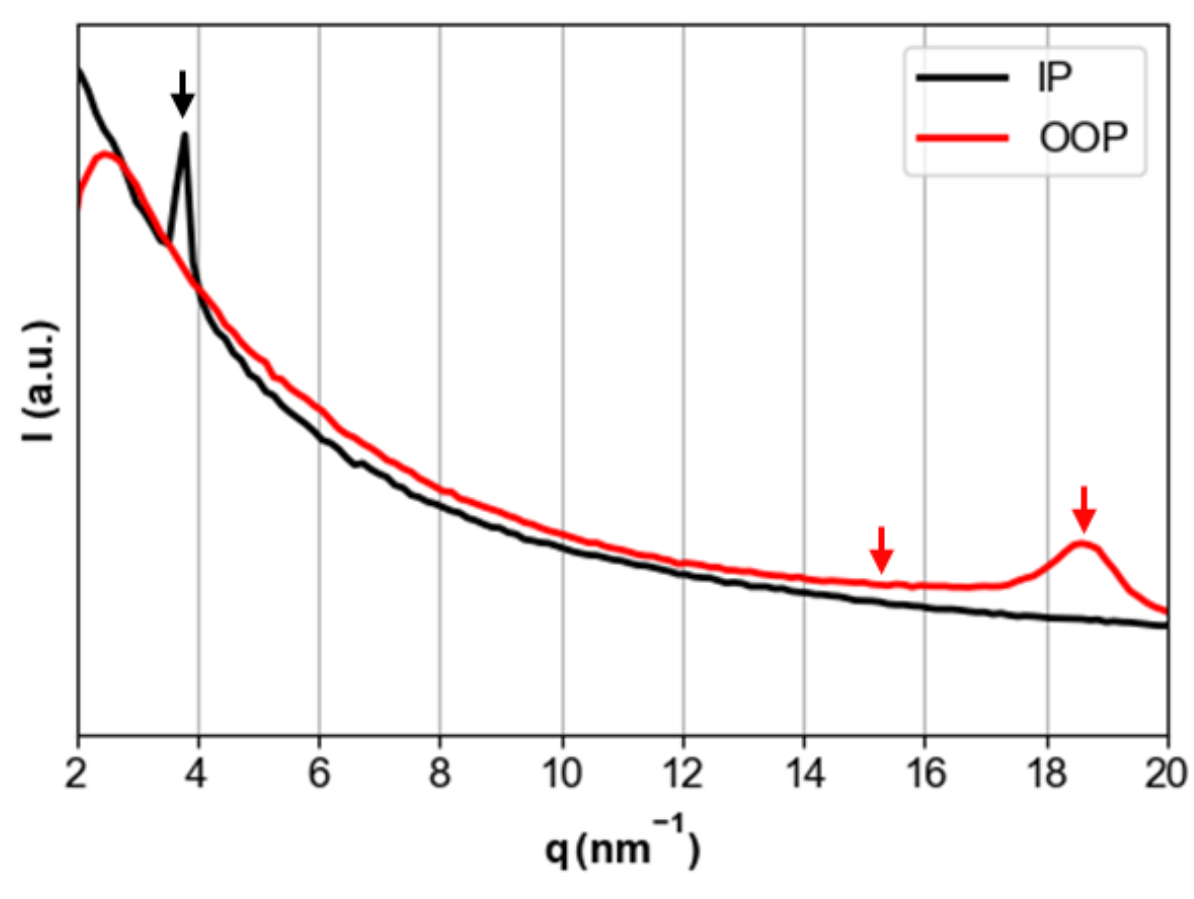

**IEICO-4F**

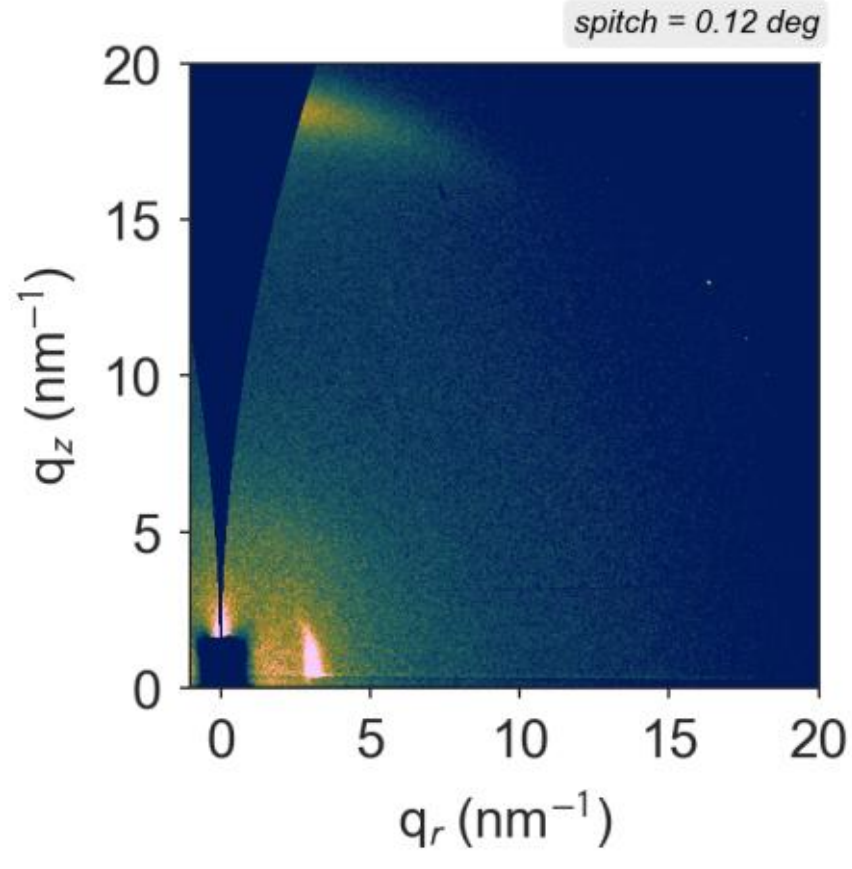

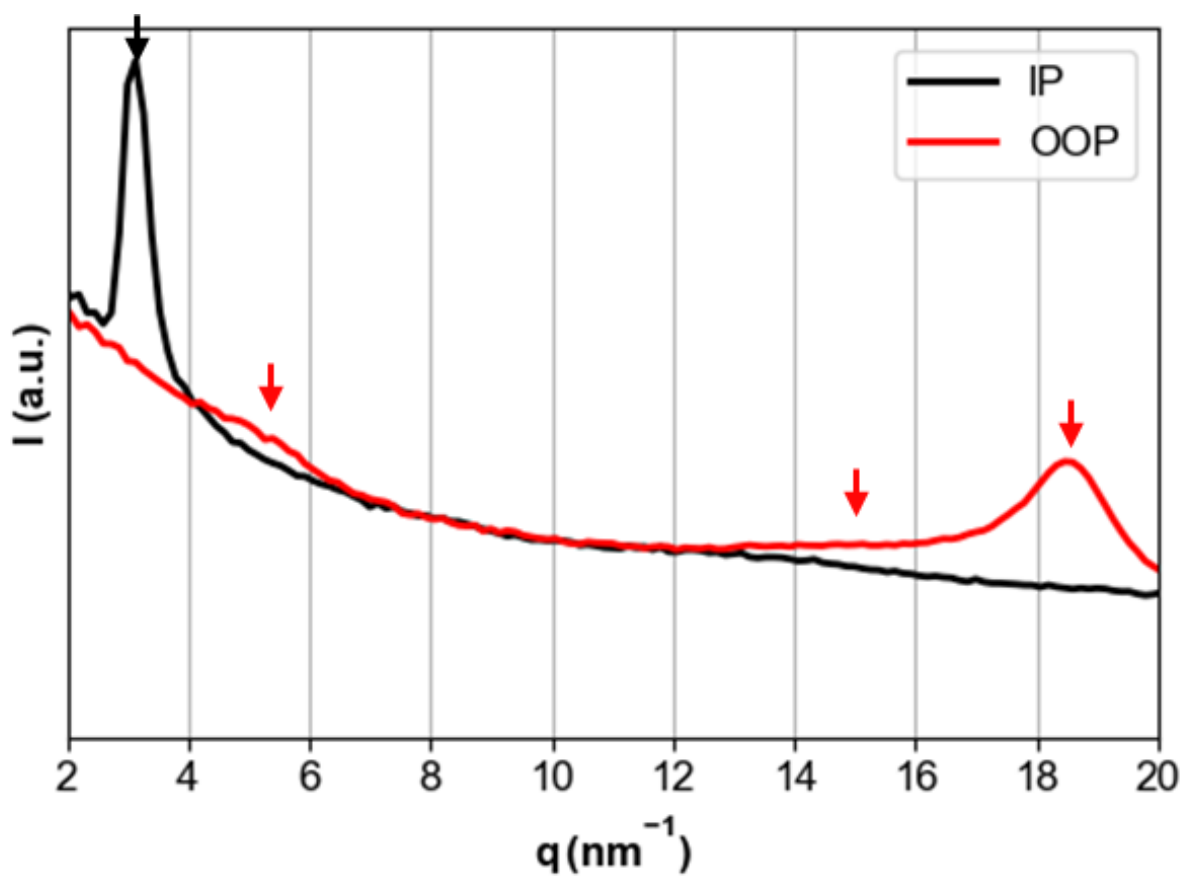

**PTB7-Th:COTIC-4F**

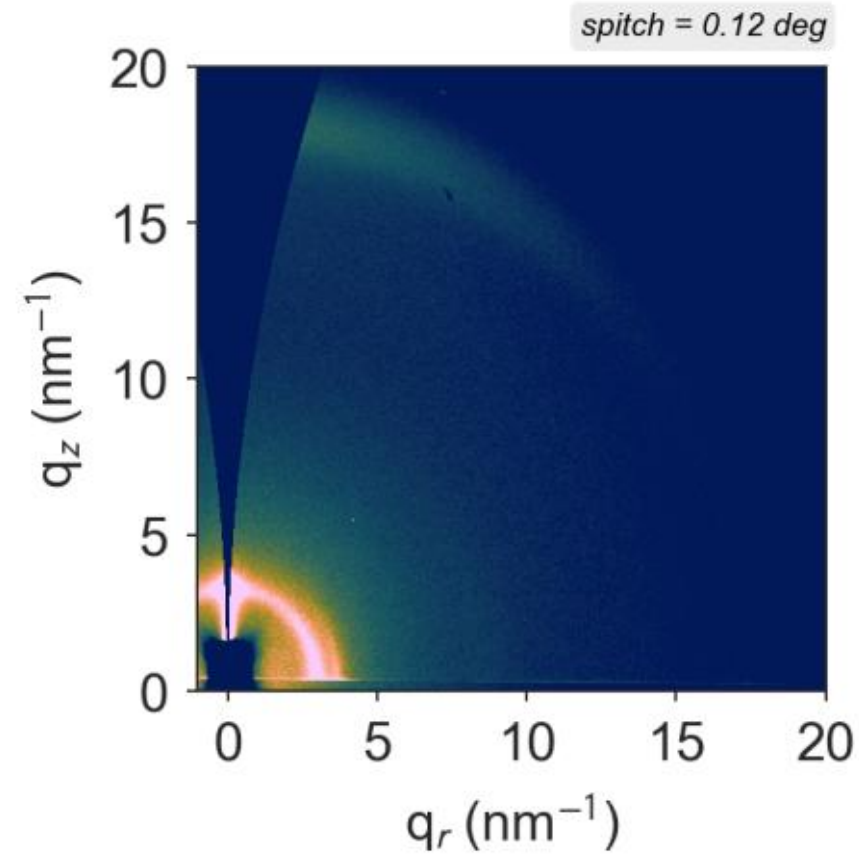

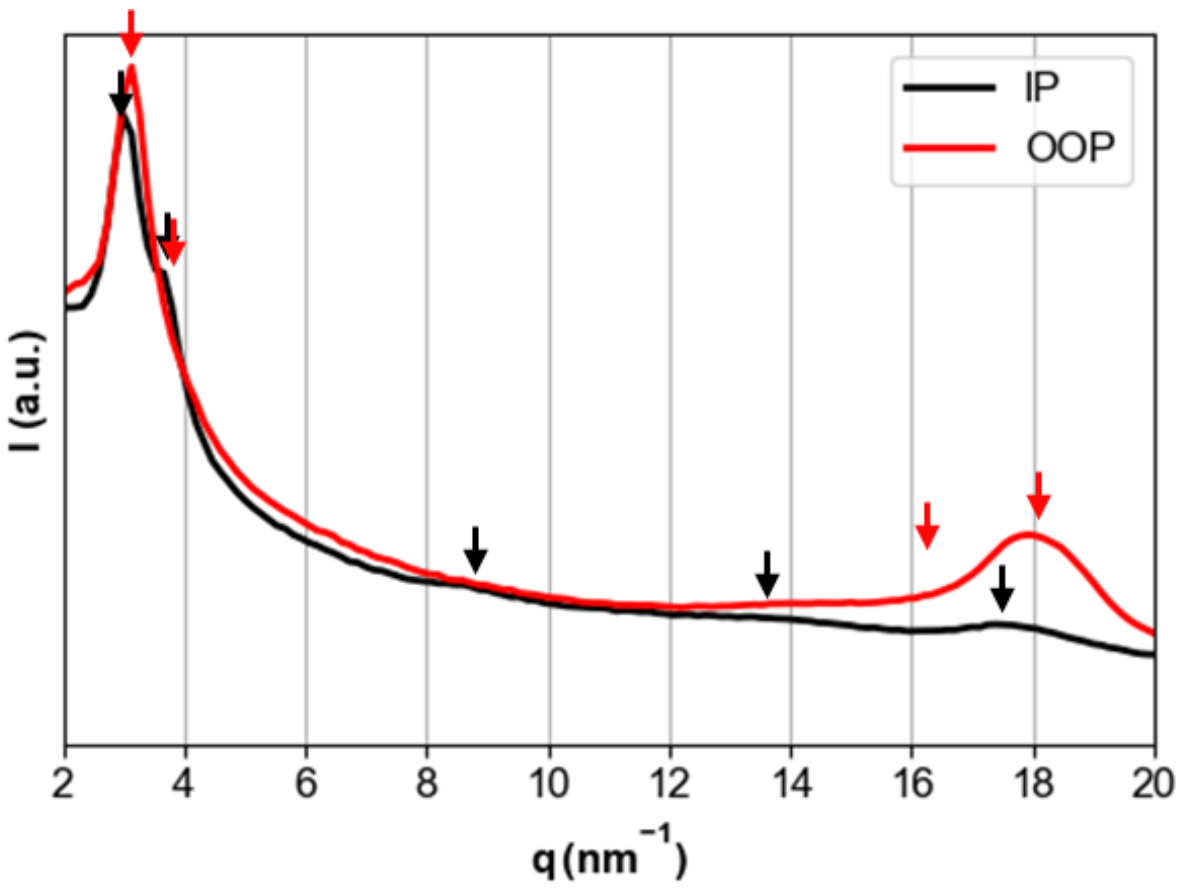

**PTB7-Th:BTP-eC9**

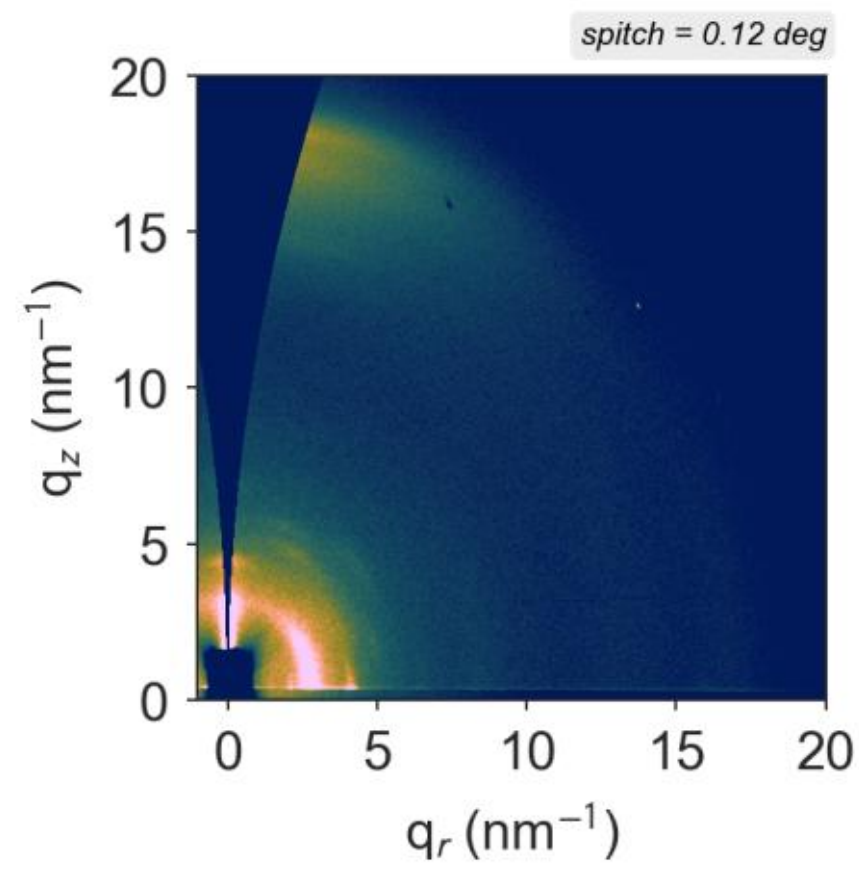

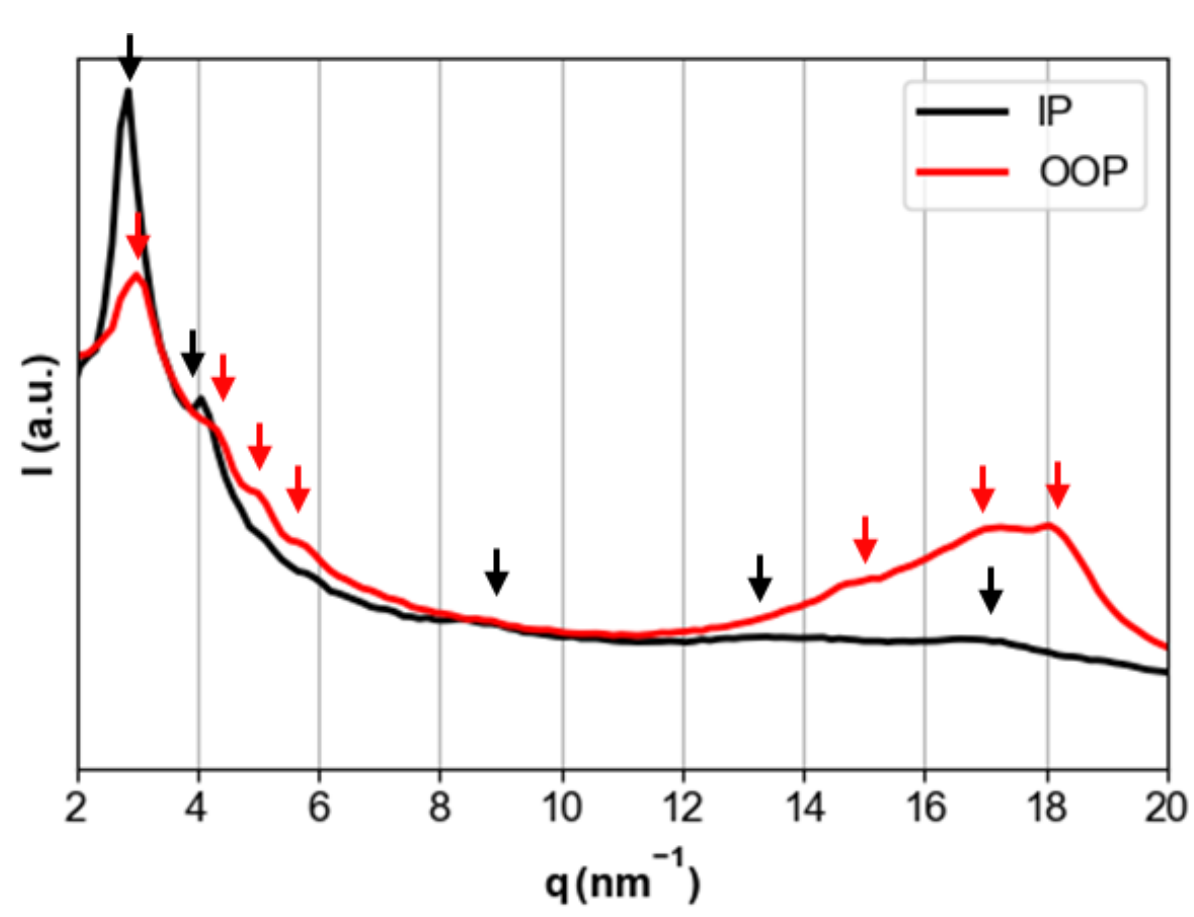

**PTB7-Th:IEICO-4F**

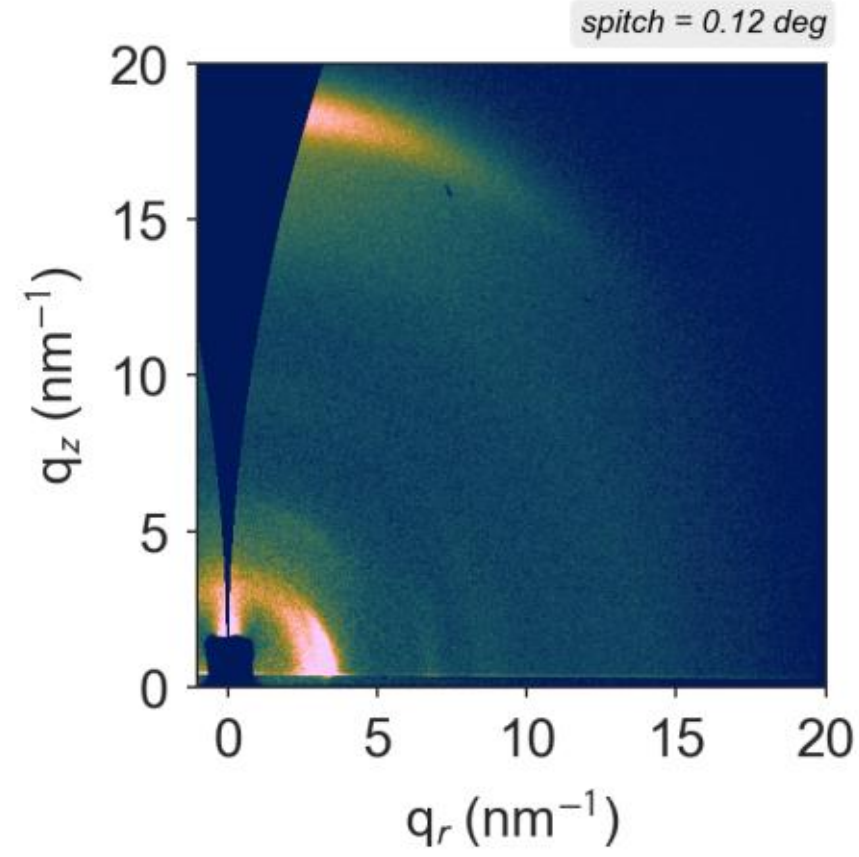

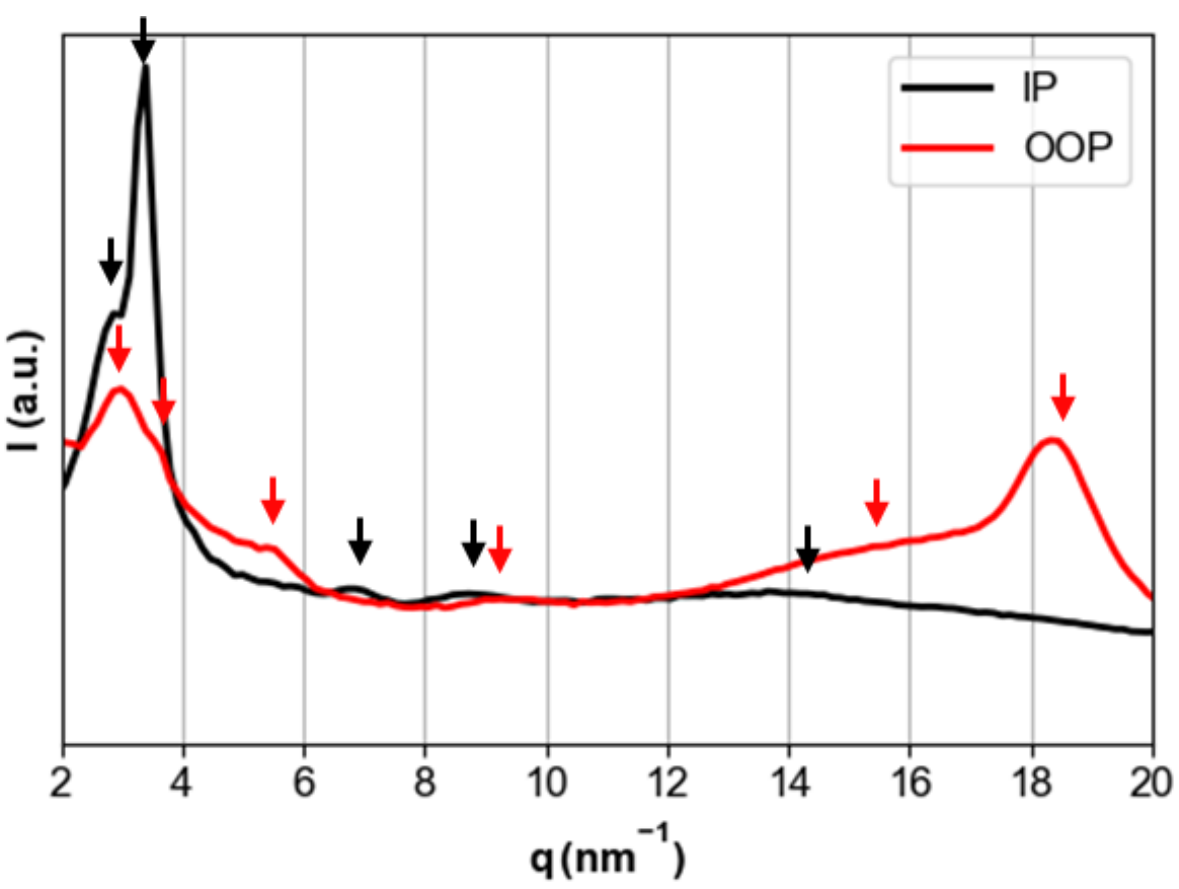

**PTB7-Th:COTIC-4F:BTP-eC9 (1:1.2:0.3)**

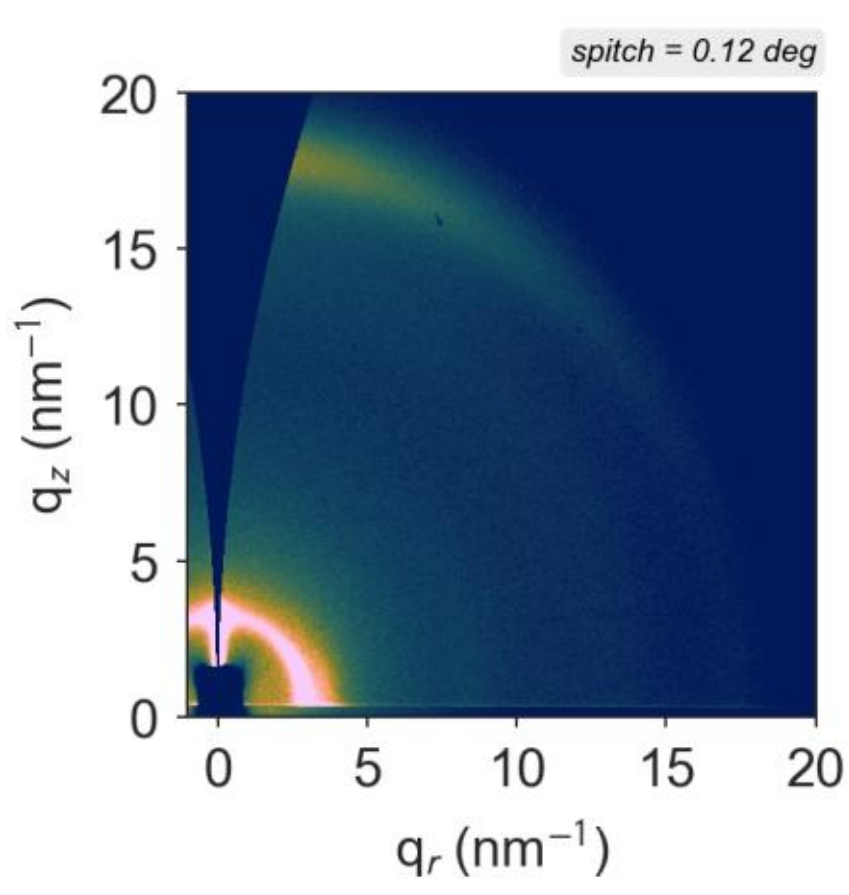

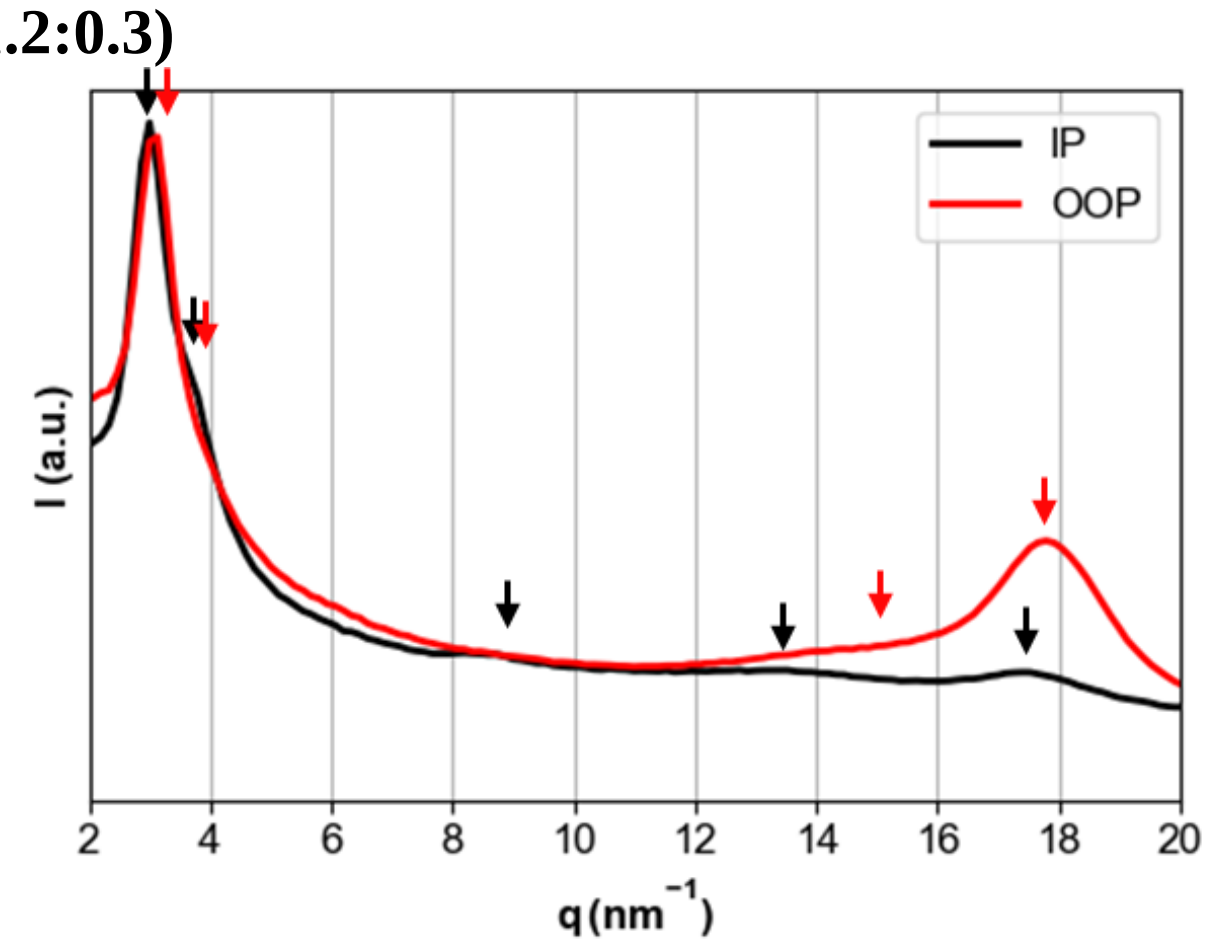

**PTB7-Th:COTIC-4F:BTP-eC9 (1:0.9:0.6)**

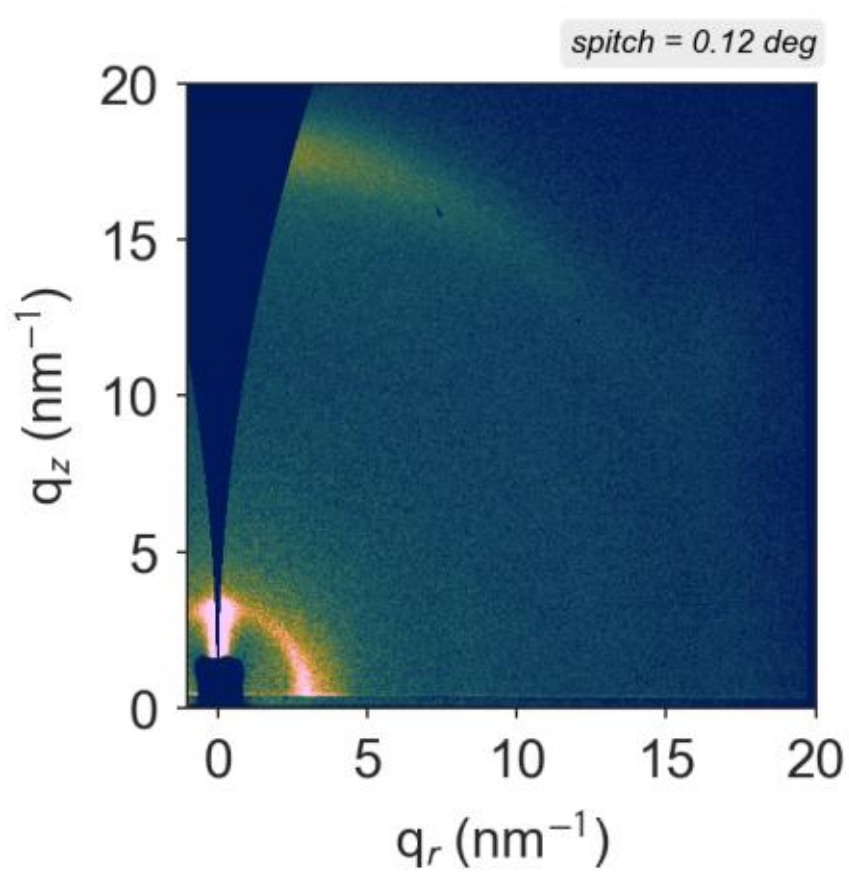

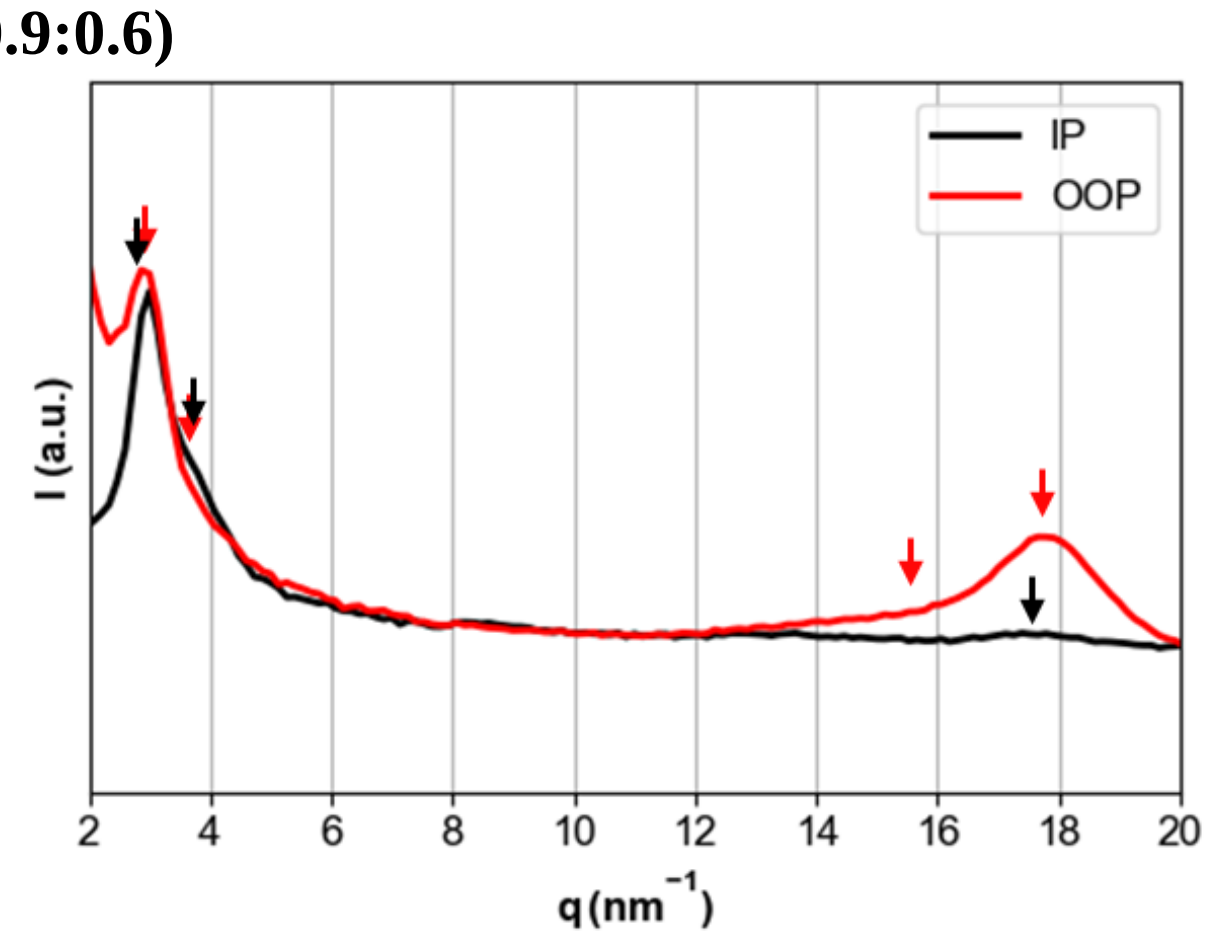

**PTB7-Th:COTIC-4F:BTP-eC9 (1:0.6:0.9)**

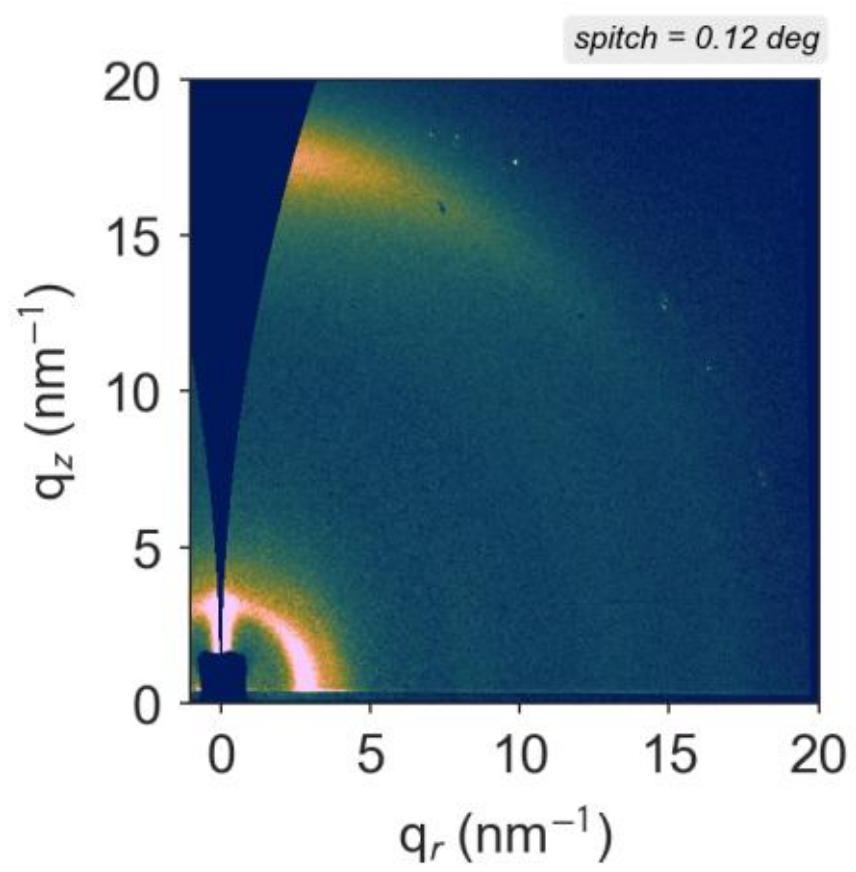

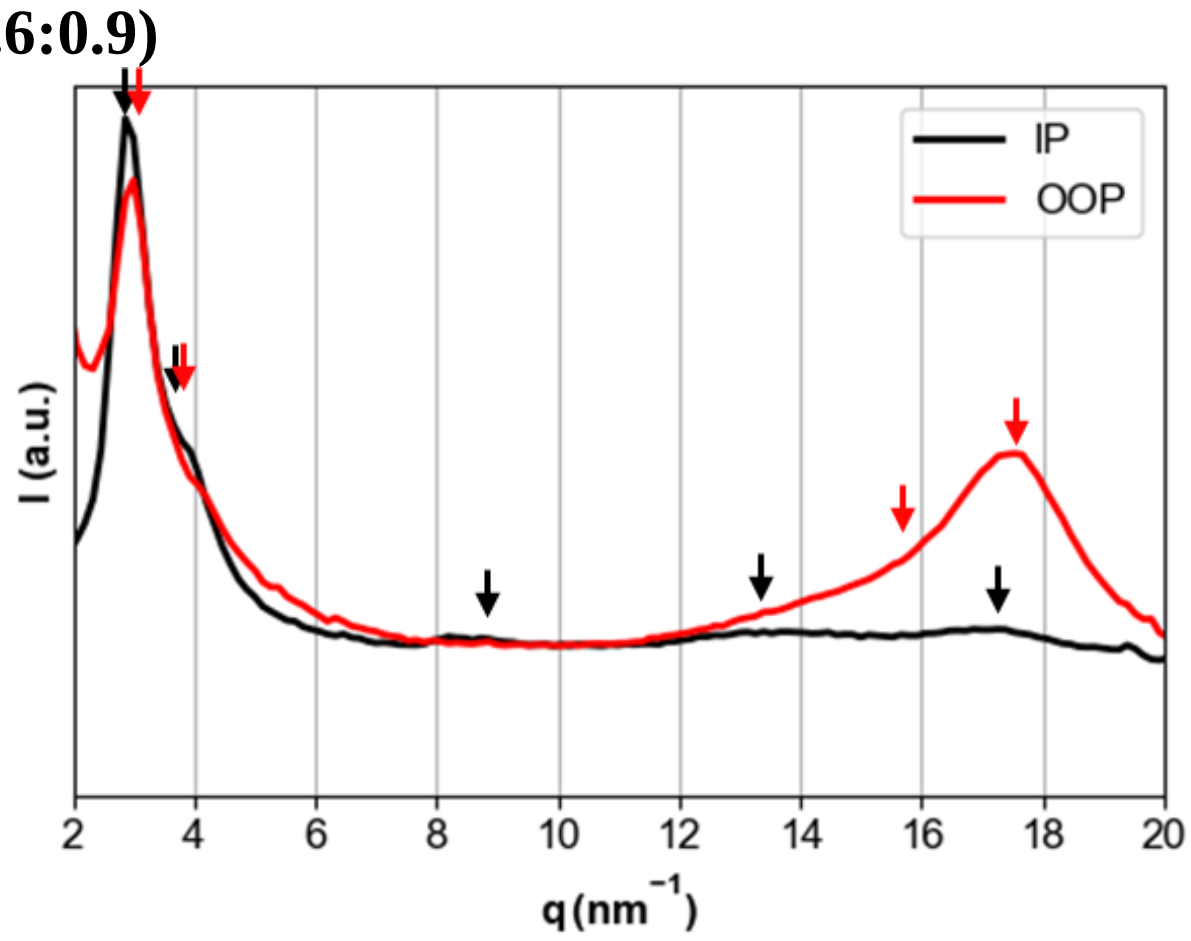

**PTB7-Th:COTIC-4F:BTP-eC9 (1:0.3:1.2)**

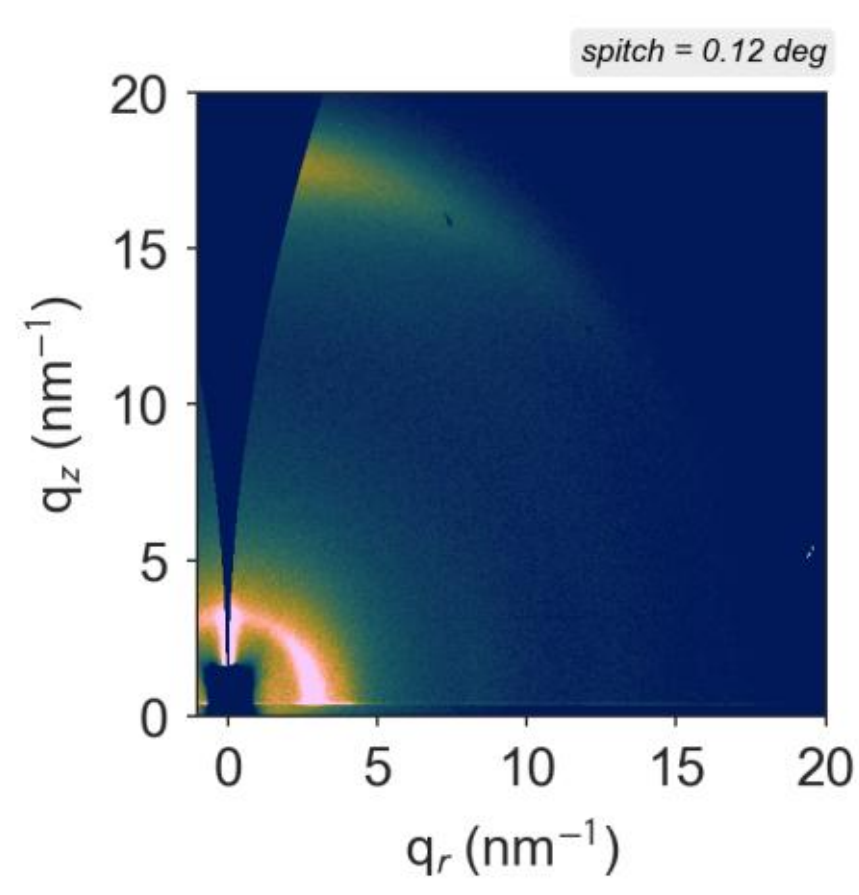

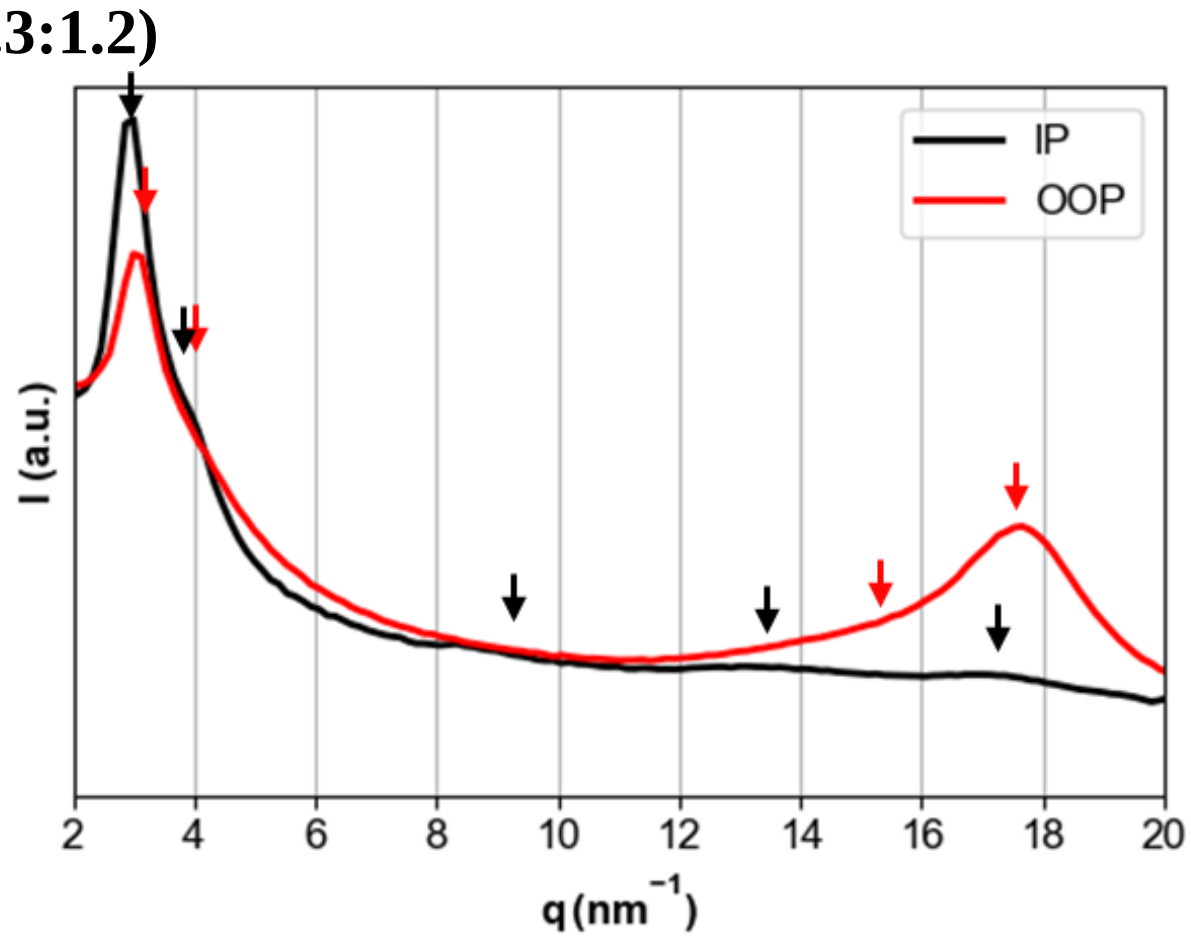

**PTB7-Th:COTIC-4F:IEICO-4F (1:0.75:0.75)**

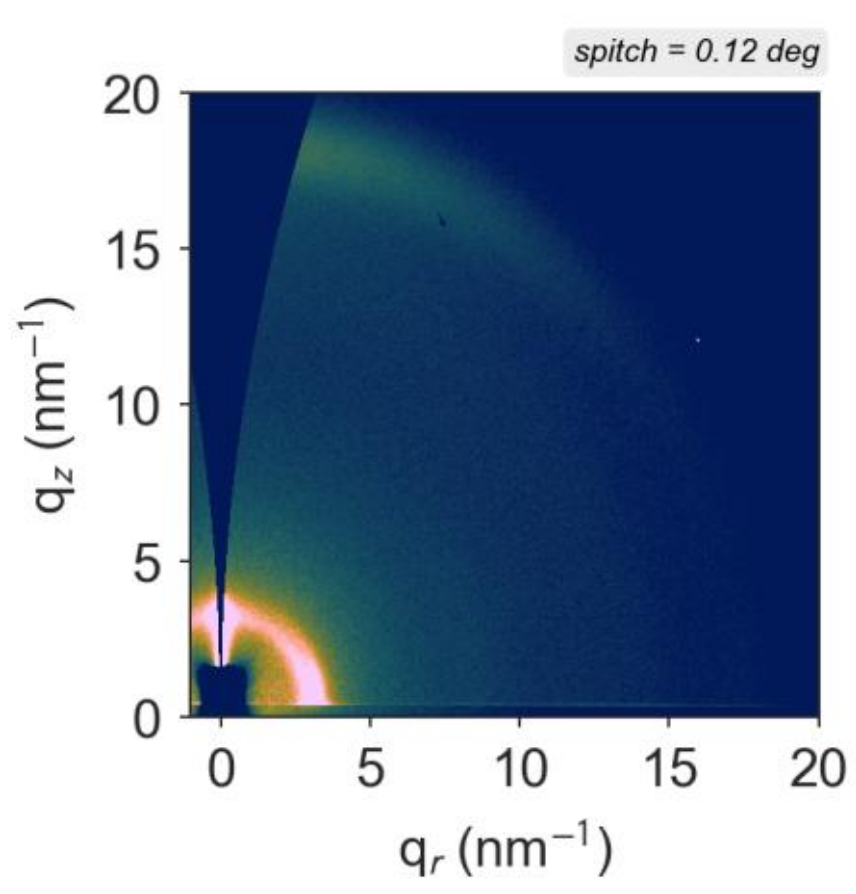

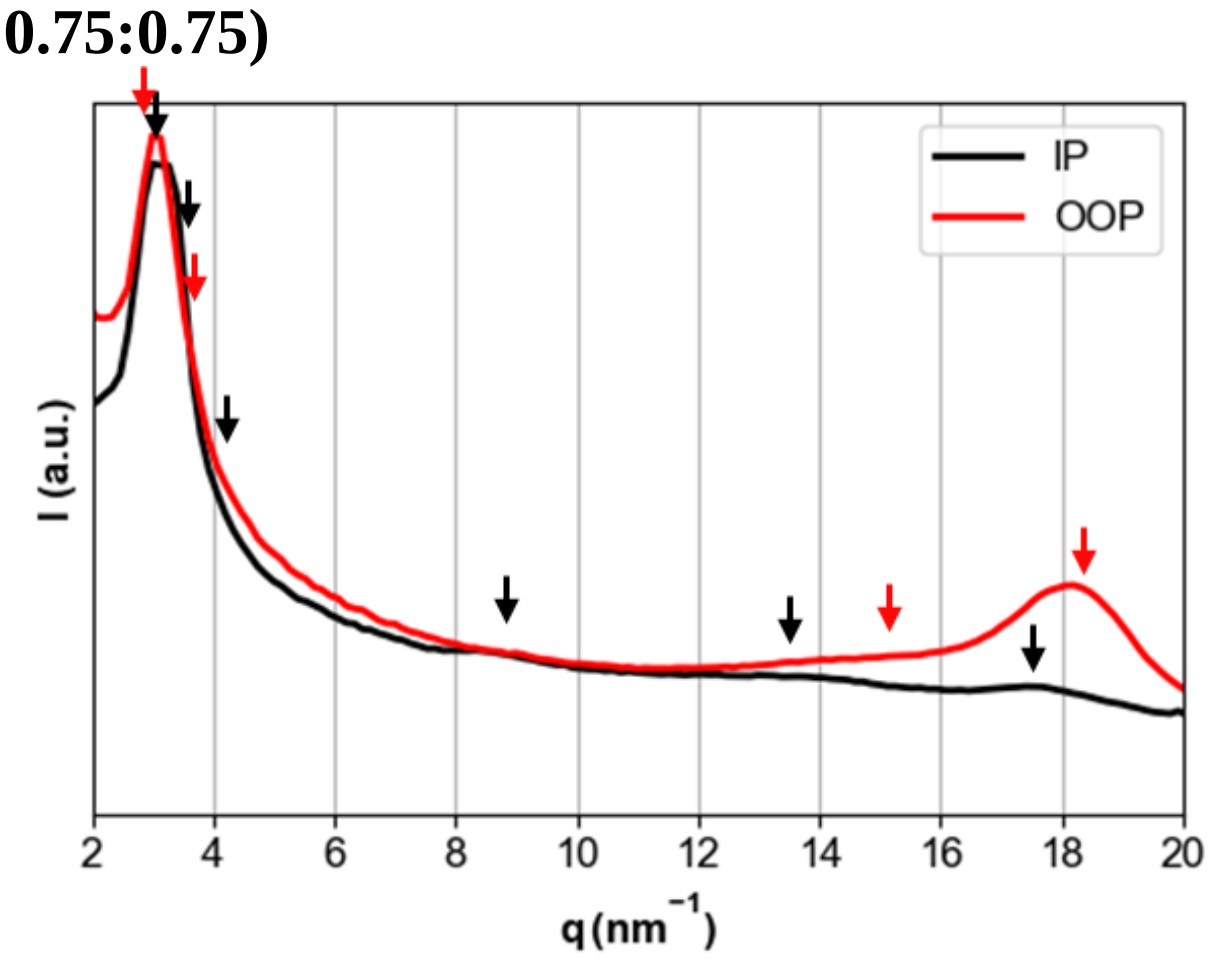

| **PTB7-Th** | IP | OOP | |
|---|---|---|---|
| Height | 0.89 | 0.18 | 0.30 |
| q (nm$^{-1}$) | 2.79 | 3.63 | 16.19 |
| FWHM | 1.67 | 1.89 | 3.93 |
| d (nm) | 2.25 | 1.73 | 0.39 |
| $L_c$ (nm) | 3.38 | 2.99 | 1.44 |
| g (%) | 30.88 | 28.83 | 19.65 |

| **COTIC-4F** | IP | OOP | |
|---|---|---|---|
| Height | 1.12 | 0.21 | 0.47 |
| q (nm$^{-1}$) | 3.77 | 15.40 | 18.60 |
| FWHM | 0.21 | 9.35 | 1.36 |
| d (nm) | 1.67 | 0.41 | 0.34 |
| $L_c$ (nm) | 27.01 | 0.60 | 4.15 |
| g (%) | 9.41 | 31.09 | 10.80 |

| **IEICO-4F** | IP | OOP | | |
|---|---|---|---|---|
| Height | 1.73 | 0.14 | 0.15 | 0.58 |
| q (nm$^{-1}$) | 3.15 | 5.03 | 16.02 | 18.52 |
| FWHM | 0.48 | 2.23 | 5.43 | 1.69 |
| d (nm) | 2.00 | 1.25 | 0.39 | 0.34 |
| $L_c$ (nm) | 11.74 | 2.53 | 1.04 | 3.34 |
| g (%) | 15.61 | 26.59 | 23.23 | 12.05 |

| **PTB7-Th:COTIC-4F** | IP | | | | | OOP | | | |
|---|---|---|---|---|---|---|---|---|---|
| Height<br>q ($nm^{-1}$)<br>FWHM | 2.43<br>3.03<br>0.67 | 1.02<br>3.70<br>0.65 | 0.07<br>8.79<br>1.74 | 0.28<br>13.68<br>9.14 | 0.16<br>17.80<br>2.03 | 2.87<br>3.13<br>0.75 | 0.21<br>3.96<br>0.88 | 0.43<br>15.20<br>8.15 | 0.85<br>18.08<br>2.14 |
| d (nm)<br>$L_c$ (nm)<br>g (%) | 2.07<br>8.41<br>18.78 | 1.70<br>8.69<br>16.72 | 0.71<br>3.24<br>17.77 | 0.46<br>0.62<br>32.62 | 0.35<br>2.79<br>13.46 | 2.01<br>7.51<br>19.57 | 1.59<br>6.43<br>18.80 | 0.41<br>0.69<br>29.21 | 0.35<br>2.65<br>13.72 |

| **PTB7-Th:BTP-eC9** | IP | | | | | | |
|---|---|---|---|---|---|---|---|
| Height<br>q ($nm^{-1}$)<br>FWHM | 3.12<br>2.86<br>0.66 | 0.95<br>3.98<br>1.34 | 0.09<br>8.80<br>1.75 | 0.25<br>13.67<br>5.99 | 0.23<br>17.10<br>3.88 | | |
| d (nm)<br>$L_c$ (nm)<br>g (%) | 2.20<br>8.55<br>19.20 | 1.58<br>4.23<br>23.13 | 0.71<br>3.23<br>17.80 | 0.46<br>0.94<br>26.41 | 0.37<br>1.46<br>18.99 | | |
| | OOP | | | | | | |
| Height<br>q ($nm^{-1}$)<br>FWHM | 0.86<br>3.03<br>1.69 | 0.05<br>3.64<br>0.19 | 0.24<br>5.01<br>2.31 | 0.13<br>5.74<br>0.61 | 0.80<br>15.65<br>5.72 | 0.54<br>17.08<br>1.97 | 0.72<br>18.22<br>1.53 |
| d (nm)<br>$L_c$ (nm)<br>g (%) | 2.08<br>3.35<br>29.78 | 1.73<br>29.55<br>9.15 | 1.25<br>2.45<br>27.07 | 1.09<br>9.25<br>13.01 | 0.40<br>0.99<br>24.12 | 0.37<br>2.87<br>13.55 | 0.34<br>3.71<br>11.54 |

| **PTB7-Th:IEICO-4F** | IP | | | | | |
|---|---|---|---|---|---|---|
| Height<br>q ($nm^{-1}$)<br>FWHM | 0.80<br>2.77<br>0.90 | 1.85<br>3.38<br>0.40 | 0.04<br>6.85<br>0.49 | 0.04<br>8.85<br>1.02 | 0.19<br>14.02<br>9.54 | |
| d (nm)<br>$L_c$ (nm)<br>g (%) | 2.27<br>6.31<br>22.69 | 1.86<br>14.08<br>13.76 | 0.92<br>11.45<br>10.71 | 0.71<br>5.53<br>13.56 | 0.45<br>0.59<br>32.91 | |
| | OOP | | | | | |
| Height<br>q ($nm^{-1}$)<br>FWHM | 0.86<br>3.03<br>1.69 | 0.05<br>3.64<br>0.19 | 0.24<br>5.01<br>2.31 | 0.13<br>9.42<br>5.66 | 0.34<br>15.88<br>5.65 | 0.63<br>18.42<br>1.61 |
| d (nm)<br>$L_c$ (nm)<br>g (%) | 2.08<br>3.35<br>29.78 | 1.73<br>29.55<br>9.15 | 1.25<br>2.45<br>27.07 | 0.67<br>1.00<br>30.94 | 0.40<br>1.00<br>23.79 | 0.34<br>3.50<br>11.81 |

| **PTB7-Th:COTIC-4F: BTP-eC9 (1:1.2:0.3)** | IP | | | | | OOP | | | |
|---|---|---|---|---|---|---|---|---|---|
| Height | 3.66 | 1.35 | 0.09 | 0.27 | 0.24 | 3.63 | 0.34 | 0.56 | 1.39 |
| q ($nm^{-1}$) | 2.99 | 3.70 | 8.74 | 13.60 | 17.58 | 3.08 | 3.94 | 15.07 | 17.88 |
| FWHM | 0.66 | 1.08 | 1.47 | 7.02 | 2.11 | 0.74 | 1.08 | 7.26 | 2.15 |
| d (nm) | 2.10 | 1.70 | 0.72 | 0.46 | 0.36 | 2.04 | 1.59 | 0.42 | 0.35 |
| $L_c$ (nm) | 8.58 | 5.24 | 3.86 | 0.81 | 2.68 | 7.66 | 5.21 | 0.78 | 2.64 |
| g (%) | 18.74 | 21.55 | 16.33 | 28.65 | 13.82 | 19.53 | 20.92 | 27.70 | 13.82 |

| **PTB7-Th:COTIC-4F: BTP-eC9 (1:0.9:0.6)** | IP | | | OOP | | | |
|---|---|---|---|---|---|---|---|
| Height | 1.26 | 0.78 | 0.04 | 0.87 | 0.07 | 0.12 | 0.61 |
| q ($nm^{-1}$) | 2.97 | 3.65 | 17.57 | 2.98 | 3.71 | 15.11 | 17.85 |
| FWHM | 0.62 | 1.42 | 1.37 | 0.62 | 1.05 | 4.49 | 2.24 |
| d (nm) | 2.12 | 1.72 | 0.36 | 2.11 | 1.70 | 0.42 | 0.35 |
| $L_c$ (nm) | 9.14 | 3.98 | 4.13 | 9.16 | 5.40 | 1.26 | 2.53 |
| g (%) | 18.21 | 24.90 | 11.14 | 18.17 | 21.21 | 21.73 | 14.12 |

| **PTB7-Th:COTIC-4F: BTP-eC9 (1:0.6:0.9)** | IP | | | | | OOP | | | |
|---|---|---|---|---|---|---|---|---|---|
| Height | 1.68 | 0.62 | 0.05 | 0.16 | 0.14 | 1.23 | 0.52 | 0.22 | 0.69 |
| q ($nm^{-1}$) | 2.91 | 3.72 | 8.72 | 13.49 | 17.36 | 2.97 | 3.74 | 15.68 | 17.52 |
| FWHM | 0.63 | 1.44 | 1.97 | 6.17 | 3.51 | 0.69 | 2.64 | 5.80 | 2.42 |
| d (nm) | 2.16 | 1.69 | 0.72 | 0.47 | 0.36 | 2.12 | 1.68 | 0.40 | 0.36 |
| $L_c$ (nm) | 8.97 | 3.92 | 2.87 | 0.92 | 1.61 | 8.16 | 2.14 | 0.97 | 2.34 |
| g (%) | 18.56 | 24.86 | 18.97 | 26.98 | 17.95 | 19.27 | 33.51 | 24.26 | 14.83 |

| **PTB7-Th:COTIC-4F: BTP-eC9 (1:0.3:1.2)** | IP | | | | | OOP | | | |
|---|---|---|---|---|---|---|---|---|---|
| Height | 2.97 | 1.09 | 0.10 | 0.24 | 0.20 | 1.62 | 0.88 | 0.41 | 1.20 |
| q ($nm^{-1}$) | 2.94 | 3.75 | 8.89 | 13.50 | 17.34 | 3.05 | 3.76 | 15.93 | 17.70 |
| FWHM | 0.67 | 1.41 | 2.38 | 5.80 | 3.64 | 0.76 | 2.64 | 5.47 | 2.36 |
| d (nm) | 2.14 | 1.68 | 0.71 | 0.47 | 0.36 | 2.06 | 1.67 | 0.39 | 0.36 |
| $L_c$ (nm) | 8.47 | 4.01 | 2.37 | 0.98 | 1.55 | 7.42 | 2.15 | 1.03 | 2.39 |
| g (%) | 19.00 | 24.47 | 20.65 | 26.14 | 18.28 | 19.93 | 33.40 | 23.37 | 14.58 |

| PTB7-Th:COTIC-4F: IEICO-4F (1:0.75:0.75) | IP | | | | | |
|---|---|---|---|---|---|---|
| Height<br>q (nm$^{-1}$)<br>FWHM | 2.05<br>2.97<br>0.71 | 1.74<br>3.37<br>0.54 | 0.11<br>4.07<br>0.80 | 0.06<br>8.83<br>1.47 | 0.22<br>13.76<br>7.63 | 0.13<br>17.78<br>2.06 |
| d (nm)<br>$L_c$ (nm)<br>g (%) | 2.12<br>7.99<br>19.49 | 1.86<br>10.52<br>15.93 | 1.55<br>7.11<br>17.64 | 0.71<br>3.86<br>16.26 | 0.46<br>0.74<br>29.70 | 0.35<br>2.75<br>13.57 |
| | OOP | | | | | |
| Height<br>q (nm$^{-1}$)<br>FWHM | 2.03<br>3.06<br>0.74 | 0.59<br>3.48<br>1.26 | 0.31<br>15.85<br>5.91 | 0.74<br>18.21<br>2.16 | | |
| d (nm)<br>$L_c$ (nm)<br>g (%) | 2.05<br>7.59<br>19.67 | 1.81<br>4.47<br>24.07 | 0.40<br>0.96<br>24.36 | 0.34<br>2.62<br>13.74 | | |

Linecuts were fit with Fityk.[1] The background is fit with a linear and an exponential decay function. The peaks are fit with pseudo-Voigt functions, retrieving their height, full width half maximum (FWHM), and the central peak position as the characteristic scattering vector ($q$). $q$ relates with the real-space $d$-spacing, $d = \frac{2\pi}{q}$. Using these characteristic peak parameters, we can obtain information about the crystal dimensions and quality of the microstructure exploiting additional metrics.[2] The X-ray coherence length ($L_c$) can be calculated with Scherrer's equation using a shape factor $K = 0.9$

$$L_C = \frac{2\pi K}{FWHM}.$$

$L_c$ gives information about the length over which coherence of scattered X-rays occurs. Typically, in semiconducting polymers $L_c$ is governed by cumulative disorder in the crystal lattice. Therefore, larger $L_c$ values indicate that less disordered crystalline domains exist, thus likely improving charge transport characteristics. On the other hand, and with the assumption that cumulative disorder dominates peak broadening, the degree of disorder of the (paracrystalline) lattice is estimated by the lattice distortion parameter (g), calculated as (for first-order diffraction peaks)

$$g(\%) = \sqrt{\frac{FWHM}{2\pi q}},$$

with lower values of g indicating less degree of disorder in the corresponding paracrystalline lattice.

Typically, lamellar stacking (*h*00) corresponding to the polymers is appreciated in the low-q region of the IP direction (q < 4). π-π stacking (0*k*0) of the NFA is observed in the high-q region of the OOP direction (q > 15).

## S4. RAINBOW simulations and experimental proof-of-concept

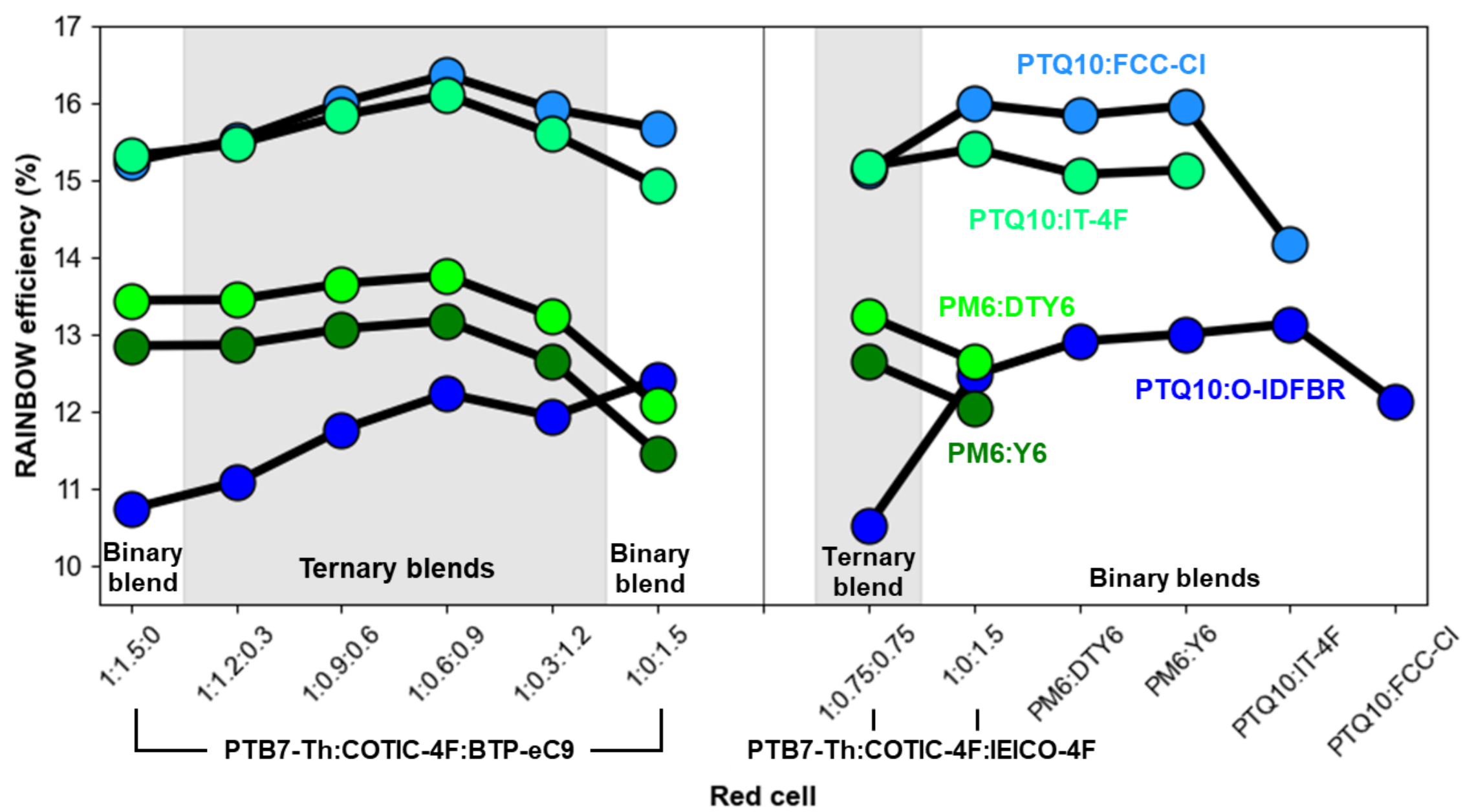

**Figure S5.** 2-junction results with other binary systems used as red cells.

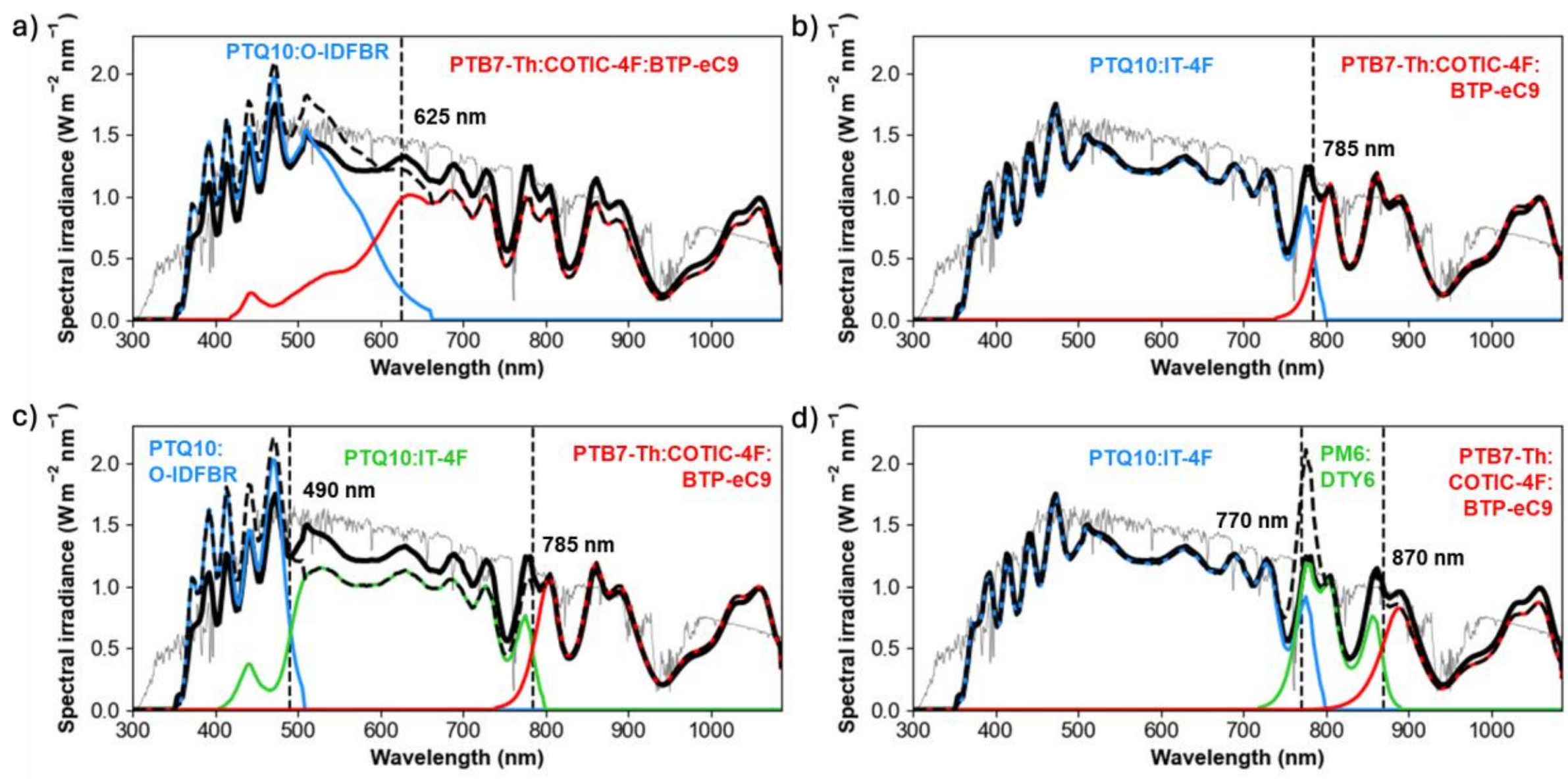

**Figure S6.** Partial spectra used for different materials in 2- and 3-junction configurations.

**Table S2.** 3-junction simulations results with a red cell of PTB7-Th:COTIC-4F:BTP-eC9 in different ratios.

| Blue cell | Green cell / Red cell | PTB7-Th: BTP-eC9 | PTB7-Th: IEICO-4F | PM6: DTY6 | PM6: Y6 | PTQ10: IT-4F | PTQ10: FCC-Cl |
|---|---|---|---|---|---|---|---|
| PTQ10:FCC-Cl | 1:1.5:0 | 17.22 | 17.24 | 17.37 | 17.47 | 16.7 | |
| | 1:1.2:0.3 | 17.23 | 17.2 | 17.38 | 17.48 | 16.86 | |
| | 1:0.9:0.6 | 17.45 | 17.38 | 17.59 | 17.69 | 17.23 | |
| | 1:0.6:0.9 | 17.56 | 17.41 | 17.71 | 17.8 | 17.5 | |
| | 1:0.3:1.2 | 17.02 | 16.96 | 17.16 | 17.26 | 16.99 | |
| | 1:0:1.5 | | | | | 16.33 | |
| PTQ10:IT-4F | 1:1.5:0 | 16.46 | 16.64 | 16.57 | 16.61 | | |
| | 1:1.2:0.3 | 16.47 | 16.6 | 16.58 | 16.62 | | |
| | 1:0.9:0.6 | 16.69 | 16.78 | 16.79 | 16.84 | | |
| | 1:0.6:0.9 | 16.8 | 16.81 | 16.89 | 16.94 | | |
| | 1:0.3:1.2 | 16.26 | 16.36 | 16.36 | 16.41 | | |
| | 1:0:1.5 | | | | | | |
| PTQ10:O-IDFBR | 1:1.5:0 | 13.74 | 14.47 | 14.49 | 15.63 | 15.34 | 15.34 |
| | 1:1.2:0.3 | 13.7 | 14.47 | 14.5 | 15.79 | 15.61 | 15.61 |
| | 1:0.9:0.6 | 18.89 | 14.69 | 14.72 | 16.16 | 16.11 | 16.11 |
| | 1:0.6:0.9 | 13.92 | 14.79 | 14.82 | 16.44 | 16.48 | 16.48 |
| | 1:0.3:1.2 | 13.46 | 14.25 | 14.28 | 15.93 | 16.04 | 16.04 |
| | 1:0:1.5 | | | | 15.26 | 15.79 | 15.79 |

**Table S3.** Simulation and experimental values of sub-cells in 2-junction and 3-junction under 1 sun and partial spectra illumination. Blue cell is PTQ10:FCC-Cl, green cell is PM6:DTY6, and red cell is PTB7-Th:COTIC-4F:BTP-eC9 (1:0.6:0.9).

| | Sim / Exp | $V_{OC}$ (V) | $J_{SC}$ (mA cm$^{-2}$) | FF (%) | PCE (%) |
|---|---|---|---|---|---|
| BLUE 1 sun | Sim | 1.06 | -16.35 | 69.94 | 12.08 |
| | Exp | 1.08 | -14.90 | 74.45 | 11.99 |
| GREEN 1 sun | Sim | 0.82 | -21.12 | 67.82 | 11.81 |
| | Exp | 0.83 | -21.92 | 68.53 | 12.54 |
| RED 1 sun | Sim | 0.60 | -22.69 | 64.24 | 8.71 |
| | Exp | 0.60 | -23.05 | 62.23 | 8.60 |
| BLUE 2J < 725 | Sim | 1.06 | -16.16 | 69.94 | 11.98 |
| | Exp | 1.08 | -14.49 | 73.69 | 11.54 |
| BLUE 3J < 715 | Sim | 1.06 | -15.96 | 69.94 | 11.83 |
| | Exp | 1.08 | -13.81 | 73.68 | 10.96 |
| GREEN 3J 715 - 865 | Sim | 0.82 | -6.47 | 67.82 | 3.62 |
| | Exp | 0.81 | -7.54 | 68.43 | 4.18 |
| RED 2J > 725 | Sim | 0.60 | -11.38 | 64.24 | 4.38 |
| | Exp | 0.58 | -11.64 | 64.36 | 4.37 |
| RED 3J > 865 | Sim | 0.60 | -5.85 | 64.20 | 2.20 |
| | Exp | 0.57 | -5.96 | 65.02 | 2.26 |

## S5. Detailed balance limit of multijunction OPV devices

The physical model was used to scan a range of devices with different energy gaps (corresponding to the LE state energy), in a range from 0.8 eV to 4.0 eV with 5 meV steps. All other model parameters were kept constant and are summarised in Table S4.

The device parameters of each slice point were simulated using the full solar spectrum, and the output values of $J_{\text{full}}(V)$, $J_{\text{dark}}(V)$, and the absorption spectrum were stored.

In the next step the values of each possible spectral "slice" were calculated based on the superposition principle (i.e. assuming that the photocurrent scales linearly with the absorbed photon flux). First the scaling factor $\gamma$ was calculated by integrating the product of the absorption and the solar spectrum within the spectral slice boundaries yielding the generation of the slice ($G_{\text{slice}}$) and normalising by the total generation from the full solar spectrum ($G_{\text{total}}$):

$$\gamma = G_{\text{slice}}/G_{\text{total}}$$

From this the generated current and the maximum power of each slice were calculated according to:

$$J_{\text{slice}}(V) = \gamma \times [J_{\text{full}}(V) - J_{\text{dark}}(V)] + J_{\text{dark}}(V)$$

$$P_{\text{max,slice}} = -min(J_{\text{slice}}(V) \times V)$$

These values were stored in a large look-up table (~2 GB) for each possible combination of slice limits (λ_start to λ_end).

Finally, we used dynamic programming to recursively find the best spectral splitting points instead of checking every possible combination of slices (which would result in computationally impossible combinatoric scaling). Quadratic scaling with the number of possible split positions was achieved with this approach, using the formula:

$$P_{\text{max}}(n_{cells}, i) = \max_k [P_{\text{max}}(i, k) + P_{\text{max}}(n_{cells} - 1, k)]$$

where $P_{\text{max}}(i, k)$ is the maximum power for the slice spanning from *i* to *k*.

The optimal cell configurations and cell parameters depending on the number of subcells are shown in Table S5.

**Table S4**. Model parameters used in the OPV device simulations.

| **Parameter** | **Value** | **Unit** |
|---|---|---|
| Thickness (d) | 100 | nm |
| Temperature (T) | 300 | K |
| Mobility (μ) | 0.001 | cm2 / Vs |
| Effective density of states for charge carriers (NC) | 5e19 | $m^{-3}$ |
| Oscillator Strength of the LE (f_LE) | 3 | (unitless) |
| Oscillator Strength of the CT (f_CT) | 0.01 | (unitless) |
| Disorder of the LE (σ_LE) | 1e-4 | eV |
| Disorder of the CT (σ_CT) | 1e-4 | eV |
| Inner and Outer Reorganisation Energy of the LE (λ_LE_inner and λ_LE_outer) | 0.075 | eV |
| Inner and Outer Reorganisation Energy of the CT (λ_inner_CT and λ_outer_LE) | 0.075 | eV |
| Density of states of LE states | 8e27 | $m^{-3}$ |
| Ratio of the number of CT to LE states | 0.1 | (unitless) |
| Electronic coupling matrix element for the CT dissociation | 0.004 | eV |
| Reorganisation energy for CT dissociation | 0.11 | eV |
| Electronic coupling matrix element for the LE dissociation | 0.03 | eV |
| Reorganisation energy for LE | 0.54 | eV |
| Energy Offset between LE and CT | 0.1 | eV |
| Energy Offset between CT and CS | 0.1 | eV |

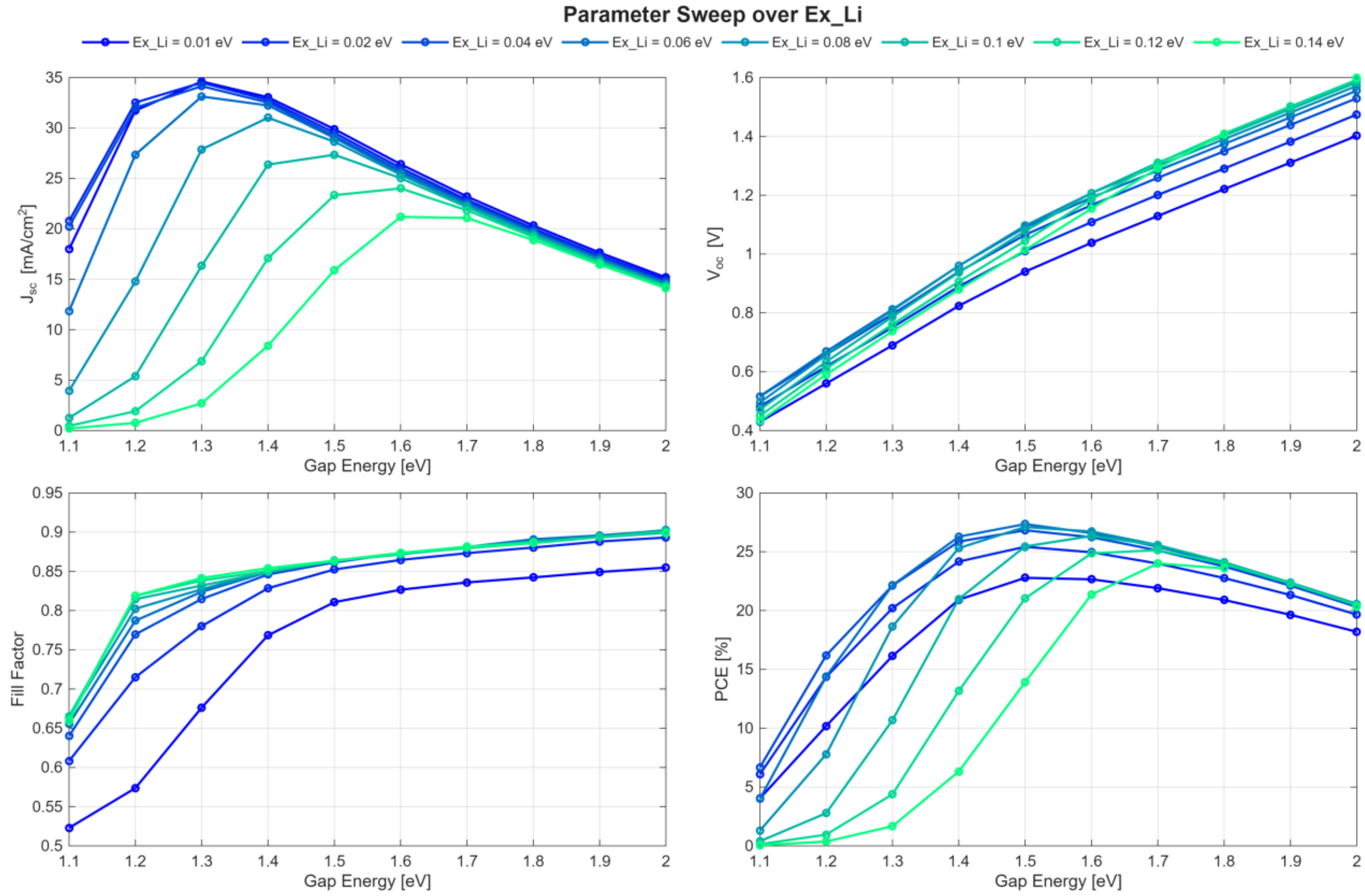

**Figure S7.** Device performance parameters from the OPV device model, showing the detailed balance limit as a function of bandgap for different values of reorganization energy.

**Table S5**. Parameters of each junction in the multi-junction cells and the combined PCE.

1 Cell:

| Cell | From | To | Bandgap | Voc | Jsc | FF | PCE | Gen_sun [1] | Gen_cell [2] |
|---|---|---|---|---|---|---|---|---|---|
| | (eV) | (eV) | (eV) | (V) | (mA/cm2) | (%) | (%) | (% of 1 sun) | (%) |
| 1 | 0 | Inf | 1.505 | 1.10 | -28.60 | 86.34 | 27.24 | 44.18 | 99.98 |
| **SUM** | | | | | | | **27.24** | **44.18** | |

[1] Gen_sun = The ratio of the slice's exciton generation rate to the total incident solar flux

[2] Gen_cell = The ratio of excitons generated within the spectral slice to the maximum possible generation if the cell received the entire solar spectrum instead of just the spectral slice.

2 Cells:

| Cell | From | To | Bandgap | Voc | Jsc | FF | PCE | Gen_sun [1] | Gen_cell [2] |
|---|---|---|---|---|---|---|---|---|---|
| | (eV) | (eV) | (eV) | (V) | (mA/cm2) | (%) | (%) | (% of 1 sun) | |
| 1 | 0 | 1.995 | 1.425 | 0.98 | -17.44 | 85.84 | 14.65 | 27.29 | 56.16 |
| 2 | 1.995 | inf | 2.030 | 1.59 | -13.39 | 90.45 | 19.31 | 20.90 | 96.60 |
| **SUM** | | | | | | | **33.96** | **48.19** | |

3 Cells:

| Cell | From | To | Bandgap | Voc | Jsc | FF | PCE | Gen_sun [1] | Gen_cell [2] |
|---|---|---|---|---|---|---|---|---|---|
| | (eV) | (eV) | (eV) | (V) | (mA/cm2) | (%) | (%) | (% of 1 sun) | (% of cell) |
| 1 | 0 | 1.710 | 1.390 | 0.92 | -11.40 | 85.63 | 8.94 | 18.14 | 36.08 |
| 2 | 1.710 | 2.305 | 1.740 | 1.31 | -12.19 | 88.97 | 14.23 | 18.85 | 57.46 |
| 3 | 2.305 | Inf | 2.330 | 1.87 | -8.03 | 91.74 | 13.76 | 12.67 | 95.05 |
| **SUM** | | | | | | | **36.93** | **49.66** | |

4 Cells:

| Cell | From | To | Bandgap | Voc | Jsc | FF | PCE | Gen_sun [1] | Gen_cell [2] |
|---|---|---|---|---|---|---|---|---|---|
| | (eV) | (eV) | (eV) | (V) | (mA/cm2) | (%) | (%) | (% of 1 sun) | (%) |
| 1 | 0 | 1.405 | 1.260 | 0.70 | -4.59 | 83.00 | 2.68 | 9.63 | 16.90 |
| 2 | 1.405 | 1.795 | 1.450 | 1.00 | -11.41 | 86.46 | 9.85 | 17.74 | 37.55 |
| 3 | 1.795 | 2.355 | 1.820 | 1.38 | -10.66 | 89.42 | 13.20 | 16.53 | 56.09 |
| 4 | 2.355 | inf | 2.375 | 1.91 | -7.35 | 91.92 | 12.89 | 11.62 | 94.31 |
| **SUM** | | | | | | | **38.63** | **55.12** | |

5 Cells:

| Cell | From | To | Bandgap | Voc | Jsc | FF | PCE | Gen_sun [1] | Gen_cell [2] |
|---|---|---|---|---|---|---|---|---|---|
| | (eV) | (eV) | (eV) | (V) | (mA/cm2) | (%) | (%) | (% of 1 sun) | (%) |
| 1 | 0 | 1.360 | 1.245 | 0.678 | -3.61 | 82.6 | 2.02 | 8.22 | 14.18 |
| 2 | 1.360 | 1.630 | 1.405 | 0.927 | -8.35 | 86.0 | 6.66 | 13.17 | 26.56 |
| 3 | 1.630 | 2.025 | 1.650 | 1.219 | -9.77 | 88.3 | 10.51 | 15.08 | 41.00 |
| 4 | 2.025 | 2.545 | 2.040 | 1.585 | -7.65 | 90.6 | 10.99 | 11.94 | 56.05 |
| 5 | 2.545 | inf | 2.560 | 2.076 | -5.08 | 92.6 | 9.78 | 8.10 | 92.86 |
| **SUM** | | | | | | | **39.96** | **56.51** | |

6 Cells:

| Cell | From | To | Bandgap | Voc | Jsc | FF | PCE | Gen_sun [1] | Gen_cell [2] |
|---|---|---|---|---|---|---|---|---|---|
| | (eV) | (eV) | (eV) | (V) | (mA/cm2) | (%) | (%) | (% of 1 sun) | (%) |
| 1 | 0 | 1.335 | 1.24 | 0.67 | -3.11 | 82.54 | 1.71 | 7.33 | 12.56 |
| 2 | 1.335 | 1.53 | 1.38 | 0.88 | -6.04 | 85.70 | 4.57 | 9.68 | 19.08 |
| 3 | 1.53 | 1.82 | 1.55 | 1.11 | -8.08 | 87.56 | 7.85 | 12.45 | 29.81 |
| 4 | 1.82 | 2.19 | 1.84 | 1.39 | -7.61 | 89.67 | 9.48 | 11.80 | 40.89 |
| 5 | 2.19 | 2.645 | 2.20 | 1.73 | -5.68 | 91.42 | 8.99 | 8.92 | 53.82 |
| 6 | 2.645 | inf | 2.65 | 2.16 | -4.07 | 92.98 | 8.19 | 6.52 | 91.24 |
| **SUM** | | | | | | | **40.79** | **56.70** | |

7 Cells:

| Cell | From | To | Bandgap | Voc | Jsc | FF | PCE | Gen_sun [1] | Gen_cell [2] |
|---|---|---|---|---|---|---|---|---|---|
| | (eV) | (eV) | (eV) | (V) | (mA/cm2) | (%) | (%) | (% of 1 sun) | (%) |
| 1 | 0 | 1.33 | 1.240 | 0.67 | -3.05 | 82.54 | 1.68 | 7.17 | 12.29 |
| 2 | 1.33 | 1.505 | 1.375 | 0.87 | -5.39 | 85.68 | 4.03 | 8.69 | 17.06 |
| 3 | 1.505 | 1.75 | 1.525 | 1.08 | -7.04 | 87.50 | 6.64 | 10.86 | 25.21 |
| 4 | 1.75 | 2.04 | 1.760 | 1.31 | -6.74 | 89.30 | 7.90 | 10.44 | 32.66 |
| 5 | 2.04 | 2.4 | 2.050 | 1.58 | -5.66 | 90.90 | 8.15 | 8.84 | 42.11 |
| 6 | 2.4 | 2.825 | 2.405 | 1.92 | -4.15 | 92.27 | 7.35 | 6.57 | 56.26 |
| 7 | 2.825 | inf | 2.830 | 2.32 | -2.59 | 93.41 | 5.60 | 4.17 | 89.07 |
| **SUM** | | | | | | | **41.35** | **56.73** | |

8 Cells:

| Cell | From | To | Bandgap | Voc | Jsc | FF | PCE | Gen_sun [1] | Gen_cell [2] |
|---|---|---|---|---|---|---|---|---|---|
| | (eV) | (eV) | (eV) | (V) | (mA/cm2) | (%) | (%) | (% of 1 sun) | (%) |
| 1 | 0 | 1.33 | 1.240 | 0.67 | -3.05 | 82.54 | 1.68 | 7.17 | 12.29 |
| 2 | 1.33 | 1.505 | 1.375 | 0.87 | -5.39 | 85.68 | 4.03 | 8.69 | 17.06 |
| 3 | 1.505 | 1.72 | 1.525 | 1.07 | -6.24 | 87.48 | 5.87 | 9.63 | 22.34 |
| 4 | 1.72 | 1.96 | 1.725 | 1.27 | -5.90 | 89.16 | 6.71 | 9.13 | 27.29 |
| 5 | 1.96 | 2.23 | 1.965 | 1.50 | -4.92 | 90.42 | 6.67 | 7.66 | 32.11 |
| 6 | 2.23 | 2.555 | 2.235 | 1.75 | -4.13 | 91.67 | 6.63 | 6.48 | 41.43 |
| 7 | 2.555 | 2.95 | 2.555 | 2.05 | -3.07 | 92.78 | 5.85 | 4.89 | 55.49 |
| 8 | 2.95 | inf | 2.950 | 2.42 | -1.89 | 93.69 | 4.30 | 3.07 | 87.98 |
| **SUM** | | | | | | | **41.74** | **56.72** | |

9 Cells:

| Cell | From | To | Bandgap | Voc | Jsc | FF | PCE | Gen_sun [1] | Gen_cell [2] |
|---|---|---|---|---|---|---|---|---|---|
| | (eV) | (eV) | (eV) | (V) | (mA/cm2) | (%) | (%) | (% of 1 sun) | (%) |
| 1 | 0 | 1.32 | 1.235 | 0.66 | -2.99 | 82.39 | 1.63 | 7.30 | 12.44 |
| 2 | 1.32 | 1.465 | 1.365 | 0.85 | -4.24 | 85.52 | 3.08 | 6.90 | 13.43 |
| 3 | 1.465 | 1.63 | 1.480 | 1.01 | -5.04 | 87.08 | 4.45 | 7.81 | 17.13 |
| 4 | 1.63 | 1.825 | 1.640 | 1.19 | -5.20 | 88.52 | 5.47 | 8.03 | 21.56 |
| 5 | 1.825 | 2.045 | 1.825 | 1.36 | -4.93 | 89.72 | 6.04 | 7.64 | 26.12 |
| 6 | 2.045 | 2.3 | 2.045 | 1.57 | -4.24 | 90.84 | 6.05 | 6.62 | 31.30 |
| 7 | 2.3 | 2.58 | 2.300 | 1.81 | -3.34 | 91.98 | 5.57 | 5.27 | 37.55 |
| 8 | 2.58 | 2.95 | 2.575 | 2.07 | -2.81 | 92.86 | 5.40 | 4.48 | 52.91 |
| 9 | 2.95 | inf | 2.950 | 2.42 | -1.89 | 93.69 | 4.30 | 3.07 | 87.98 |
| **SUM** | | | | | | | **41.99** | **57.11** | |

10 Cells:

| Cell | From | To | Bandgap | Voc | Jsc | FF | PCE | Gen_sun [1] | Gen_cell [2] |
|---|---|---|---|---|---|---|---|---|---|
| | (eV) | (eV) | (eV) | (V) | (mA/cm2) | (%) | (%) | (% of 1 sun) | (%) |
| 1 | 0 | 1.32 | 1.235 | 0.66 | -2.99 | 82.39 | 1.63 | 7.30 | 12.44 |
| 2 | 1.32 | 1.465 | 1.365 | 0.85 | -4.24 | 85.52 | 3.08 | 6.90 | 13.43 |
| 3 | 1.465 | 1.63 | 1.480 | 1.01 | -5.04 | 87.08 | 4.45 | 7.81 | 17.13 |
| 4 | 1.63 | 1.825 | 1.640 | 1.19 | -5.20 | 88.52 | 5.47 | 8.03 | 21.56 |
| 5 | 1.825 | 2.03 | 1.825 | 1.36 | -4.63 | 89.72 | 5.66 | 7.17 | 24.51 |
| 6 | 2.03 | 2.25 | 2.030 | 1.55 | -3.79 | 90.84 | 5.35 | 5.92 | 27.37 |
| 7 | 2.25 | 2.52 | 2.250 | 1.76 | -3.46 | 91.78 | 5.59 | 5.44 | 35.61 |
| 8 | 2.52 | 2.825 | 2.515 | 2.01 | -2.74 | 92.72 | 5.11 | 4.36 | 45.72 |
| 9 | 2.825 | 3.235 | 2.820 | 2.30 | -1.72 | 93.44 | 3.69 | 2.77 | 57.68 |
| 10 | 3.235 | inf | 3.235 | 2.68 | -0.85 | 94.34 | 2.15 | 1.40 | 88.78 |
| **SUM** | | | | | | | **42.19** | **57.09** | |

11 Cells:

| Cell | From | To | Bandgap | Voc | Jsc | FF | PCE | Gen_sun [1] | Gen_cell [2] |
|---|---|---|---|---|---|---|---|---|---|
| | (eV) | (eV) | (eV) | (V) | (mA/cm2) | (%) | (%) | (% of 1 sun) | (%) |
| 1 | 0 | 1.21 | 1.195 | 0.58 | -1.19 | 81.26 | 0.56 | 4.21 | 6.87 |
| 2 | 1.21 | 1.335 | 1.260 | 0.69 | -2.38 | 83.16 | 1.36 | 4.98 | 8.75 |
| 3 | 1.335 | 1.47 | 1.370 | 0.86 | -4.25 | 85.65 | 3.12 | 6.87 | 13.44 |
| 4 | 1.47 | 1.63 | 1.485 | 1.02 | -4.86 | 87.15 | 4.31 | 7.52 | 16.61 |
| 5 | 1.63 | 1.825 | 1.640 | 1.19 | -5.20 | 88.52 | 5.47 | 8.03 | 21.56 |
| 6 | 1.825 | 2.03 | 1.825 | 1.36 | -4.63 | 89.72 | 5.66 | 7.17 | 24.51 |
| 7 | 2.03 | 2.25 | 2.030 | 1.55 | -3.79 | 90.84 | 5.35 | 5.92 | 27.37 |
| 8 | 2.25 | 2.52 | 2.250 | 1.76 | -3.46 | 91.78 | 5.59 | 5.44 | 35.61 |
| 9 | 2.52 | 2.825 | 2.515 | 2.01 | -2.74 | 92.72 | 5.11 | 4.36 | 45.72 |
| 10 | 2.825 | 3.235 | 2.820 | 2.30 | -1.72 | 93.44 | 3.69 | 2.77 | 57.68 |
| 11 | 3.235 | inf | 3.235 | 2.68 | -0.85 | 94.34 | 2.15 | 1.40 | 88.78 |
| **SUM** | | | | | | | **42.38** | **58.68** | |

12 Cells:

| Cell | From | To | Bandgap | Voc | Jsc | FF | PCE | Gen_sun [1] | Gen_cell [2] |
|---|---|---|---|---|---|---|---|---|---|
| | (eV) | (eV) | (eV) | (V) | (mA/cm2) | (%) | (%) | (%) | (%) |
| 1 | 0.8 | 1.21 | 1.195 | 0.58 | -1.19 | 81.26 | 0.56 | 4.21 | 6.87 |
| 2 | 1.21 | 1.335 | 1.260 | 0.69 | -2.38 | 83.16 | 1.36 | 4.98 | 8.75 |
| 3 | 1.335 | 1.47 | 1.370 | 0.86 | -4.25 | 85.65 | 3.12 | 6.87 | 13.44 |
| 4 | 1.47 | 1.625 | 1.485 | 1.02 | -4.82 | 87.15 | 4.28 | 7.46 | 16.48 |
| 5 | 1.625 | 1.795 | 1.625 | 1.17 | -4.56 | 88.41 | 4.71 | 7.03 | 18.52 |
| 6 | 1.795 | 1.985 | 1.795 | 1.33 | -4.44 | 89.46 | 5.30 | 6.88 | 22.56 |
| 7 | 1.985 | 2.18 | 1.985 | 1.51 | -3.60 | 90.76 | 4.92 | 5.61 | 24.20 |
| 8 | 2.18 | 2.385 | 2.175 | 1.69 | -2.98 | 91.57 | 4.61 | 4.68 | 27.13 |
| 9 | 2.385 | 2.605 | 2.380 | 1.88 | -2.45 | 92.33 | 4.25 | 3.87 | 31.71 |
| 10 | 2.605 | 2.88 | 2.595 | 2.08 | -2.18 | 92.99 | 4.23 | 3.48 | 42.84 |
| 11 | 2.88 | 3.25 | 2.875 | 2.35 | -1.42 | 93.53 | 3.13 | 2.30 | 54.93 |
| 12 | 3.25 | 4 | 3.245 | 2.69 | -0.82 | 94.35 | 2.09 | 1.36 | 88.20 |
| **SUM** | | | | | | | **42.55** | **58.74** | |

**Supplementary References**